\documentclass[iop]{emulateapj}

\usepackage{lineno}
\usepackage{color}
\usepackage{hyperref}  
\usepackage{breakurl}  

\newcommand\ms{\ifmmode{\rm m\thinspace s^{-1}}\else m\thinspace s$^{-1}$\fi}
\newcommand\kms{\ifmmode{\rm km\thinspace s^{-1}}\else km\thinspace s$^{-1}$\fi}
\newcommand{\vni}[1]{\vskip 5pt\noindent{\bf #1}}

\shortauthors{Torres}
\shorttitle{Spectroscopic Survey of the Pleiades}

\begin{document} 
\submitted{Accepted for publication in The Astrophysical Journal}

\title{Long-term Spectroscopic Survey of the Pleiades Cluster: The Binary Population}

\author{Guillermo Torres}
\author{David W.\ Latham}
\author{Samuel N.\ Quinn}

\affiliation{Center for Astrophysics $\vert$ Harvard \& Smithsonian,
  60 Garden St., Cambridge, MA 02138, USA; gtorres@cfa.harvard.edu}

\begin{abstract} 

We present the results of a spectroscopic monitoring program of the
Pleiades region aimed at completing the census of spectroscopic
binaries in the cluster, extending it to longer periods than
previously reachable.  We gathered 6104 spectra of 377 stars between
1981 and 2021, and merged our radial velocities with 1151 measurements
from an independent survey by others started three years earlier.
With the combined data spanning more than 43 yr we have determined
orbits for some 30 new binary and multiple systems, more than doubling
the number previously known in the Pleiades. The longest period is
36.5~yr. A dozen additional objects display long-term trends in their
velocities, implying even longer periods. We examine the collection of
orbital elements for cluster members, and find that the shape of the
incompleteness-corrected distribution of periods (up to $10^4$ days)
is similar to that of solar-type binaries in the field, while that of
the eccentricities is different.  The mass-ratio distribution is
consistent with being flat. The binary frequency in the Pleiades for
periods up to $10^4$ days is $25 \pm 3$\%, after corrections for
undetected binaries, which is nearly double that of the field up to
the same period.  The total binary frequency including known
astrometric binaries is at least 57\%. We estimate the internal radial
velocity dispersion in the cluster to be $0.48 \pm 0.04$~\kms. We
revisit the determination of the tidal circularization period, and
confirm its value to be $7.2 \pm 1.0$~days, with an improved precision
compared to an earlier estimate.

\end{abstract}

\section{Introduction}
\label{sec:introduction}

Radial velocity (RV) studies of stars in the Pleiades cluster have
been carried out for well over a century, predating the classical
proper motion studies that were central to establishing the cluster's
membership \citep[e.g.,][]{Trumpler:1921, Hertzsprung:1947,
  Artyukhina:1969}. The earliest attempt to measure radial velocities
in the Pleiades seems to be that of \cite{Pickering:1896}, who
reported only briefly that observations with an objective prism
suggested the relative motions of the seven brightest members in the
group probably do not exceed 30~\kms. Other, more successful efforts
in the first few years of the 20th century include those of
\cite{Adams:1904}, \cite{Jung:1914}, and \cite{Hartmann:1914}. Since
then, numerous investigations of increasing measurement precision have
been made to establish the mean velocity of the cluster and its
internal dispersion, to aid in revealing its structure, and to search
for spectroscopic binaries.  Radial-velocity studies addressing one or
more these issues include those by \cite{Frost:1926},
\cite{Smith:1944}, \cite{Abt:1965}, \cite{Pearce:1975},
\cite{Liu:1991}, \cite{Morse:1991}, and \cite{Mermilliod:2009}, among
others.

Naturally, the earlier studies focused on the brighter stars in the
Pleiades, which are of spectral type B. Those happen to be the more
difficult objects because their spectral lines are usually broadened
due to rapid rotation, resulting in poor velocity
precision. Velocities for later-type stars of spectral type A and F
only began to appear with the work of \cite{Smith:1944}, which, like
most subsequent programs, ran for only one or two years or gathered
relatively few observations per star.  It is not surprising, then,
that no spectroscopic binaries in the Pleiades were known until the
late 1950's, when the first one was discovered: the double-lined
system HD\,23642 \citep{Pearce:1958, Abt:1958}, which is now known to
be eclipsing \citep{Miles:1999, Torres:2003}. More extensive studies
designed to look for spectroscopic binaries among the B, A, and F
stars were undertaken by \cite{Abt:1965} and \cite{Pearce:1975}.
Altogether they reported spectroscopic orbits for eight objects,
although unfortunately five of them have been shown to be spurious
\citep[see][]{Torres:2020b} and two others are non-members. The
challenging history of binary discovery in the Pleiades up to that
time is attributable to the difficulty of the measurements, as even
stars as late as mid F can still rotate quite rapidly in the cluster,
up to 100~\kms\ in some cases.  One other spectroscopic binary,
HD\,23631, was reported by \cite{Conti:1968}, and is a metallic-line
star, so its rotation is relatively slow.

Significant progress came with the long-term observing program carried
out by J.\ C.\ Mermilliod and collaborators \citep{Rosvick:1992,
  Mermilliod:1992a, Mermilliod:1997, Raboud:1998, Mermilliod:2009},
which ran for 20 years between the beginning of 1978 and the end of
1997. This study used the northern CORAVEL instrument
\citep{Baranne:1979} on the Swiss 1m telescope at the Haute-Provence
Observatory (France), and observed some 270 stars of spectral type
F5--K0, complementing earlier efforts that had focused on the hotter
objects.\footnote{A second CORAVEL instrument was installed on the
  Danish 1.54m telescope at La Silla Observatory (Chile), but was not
  used to observe the Pleiades. All references to CORAVEL in this
  paper will be understood hereafter to refer to the northern
  instrument.}  The CORAVEL program led to the discovery and
characterization of about a dozen new spectroscopic binaries with
orbital periods up to about two years.

An independent, long-term radial-velocity monitoring program at the
Center for Astrophysics (CfA) has been running for 39 years, beginning
in 1982, and is ripe for analysis. It is the subject of this paper. It
overlaps substantially both in time and in the sample of targets with
the CORAVEL effort, but continued for more than 20 years after the end
of those observations. It was designed to complete the census of
short-period binaries in the Pleiades that may have been missed by
earlier studies, and to extend the coverage to much longer orbital
periods, reaching into the regime in which some of them could be
spatially resolved by imaging techniques. This could enable dynamical
mass measurements that are exceedingly rare in the cluster.  The
sample of stars observed at the CfA is also larger (roughly 380
objects) and covers a wider range of spectral types than earlier
studies, from mid-B to early M. As we report below, the survey more
than doubles the number of known spectroscopic binaries in the cluster
with orbital solutions, several of them benefiting from the addition
of the CORAVEL velocities, which are of similar precision as
ours. Indeed, the combined CfA and CORAVEL data sets, which span more
than 43 years, permit a more comprehensive study of the binary
population in the Pleiades, including a redetermination of the binary
frequency, the distribution of orbital elements, the internal velocity
dispersion in the cluster, and a more robust determination of the
tidal circularization period. In this work we will therefore merge the
two samples together, with the goal of addressing all of these issues.

Most of the early-type stars of spectral type B and A on the CfA
observing list are rotating too rapidly for the standard
cross-correlation techniques that we apply in this work to yield
meaningful RVs. For this reason, results for 33 of those stars based
on a different methodology for measuring velocities have been
published separately \citep{Torres:2020b}, and include the discovery
of three new spectroscopic binaries with orbital solutions reported in
that work.  Nevertheless, these early-type stars are part of the
original sample, so they will be included in portions of the analysis
in this paper, which we have organized as follows.

In Section~\ref{sec:sample} we describe the CfA sample and how it was
assembled, and discuss membership in the cluster.  We also summarize
the properties of the CORAVEL sample that we incorporate into the
subsequent analysis. The spectroscopic observations and techniques for
radial velocity determination are described in
Section~\ref{sec:observations}, and in Section~\ref{sec:teffvsini} we
present our effective temperature and projected rotational velocity
estimates for all stars. The derivation of orbital elements for the
spectroscopic binaries and the discussion of a few triple systems is
found in Section~\ref{sec:orbits}. It is followed by a description of
systems that only show long-term trends, and additional binaries with
orbital elements published by others (Sections~\ref{sec:longperiod}
and \ref{sec:otherpublished}). The next two sections discuss our
criteria for radial velocity variability and our detection
completeness. Then in Section~\ref{sec:distribution} we analyze the
distributions of orbital periods, eccentricities, and mass ratios for
the binaries with orbits. With the larger number of binaries resulting
from this work, Section~\ref{sec:binaryfrequency} discusses the binary
frequency in the cluster, comparing it with estimates in other
populations. Astrometric binaries among the stars in our sample are
reported in Section~\ref{sec:astrometric}, and in
Section~\ref{sec:dispersion} we use the collection of RVs to
redetermine the mean velocity of the cluster, and for a detailed study
of the internal velocity dispersion. Section~\ref{sec:circularization}
then revisits the eccentricity versus log period diagram in the
Pleiades, which is a powerful indicator of the effectiveness of tidal
forces in binaries. The prospects for mass determinations for the
binaries in the Pleiades using astrometric information that will be
available at the conclusion of the Gaia mission are discussed in
Section~\ref{sec:masses}.  We summarize our findings in
Section~\ref{sec:summary}.

\section{Sample Selection}
\label{sec:sample}

The list of targets for the CfA survey of the Pleiades area was drawn
initially from an unpublished catalog of proper motion members and
suspected members assembled by John Stauffer and Charles Prosser at
the Center for Astrophysics, which was in turn compiled from various
classical sources \citep{Hertzsprung:1947, Artyukhina:1969, Haro:1982,
  vanLeeuwen:1986}. This list evolved over time to include more
confirmed members from other sources \citep{Jones:1981, Stauffer:1991,
  Prosser:1991, Stauffer:2007, Kamai:2014} in order to improve
completeness down to about $V = 14$, which was considered the
practical limit given the telescopes available for this work.
Brighter stars of spectral type B that are typically rotating very
rapidly were added later, when it was realized that the more advanced
instrumentation described below that we began using in 2009 would
allow meaningful velocities to be determined. The final sample of 377
stars covers the spectral type range from mid B to early M, and an
area of roughly $10\arcdeg \times 10\arcdeg$ on the sky.  While we
cannot claim completeness down to $V = 14$ because our sample relies
on a number of external sources of candidate members consulted over
the years, we are not aware of any selection biases against stars of a
specific spectral type (i.e., mass).  We list all targets in
Table~\ref{tab:sample}, with SIMBAD identifiers, coordinates, and
other information extracted from the Gaia EDR3 catalog
\citep{Fabricius:2020}. The 33 early-type stars whose velocities have
been reported in a separate publication are flagged with a ``b'' in
the table.

\begin{deluxetable*}{rlccrccccl}
\tablecaption{List of Objects in the Pleiades Region Observed Spectroscopically at the CfA.\label{tab:sample}}
\tablehead{
\colhead{} &
\colhead{Name} &
\colhead{R.A.} &
\colhead{Dec.} &
\colhead{Gaia ID} &
\colhead{$G$} &
\colhead{$\pi_{\rm Gaia}$} &
\colhead{$\mu_{\alpha} \cos\delta$} &
\colhead{$\mu_{\delta}$} &
\colhead{Mem}
\\
\colhead{} &
\colhead{} &
\colhead{(degree)} &
\colhead{(degree)} &
\colhead{} &
\colhead{(mag)} &
\colhead{(mas)} &
\colhead{(mas yr$^{-1}$)} &
\colhead{(mas yr$^{-1}$)} &
\colhead{}
}
\startdata
1  & AK III-31  & 51.88537 & +25.89995  &  117280133627745280  & \phn9.05  & $13.737 \pm 0.021$\phn & $+16.513 \pm 0.023$ &  $-29.410 \pm 0.020$ & NM \\
2  & PELS 121   & 51.92529 & +23.80351  &   68711956249760768  & 10.13  &  $5.237 \pm 0.015$ &  \phn$+4.479 \pm 0.016$ &  $-35.678 \pm 0.013$ & NM \\
3  & PELS 1     & 52.06861 & +22.64106  &   61842035801405824  & 10.41  &  $2.281 \pm 0.049$ & $+15.006 \pm 0.050$ &  $-24.159 \pm 0.042$ & NM \\
4  & AK III-59  & 52.16877 & +25.60767  &   69221506874062720  & 11.51  &  $6.334 \pm 0.020$ & $+24.659 \pm 0.025$ &  $-28.865 \pm 0.018$ & NM \\
5  & AK III-79  & 52.35599 & +25.65215  &   69595035885590144  & \phn9.33  &  $7.318 \pm 0.017$ & $+31.053 \pm 0.019$ &  $-26.596 \pm 0.016$ & NM \\
6  & AK III-158 & 52.80562 & +26.43951  &   69659730977217792  & \phn9.12 &  $10.624 \pm 0.021$\phn &  $+15.011 \pm 0.025$ &   $-50.421 \pm 0.017$ & NM \\
7  & AK III-153 & 52.81658 & +25.25526  &   69335619861034752  & \phn8.09 &   $7.617 \pm 0.033$ &  $+22.625 \pm 0.037$ &   $-46.757 \pm 0.025$ & b
\enddata

\tablecomments{ICRS coordinates, source identifiers, $G$-band
  magnitudes, parallaxes, and proper motion components are extracted
  from the Gaia EDR3 catalog. Objects that are not considered
  members of the cluster are indicated with ``NM'' in the last column
  (see Section~\ref{sec:membership}). Early-type stars in the group of
  33 with spectroscopic results reported previously
  \citep{Torres:2020b} are indicated with a ``b'' in the last column.
  (This table is available in its entirety in machine-readable form.)}

\end{deluxetable*}

\subsection{Membership}
\label{sec:membership}

The original selection of the stars in the CfA sample (including the
early-type stars) was based largely on proper motion and other
information to establish membership, but was made without the benefit
of the much more precise astrometry now available from the Gaia
mission. We therefore expected that a reevaluation at this time would
reveal that a subset of our targets are not actual cluster
members. This is borne out by the velocity measurements presented
later, which deviate significantly in a number of cases from the
cluster's mean velocity of about 5.7~\kms.  For a better assessment of
whether our stars belong to the Pleiades, we initially relied on the
list of probable members from the Gaia team \citep{Gaia:2018b}, which
was assembled from parallax and proper motion information from the
second data release (DR2), but without making use of radial
velocities. As with any membership list, there is always the
possibility that it is both contaminated and incomplete to some
degree.\footnote{A detailed investigation of Pleiades membership based
  on the more recent early third data release (EDR3) of Gaia
  \citep{Fabricius:2020} has not yet been made as of this writing, and
  is outside the scope of this paper. However, we do not expect it to
  be radically different from DR2, as the proper motions and
  parallaxes are quite similar, only more precise in EDR3.}  We find
that 229 of our 377 stars are included in that list. Two of them
(HII\,5 and HCG\,258) have high-quality astrometry and no indication
that they are binaries, but their radial velocities are several
\kms\ lower than the cluster mean, and their parallax and p.m.\ are at
the very edges of the distributions for the cluster. We have therefore
chosen to consider them as non-members.

Conversely, 148 of our targets have no entries in the Gaia membership
list, and this includes such well known bona fide members in our
sample as Electra (HII~468), Taygeta (HII~563), and Maia (HII~785). The
explanation for these particular cases is that their extreme
brightness has seriously impacted the precision of their astrometry,
and kept them off the Gaia list.  For about half of the 148 stars
missing in that list, the astrometry and/or our radial velocity
measurements clearly indicate they are non-members.  Of the remaining
half, six have no parallax or p.m.\ measurements in the Gaia EDR3
catalog, but their RVs are consistent with membership, so we have
tentatively retained them as members. For many of the others the
parallax or p.m.\ deviations are such that the choice is not clear,
although there is reason to believe their astrometry may be
compromised, causing them to have been excluded as members in the Gaia
team's analysis.  For example, a quantity known as the renormalized
unit weight error ({\tt RUWE}) is advocated as a robust statistical
indicator of the quality of the astrometric fit, with values in the
range 1--1.4 generally being considered acceptable \citep[see,
  e.g.,][]{Lindegren:2018}. Taking the {\tt RUWE} values from Gaia EDR3, we
find a median of ${\rm {\tt RUWE}} = 1.72$ for the stars in our sample
that are not in the Gaia membership list, versus 1.08 for the ones
that are. One possible reason for the inflated {\tt RUWE} values is
unrecognized binarity, and in fact we have found that several dozen of the
stars not on the list of Gaia members are either spectroscopic
binaries, many with long orbital periods, or else they have close visual
companions. We give these stars the benefit of the doubt, and do not
exclude them as members if their radial velocity is within 2~\kms\ of
the expected value. Most of the other missing objects have astrometry
consistent with membership, once we allow for the observed trend
between the parallaxes and the p.m.\ components illustrated in
Figure~6 of \cite{Gao:2019}. That trend is the result of Gaia being able to
resolve the depth of the cluster along the line of sight (the depth is
slightly less than 1/10 of its distance).

\begin{figure}
\epsscale{1.15}
\plotone{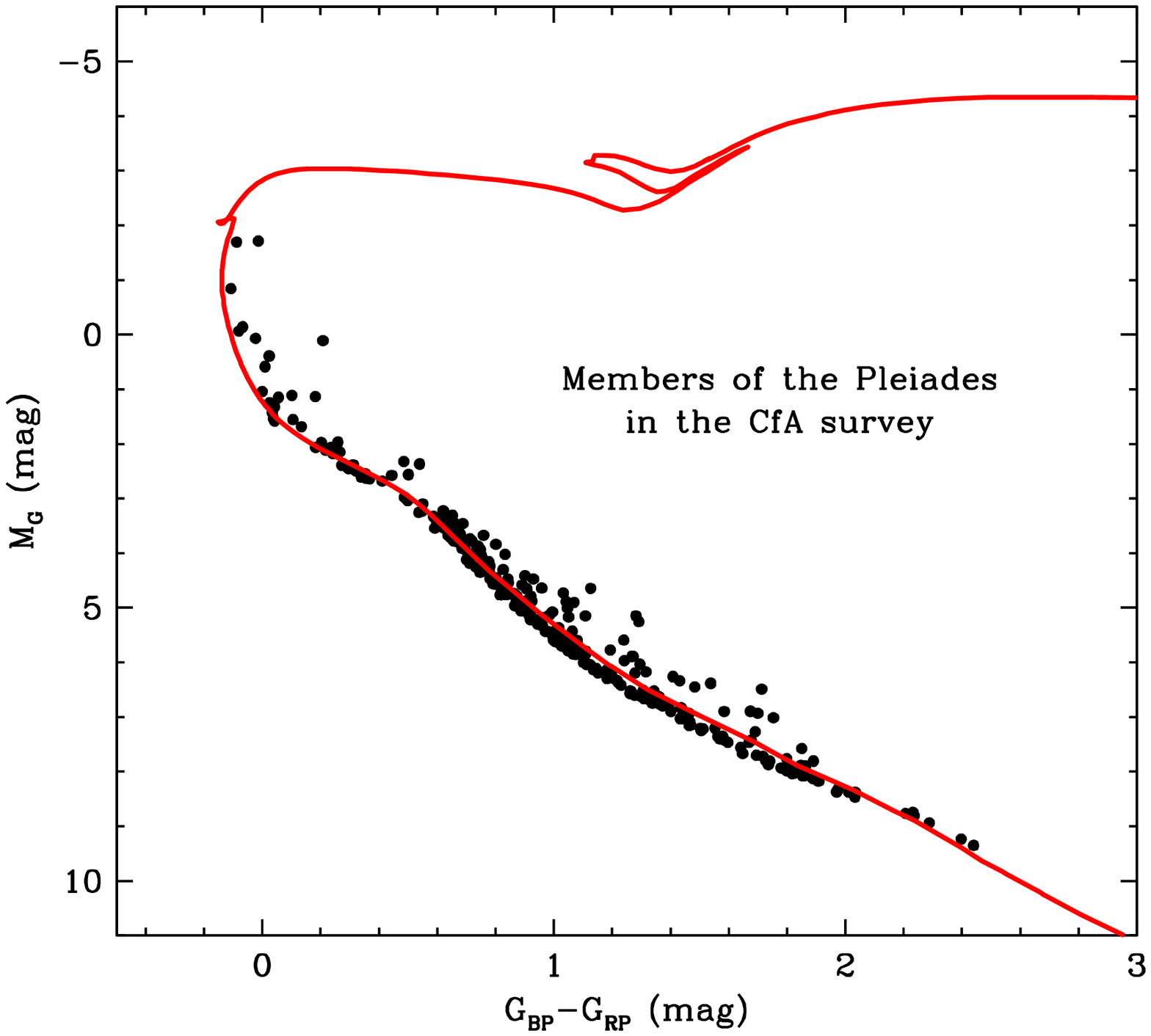}
\vskip 8pt
\plotone{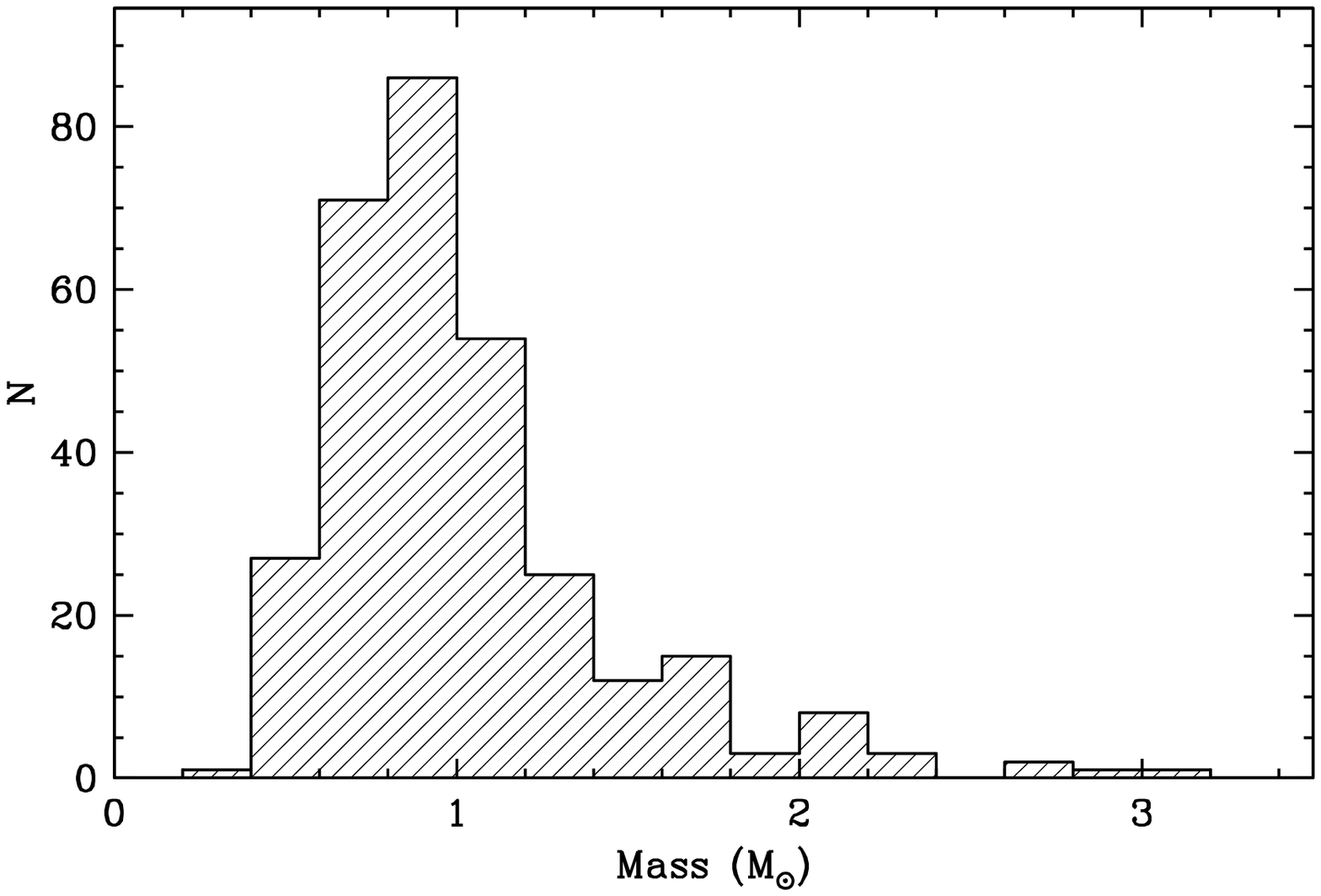}

\figcaption{\emph{Top:} Absolute $G$-band magnitude as a function of
  $G_{\rm BP}-G_{\rm RP}$ color from the Gaia EDR3 catalog for the
  Pleiades members in our sample. Also shown is a 125~Myr isochrone
  from the PARSEC~1.2S series of \cite{Chen:2014} for solar
  metallicity ($Z = 0.0152$ in these models). Interstellar extinction
  has been applied to the isochrone based on an assumed average
  reddening of $E(B-V) = 0.04$. The $M_G$ values rely on the
  individual parallax of each star. {\emph{Bottom:} Histogram of
    stellar masses, estimated from the $G_{\rm BP}-G_{\rm RP}$ colors
    and the same isochrone as above.}\label{fig:cmd}}

\end{figure}

All in all, we flagged 68 of the objects in the CfA sample as
non-members, and consider another four to be doubtful, but still
possible members (MT~41, HII~1653, TRU~S177, and AK~V-198). This leaves
309 stars out of our original sample of 377 that we consider to be
members or possible members. The non-members and doubtful members are
marked as such in Table~\ref{tab:sample}. The location of the members
in the color-magnitude diagram of the cluster is shown in
Figure~\ref{fig:cmd}, along with a 125~Myr solar-metallicity model
isochrone from the PARSEC~1.2S series \citep{Chen:2014}. The
  bottom panel shows the distribution of estimated stellar masses.

The CORAVEL program in the Pleiades observed a total of about 270
stars, and included many objects in the outer regions of the cluster
that had previously been proposed as possible members, but which had
never been observed spectroscopically. Not surprisingly, a good
fraction of those turned out to be non-members based on the CORAVEL
radial velocities or other information, and we have verified those
assessments using the parallaxes and proper motions now available from
Gaia EDR3. A total of 178 CORAVEL objects remain as members, of which
all but two (PELS~30 and PELS~39) were also observed at the CfA. For
completeness, we have chosen to add these two objects to our sample
for the analysis of the binary population in the cluster in later
sections of the paper.

\section{Observations}
\label{sec:observations}

Spectroscopic monitoring of the Pleiades region at the CfA began in
December of 1981, and was carried out with four instrument/telescope
combinations. From that initial date until February of 2009, spectra
were gathered with three nearly identical copies of the Digital
Speedometer \citep{Latham:1992a}. They were attached to the 1.5m
Tillinghast reflector at the Fred L.\ Whipple Observatory on Mount
Hopkins (AZ), the now closed 1.5m Wyeth reflector at the Oak Ridge
Observatory (MA), and the 4.5m-equivalent MMT also on Mount Hopkins,
before its conversion to a monolithic 6.5m telescope.  The Digital
Speedometers had a resolving power of $R \approx 35,\!000$, and were
equipped with photon-counting Reticon detectors that limited the
recorded output to a single echelle order about 45~\AA\ wide, centered
on the \ion{Mg}{1}\,b triplet near 5187~\AA. The signal-to-noise
ratios (S/N) of the 2441 usable spectra obtained at the three
telescopes range from 5 to about 100 per resolution element of
8.5~\kms, although at the higher levels the limitation is systematics
in the flatfield corrections rather than photon noise. Reduction
procedures have been described by \cite{Latham:1985, Latham:1992a}.

The stability of the velocity zero-point of each of these three
instruments was monitored by taking sky exposures at dusk and dawn,
and applying small velocity corrections to the raw velocities from run
to run that were typically smaller than
2~\kms\ \citep[see][]{Latham:1992a}. With these corrections the
velocities from the three telescopes were placed on the same native
CfA system, which is slightly offset from the IAU system by
0.14~\kms\ \citep{Stefanik:1999}, as determined from observations of
minor planets in the solar system. To remove this shift, a correction
of +0.14~\kms\ has been added to all our raw velocities from the
Digital Speedometers.

Beginning in October of 2009, further observations were collected with
the Tillinghast Reflector Echelle Spectrograph
\citep[TRES;][]{Furesz:2008, Szentgyorgyi:2007} on the 1.5m
Tillinghast reflector. This is a modern, bench-mounted, fiber-fed
spectrograph delivering a resolving power of $R \approx 44,\!000$, with
a CCD detector that records 51 echelle orders between 3800--9100~\AA.
We gathered 3663 spectra with this instrument through the end of the
2020/2021 observing season, with S/N ranging between 5 and 500 per
resolution element of 6.8~\kms.  Reductions were performed with a
dedicated pipeline \citep[see][]{Buchhave:2012}.  Observations of IAU
standard stars were made each run to monitor changes in the velocity
zero-point of TRES. This was done by cross-correlating the spectra of
each standard against a high-S/N observation of the same star.
Observations of asteroids were then used to translate the raw TRES
velocities to an absolute system, as done with the Digital
Speedometers.

\begin{figure}
\epsscale{1.15}
\plotone{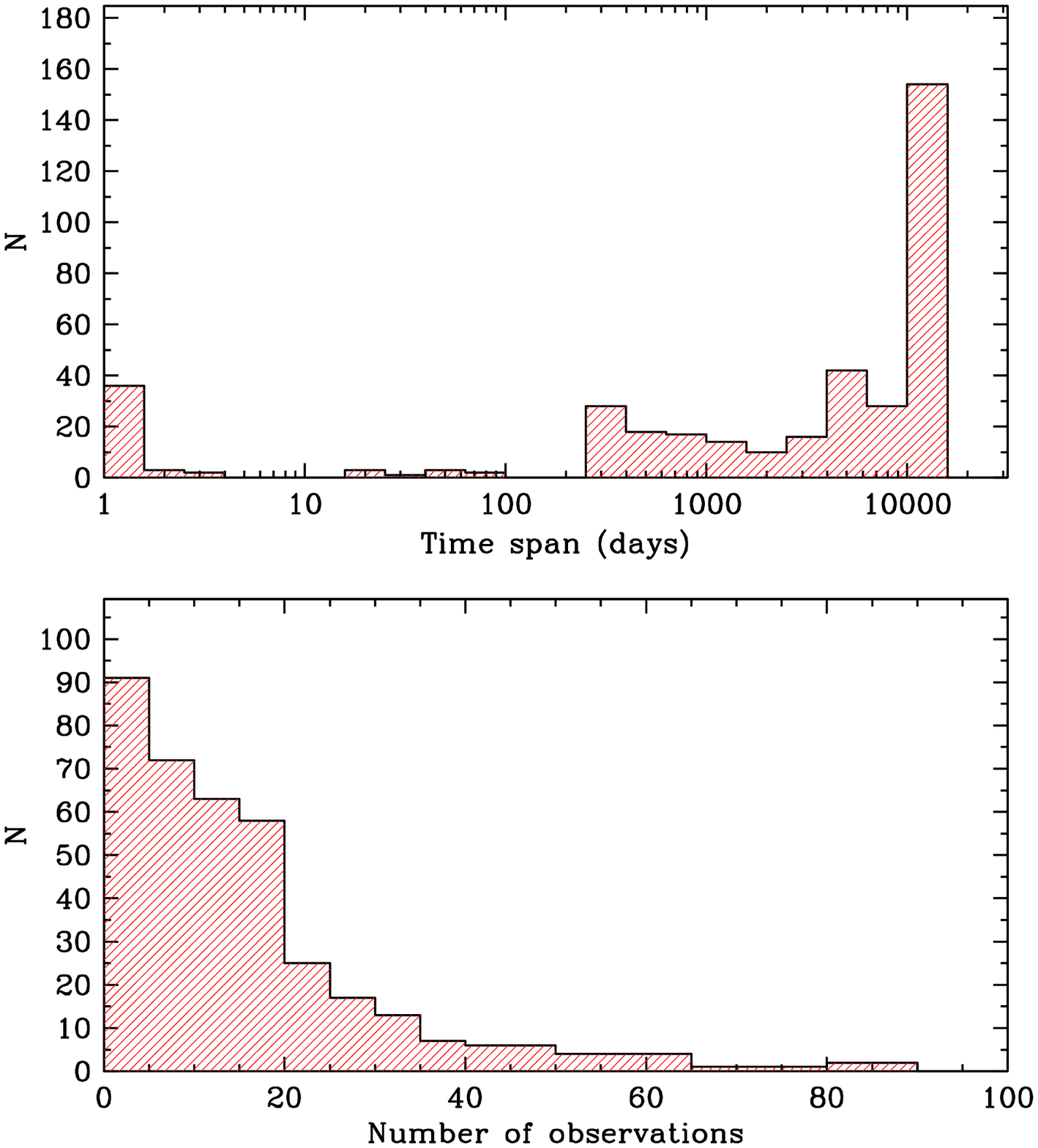}
\figcaption{Histograms of the time span and number of spectra per
  object for the 377 targets in the CfA survey of the
  Pleiades.\label{fig:nspan}}
\end{figure}

The distribution of the time span and number of observations for each
target are seen in Figure~\ref{fig:nspan}. About 40\% of our stars
have observations that span 30~yr or more, and for 60\% of the sample
the coverage exceeds 10~yr.

\subsection{Radial Velocities}
\label{sec:radialvelocities}

Radial velocities for all except the early-type stars treated
separately were obtained by cross-correlation using the {\tt XCSAO}
task running under IRAF.\footnote{IRAF is distributed by the National
  Optical Astronomy Observatory, which is operated by the Association
  of Universities for Research in Astronomy (AURA) under a cooperative
  agreement with the National Science Foundation.} This procedure also
reports an internal RV error estimate for each measurement that is a
function of the properties of the cross-correlation peak
\citep[see][]{Kurtz:1998}. For the TRES instrument, we used only the
order centered on the \ion{Mg}{1}\,b triplet so as to match the
spectral region of the Digital Speedometers. Templates were taken from
a pre-computed library of calculated spectra based on model
atmospheres by R.\ L.\ Kurucz, and a line list tuned to better match
real stars \citep[see][]{Nordstrom:1994, Latham:2002}.  The templates
are available over wide ranges in effective temperature ($T_{\rm
  eff}$), rotational broadening ($V_{\rm rot}$), metallicity ([m/H]),
and surface gravity ($\log g$). Gaussian broadening was applied to
match the resolution of each instrument. The optimal synthetic
template for each star was derived as described by
\cite{Torres:2002}. Briefly, this involved running grids of
correlations over wide intervals for the template parameters, and
selecting the ones providing the largest cross-correlation coefficient
averaged over all exposures. We assumed ${\rm [m/H]} = 0.0$, as the
composition of the Pleiades is very near solar
\citep[e.g.,][]{Taylor:2008, Soderblom:2009, Schuler:2010,
  Netopil:2016}, and $\log g = 4.5$, which is appropriate for the vast
majority of our stars.

Velocities for double-lined binaries were measured with {\tt TODCOR}
\citep{Zucker:1994}, a two-dimensional cross-correlation technique
that uses two templates, one for each component. There is one
triple-lined object in our sample (HII\,1338), and for this target we
used {\tt TRICOR}, which is an extension of {\tt TODCOR} to three
dimensions \citep{Zucker:1995}. The selection of templates for double-
and triple-lined systems followed the general procedure described
before. {\tt TODCOR} and {\tt TRICOR} also yield the light ratios
among the components, which we report below in
Section~\ref{sec:orbits}.

Because of the limited number of lines in the narrow spectral region
recorded with the Digital Speedometers (45~\AA), systematic errors in
the velocities for double-lined binaries can occur as lines of the
components shift in and out of the window in opposite directions at
different orbital phases. Experience has shown that the effect can be
several \kms\ in some cases, but is negligible for TRES, which has
more than twice the spectral coverage (100~\AA).  We corrected these
velocity errors by performing numerical simulations, as described by
\cite{Latham:1996}. See also \cite{Torres:1997}.

\begin{figure}
\epsscale{1.15}
\plotone{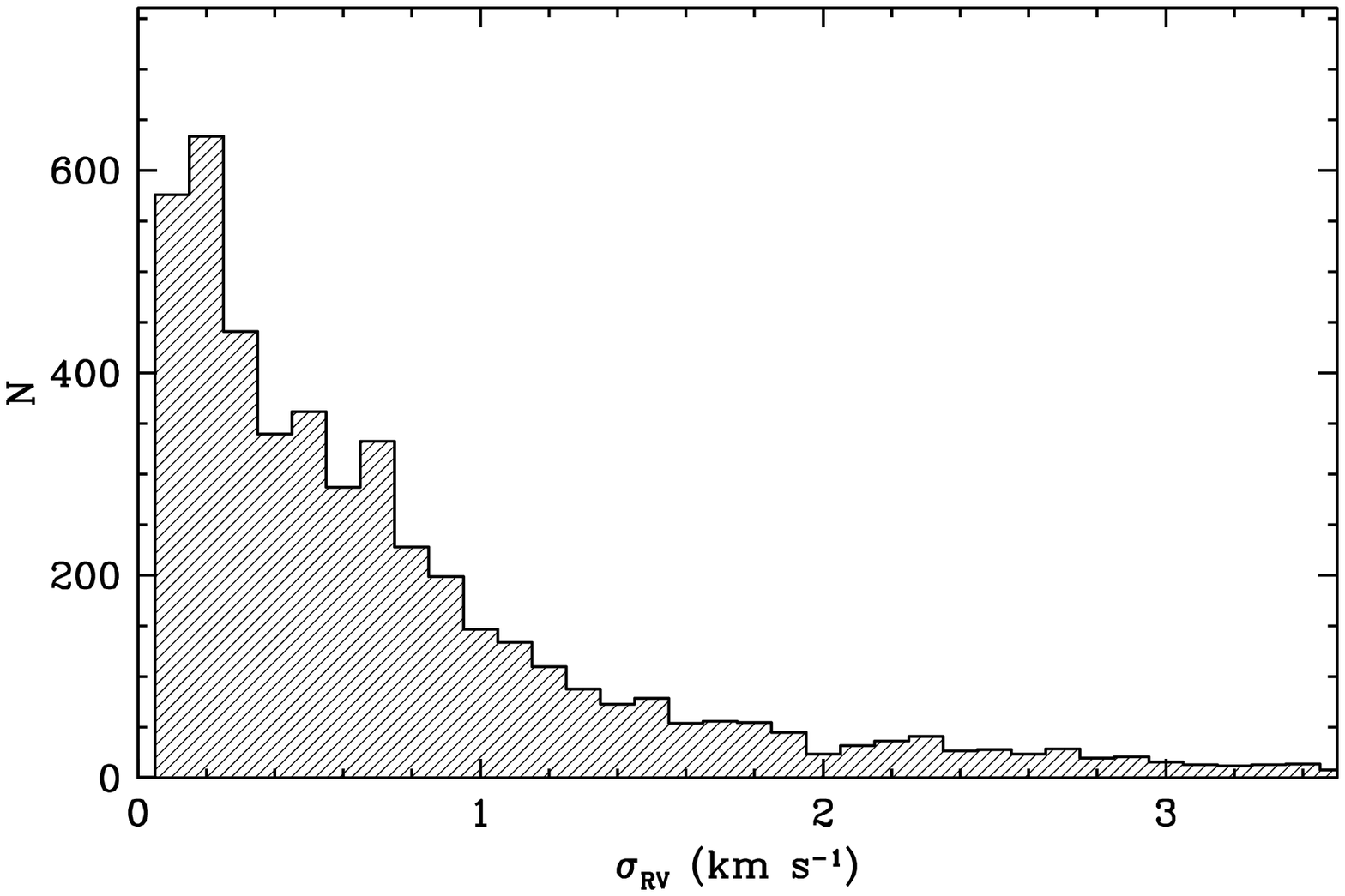}
\vskip 7pt
\plotone{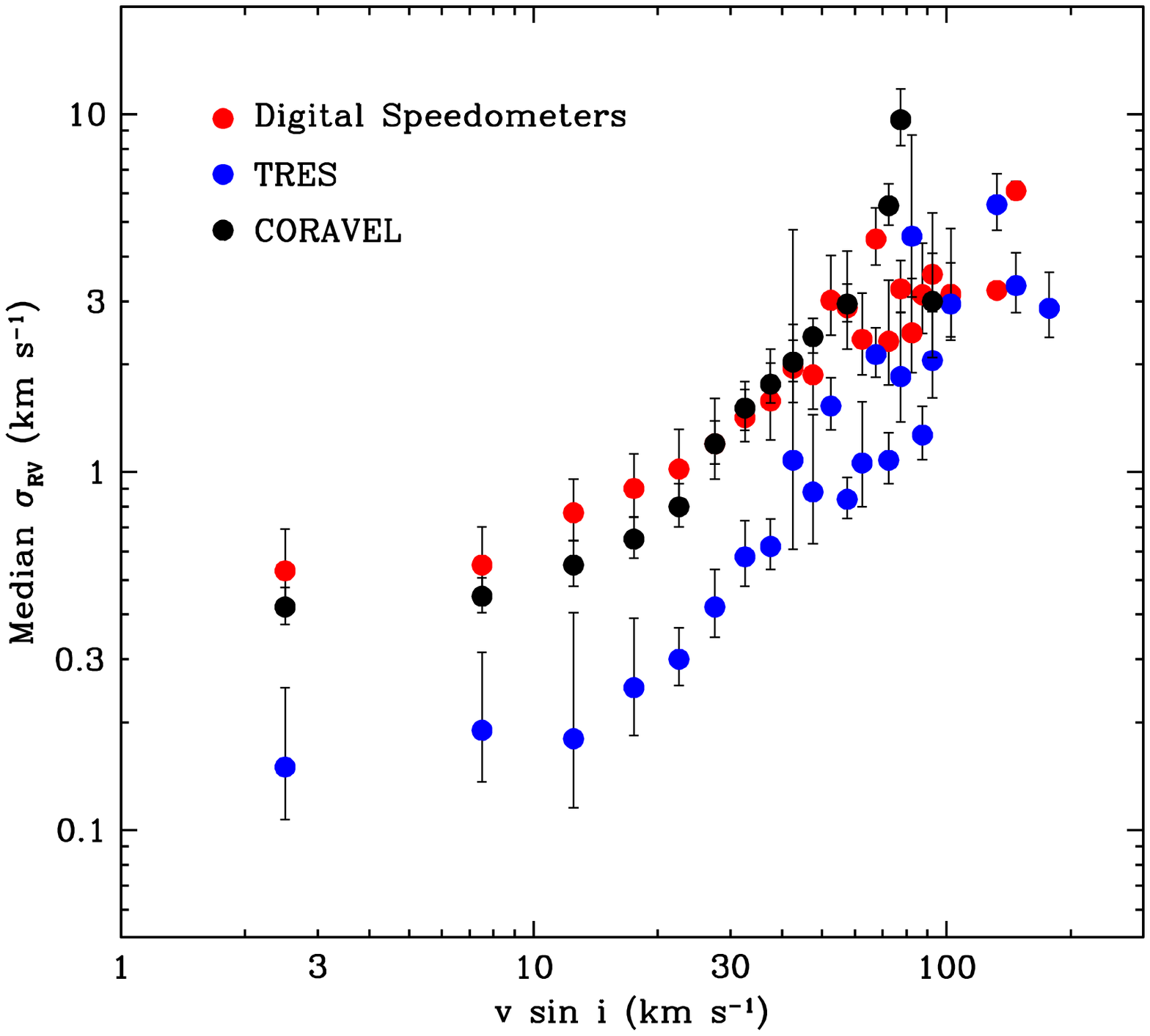}
\figcaption{\emph{Top:} Distribution of the individual internal radial
  velocity uncertainties for stars with single-lined spectra observed
  at the CfA, as reported by the IRAF task {\tt XCSAO}. \emph{Bottom:}
  Radial velocity precision for the Digital Speedometers, TRES, and
  CORAVEL as a function of rotational broadening (see
  Section~\ref{sec:teffvsini}), in bins of 5~\kms. The points show the
  median of each bin, with error bars calculated as half of the
  interquartile range.\label{fig:sigma}}
\end{figure}

For single-lined objects with sharp lines and spectral type similar to
the Sun, typical internal velocity precisions reported by {\tt XCSAO}
are about 0.5~\kms\ for the Digital Speedometers, and
$\sim$0.15~\kms\ for TRES using the \ion{Mg}{1}\,b order and the
templates described above. We note that TRES is capable of
significantly better velocity precision for certain applications such
as exoplanet follow-up, in which absolute velocities are typically not
needed.  One can then take advantage of the full wavelength coverage
afforded by the instrument, by cross-correlating multiple echelle
orders against a spectrum of the same star used as a template, to
measure relative velocities. For the present work, however, we require
absolute velocities, so we have chosen to use synthetic templates and
the single \ion{Mg}{1}\,b order, which is rich in RV information. This
is sufficient for our purposes, given that we expect many of the
Pleiades stars to exhibit activity-driven apparent variations of at
least many tens of \ms\ by virtue of the young age of the
cluster. Moreover, rapid rotation is also common among these stars,
which will further degrade the precision.  Figure~\ref{fig:sigma}
shows how the errors increase as the stars rotate more rapidly,
reaching 3--4~\kms\ at $V_{\rm rot} \approx 100~\kms$.

Individual heliocentric radial velocities on the IAU system are listed
for all single-lined objects in Table~\ref{tab:RV1},\footnote{We
  exclude from this table the velocities for HII~2147, which were
  published separately by \cite{Torres:2020}, as well as those for the
  33 rapidly-rotating stars reported by \cite{Torres:2020b}.} and for
double-lined binaries (SB2s) in Table~\ref{tab:RV2}. The latter table
incorporates the corrections for systematics mentioned above for all
velocity measurements from the Digital Speedometers. Both tables also
list the RV uncertainties, and the S/N of each exposure. Statistics
for the 377 objects in the sample are presented in
Table~\ref{tab:meanRV}, and include the time span, the number of
observations from the Digital Speedometers and TRES, the weighted
average velocity and corresponding uncertainty (including the CORAVEL
measurements; see below), and other information described later.

\setlength{\tabcolsep}{8pt}
\begin{deluxetable}{lcccc}
\tablecaption{CfA Radial Velocity Measurements for Single-lined Objects.\label{tab:RV1}}
\tablehead{
\colhead{Name} &
\colhead{HJD} &
\colhead{$RV$} &
\colhead{S/N} &
\colhead{Inst}
\\
\colhead{} &
\colhead{(2,400,000+)} &
\colhead{(\kms)} &
\colhead{} &
\colhead{}
}
\startdata
PELS 121  &   52497.8622  &  $5.67 \pm 0.50$  &  11  &  1 \\
PELS 121  &   52529.8427  &  $6.32 \pm 0.75$  &  11  &  1 \\
PELS 121  &   52583.7015  &  $5.59 \pm 0.52$  &  10  &  1 \\
PELS 121  &   56672.6857  &  $5.60 \pm 0.09$  &  54  &  2 \\
PELS 121  &   56977.8214  &  $5.60 \pm 0.13$  &  40  &  2
\enddata

\tablecomments{The heliocentric radial velocities are on the reference frame of the
minor planets in the Solar
System \citep[see][]{Stefanik:1999}. Signal-to-noise ratios in the S/N
column are per resolution element. Codes in the Inst column are 1 for
the Digital Speedometers and 2 for TRES. HII~2147 and rapidly-rotating
early type stars are not included, as their velocities have been
published separately by \cite{Torres:2020} and \cite{Torres:2020b}, respectively.
(This table is available in its entirety in machine-readable form.)}

\end{deluxetable}
\setlength{\tabcolsep}{6pt}  

\setlength{\tabcolsep}{1.5pt}
\begin{deluxetable}{lccccc}
\tablecaption{CfA Radial Velocity Measurements for Double-lined Objects.\label{tab:RV2}}
\tablehead{
\colhead{Name} &
\colhead{HJD} &
\colhead{$RV_1$} &
\colhead{$RV_2$} &
\colhead{S/N} &
\colhead{Inst}
\\
\colhead{} &
\colhead{(2,400,000+)} &
\colhead{(\kms)} &
\colhead{(\kms)} &
\colhead{} &
\colhead{}
}
\startdata
AK III-31  &  45695.6647  & \phs$48.32 \pm 1.31$ &  $-66.64 \pm  15.84$     &   9  &  1 \\
AK III-31  &  45711.6059  & \phn$-7.39 \pm 1.49$ &  \phs$20.72 \pm  18.11$  &   8  &  1 \\
AK III-31  &  45757.4924  & $-47.10 \pm 1.15$    &  \phs$98.40 \pm  13.93$  &  10  &  1 \\
AK III-31  &  52879.8425  & $-11.59 \pm 0.85$    &  \phs$25.42 \pm  10.33$  &  14  &  1 \\
AK III-31  &  52921.8160  & \phs$52.62 \pm 0.85$ &  $-78.92 \pm  10.33$     &  14  &  1
\enddata

\tablecomments{The heliocentric radial velocities are on the reference frame of the
minor planets in the Solar
System \citep[see][]{Stefanik:1999}. Signal-to-noise ratios in the S/N
column are per resolution element. Codes in the Inst column are 1 for
the Digital Speedometers and 2 for TRES. (This table is available in
its entirety in machine-readable form.)}

\end{deluxetable}
\setlength{\tabcolsep}{6pt}  

\setlength{\tabcolsep}{10pt}
\begin{deluxetable*}{llcccccccl}
\tablecaption{Radial Velocity Statistics for Objects in the Pleiades
  Region Observed Spectroscopically at the CfA.\label{tab:meanRV}}
\tablehead{
\colhead{} &
\colhead{Name} &
\colhead{Time Span} &
\colhead{$N_{\rm DS}$} &
\colhead{$N_{\rm TRES}$} &
\colhead{$N_{\rm COR}$} &
\colhead{Mean $RV$} &
\colhead{$P(\chi^2)$} &
\colhead{$e/i$} &
\colhead{Flags}
\\
\colhead{} &
\colhead{} &
\colhead{(days)} &
\colhead{} &
\colhead{} &
\colhead{} &
\colhead{(\kms)} &
\colhead{} &
\colhead{} &
\colhead{}
}
\startdata
  1* & AK III-31  &  13481 & 65 &  1 & 27 & \phs$13.72 \pm 4.67$     &   0.00000  &  75.938     &  NM, SB2 \\
  2  & PELS 121   &  12846 &  3 &  8 &  4 & \phs\phn$5.56 \pm 0.14$  &   0.92131  &  \phn0.269  &  NM \\
  3  & PELS 1     &      0 &  1 &  0 &  0 & $-60.38 \pm 2.69$        &  \nodata   & \nodata     &  NM \\
  4  & AK III-59  &   8740 &  2 &  7 &  3 & \phs\phn$5.14 \pm 0.17$  &   0.27167  &  \phn0.441  &  NM \\
  5* & AK III-79  &   6877 &  6 &  0 &  0 & \phs$10.35 \pm 1.91$     &   0.06339  &  \phn1.123  &  NM \\
  6  & AK III-158 &    20  &  4 &  0 &  0 & \phn$-5.01 \pm  2.67$    &   0.64266  &  \phn0.564  &  NM 
\enddata

\tablecomments{The columns give the time span of the CfA observations
followed by the number of radial velocity measurements from the
Digital Speedometers, TRES, and CORAVEL, and the weighted mean RV
including the CORAVEL measurements, when available. For objects with
more than one measurement we then list the $\chi^2$ probability and
the $e/i$ metric for velocity variability, described in
Section~\ref{sec:variability}. The last column indicates non-members
(NM), and known binaries with spectroscopic orbits (SB1, SB2, SB3),
astrometric orbits (AST), or long-term trends (L), as reported in
Sections~\ref{sec:orbits}, \ref{sec:longperiod},
and \ref{sec:otherpublished}. Early-type stars with spectroscopic
results published previously \citep{Torres:2020b} are included for
completeness, and are flagged with a ``b'' in the last column. Stars
with notes of interest given in the Appendix are indicated with an
asterisk following the running number before the name. (This table is
available in its entirety in machine-readable form.)}

\end{deluxetable*}
\setlength{\tabcolsep}{6pt}  

As indicated before, the CORAVEL observations have similar
uncertainties as ours (see Figure~\ref{fig:sigma}), and several of the
binary orbital solutions presented below in Section~\ref{sec:orbits}
are helped by including those velocities.  Most of those measurements
appeared in original form in a series of papers over a period of seven
years \citep{Rosvick:1992, Mermilliod:1992a, Mermilliod:1997,
  Raboud:1998}, and the final catalog of Pleiades velocities from the
CORAVEL team that included additional measurements was published later
by \cite{Mermilliod:2009}. The first three of the original sources
described how they adjusted the RV measurements for zero-point
differences to place them on the reference frame defined by
\cite{Mayor:1985}, which corresponds to the faint IAU standard system
\citep[$V > 4.3$; see][]{Stefanik:1999}. The study of
\cite{Raboud:1998} did not specify the zero-point, but indicated that
the entire CORAVEL data set was in the process of being recalibrated
for zero-point errors and color effects. By the time of the
publication by \cite{Mermilliod:2009}, the recalibration had been
completed and the full set of Pleiades velocities was placed on the
system defined by \cite{Udry:1999}, based on more precise observations
of IAU standards with a different instrument
\citep[ELODIE;][]{Baranne:1996}. The velocities of
\cite{Mermilliod:2009} supersede and sometimes augment those in the
original sources, and are systematically higher by 0.3--0.5~\kms.  The
total number of CORAVEL measurements incorporated into our analysis in
this paper is 1151. When added to our own 6104 observations from TRES
and the Digital Speedometers, this brings the total number of
individual radial-velocity epochs to 7255.

\begin{figure}
\epsscale{1.15}
\plotone{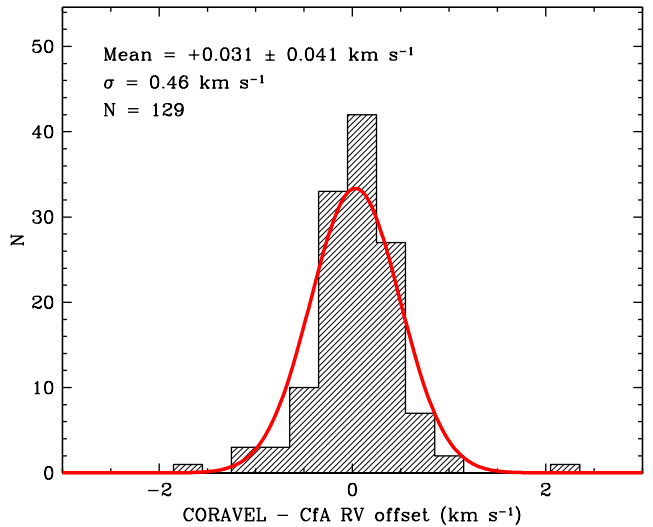}
\figcaption{Differences between the weighted mean velocities from
  CORAVEL and CfA for 129 stars in common that are not known to be
  binaries, and have 3 or more measurements each. Two outliers greater
  than 4~\kms\ in absolute value have been removed, as they may be
  previously unrecognized binaries. The curve represents a Gaussian
  fit. \label{fig:cor-cfa}}
\end{figure}

A comparison between the mean CORAVEL and CfA velocities for 129 stars
in common that are not known to be binaries, and have 3 or more
measurements in each data set, indicates a systematic difference of
only ${\rm CORAVEL-CfA} = +0.031 \pm 0.041~\kms$, with a standard
deviation of 0.46~\kms\ (see Figure~\ref{fig:cor-cfa}). There is no
significant dependence of the differences on either effective
temperature or rotational velocity. Given the excellent agreement in
the zero-points, we proceed below to combine the data with no further
adjustments. The number of CORAVEL observations for each star is
indicated in Table~\ref{tab:meanRV}.

\section{Effective Temperatures and Rotational Velocities}
\label{sec:teffvsini}

In the process of selecting the best synthetic template for the
cross-correlations, we made an estimate of the effective temperature
and projected rotational velocity of each star to a finer resolution
than the sampling of our library of templates, which is 250\,K in
temperature and is variable in $V_{\rm rot}$ (smaller steps for slow
rotators, increasing for fast rotators). We did this by interpolating
among neighboring templates independently for the spectra from the
Digital Speedometers and TRES, and found excellent agreement. In
general we have adopted the TRES results on account of the higher
quality and greater wavelength coverage of those observations.  We
report these measurements in Table~\ref{tab:teffvsini}, for both the
single-lined and double-lined objects. The uncertainties are based on
the scatter of the individual spectra, with an adopted floor of 100\,K
in temperature and 2~\kms\ in the rotational velocity when it is below
our spectral resolution. For completeness, we include the rotational
velocities for the early-type stars reproduced from the work of
\cite{Torres:2020b}. In a few cases (HII~1284, HII~1876, and TRU~S194) we
have been able to estimate the temperatures of those rapidly rotating
objects using the techniques in this paper.

\begin{figure}
\epsscale{1.15}
\plotone{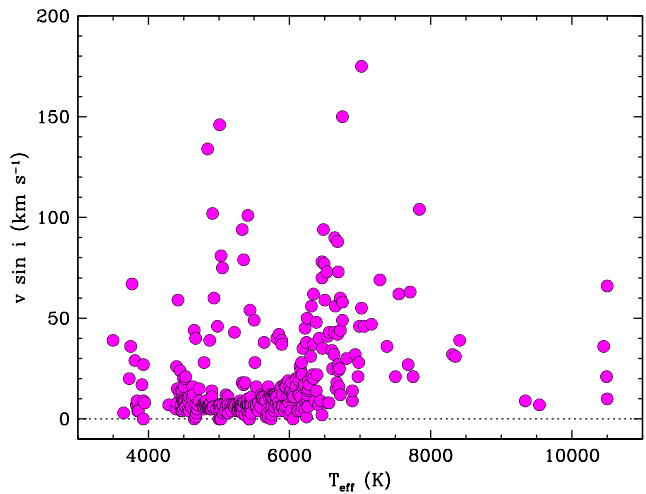}
\figcaption{Effective temperatures and projected rotational velocities
  for all objects in the CfA survey with single-lined
  spectra. \label{fig:teffvsini}}
\end{figure}

Until now we have been referring to our estimates of the projected
rotational velocity as $V_{\rm rot}$, rather than $v \sin i$. We note
that strictly speaking the measurements include other sources of line
broadening aside from rotation, such as macroturbulence ($\zeta_{\rm
  RT}$), which we do not attempt to account for. Our synthetic spectra
were calculated with a value of $\zeta_{\rm RT} = 2~\kms$, which
should be representative of solar-type and cooler stars. Earlier-type
stars, on the other hand, are expected to have larger values, but they
are also rotating more rapidly, so that rotation ends up being the
dominant broadening mechanism. Additionally, the parameter $V_{\rm
  rot}$ used for the library of synthetic templates is the equatorial
velocity for a model star viewed equator-on. There may be systematic
differences with $v \sin i$ as rotation increases and stars are viewed
more pole-on. For example, gravity darkening is not included. For our
sample, we do not expect these differences to be significant given the
reduced precision of the measurements for rapid rotators, so for the
remainder of the paper we will use the more familiar notation $v \sin
i$ for all our rotational velocity estimates.
Figure~\ref{fig:teffvsini} shows the distribution of $T_{\rm eff}$ and
$v \sin i$ for the CfA sample.  Objects reported previously by
\cite{Torres:2020b} are excluded from this figure, as they generally
have no temperature estimates. Objects hotter than about 6000~K tend
to rotate more rapidly, although a number of cooler objects can have
very large $v \sin i$ values as well. These are sometimes referred to
as ultrafast rotators \citep[see, e.g.,][]{Soderblom:1993}.

\setlength{\tabcolsep}{5pt}
\begin{deluxetable}{rlccccc}
\tablecaption{Effective Temperatures and Projected Rotational Velocities.\label{tab:teffvsini}}
\tablehead{
\colhead{} &
\colhead{Name} &
\colhead{$T_{\rm eff}$} &
\colhead{$v \sin i$} &&
\colhead{$T_{\rm eff}$} &
\colhead{$v \sin i$}
\\
\colhead{} &
\colhead{} &
\colhead{(K)} &
\colhead{(\kms)} &&
\colhead{(K)} &
\colhead{(\kms)}
}
\startdata
&& \multicolumn{2}{c}{Primary} && \multicolumn{2}{c}{Secondary} \\ [0.5ex]
\cline{3-4} \cline{6-7} \\ [-1ex]
 1 & AK III-31 & $5770 \pm 100$ & $10 \pm 2$    &&     4250*      &    6*      \\  
 2 & PELS 121  & $6120 \pm 100$ & \phn$4 \pm 2$ &&    \nodata     & \nodata    \\  
 3 & PELS 1    & $8350 \pm 150$ & $31 \pm 3$    &&    \nodata     & \nodata    \\  
 4 & AK III-59 & $5330 \pm 100$ & \phn$3 \pm 2$ &&    \nodata     & \nodata    \\  
 5 & AK III-79 & $6530 \pm 300$ & $73 \pm 4$    &&    \nodata     & \nodata        
\enddata

\tablecomments{Quantities flagged with an asterisk were
adopted based on external information. Tertiary parameters adopted for
HII~1338 are $T_{\rm eff} = 5250$~K and $v \sin i = 0$~\kms. The
temperature determination for the Am secondary of HII~1431 used ${\rm
[Fe/H]} = +0.5$. Rotational velocities for rapidly-rotating early type
stars are reproduced from the work of \cite{Torres:2020b}. Parameters
for the secondary of HII~2147 are from \cite{Torres:2020}. (This
table is available in its entirety in machine-readable form.)}

\end{deluxetable}
\setlength{\tabcolsep}{6pt}  

\section{Orbital Solutions for Spectroscopic Binaries}
\label{sec:orbits}

\subsection{Single- and Double-lined Binaries}

The combination of our radial velocity measurements with those of
\cite{Mermilliod:2009} has allowed us to discover many new
spectroscopic binaries in our sample. For the single-lined systems
(SB1) we performed weighted least-squares orbital solutions solving
for the standard elements $P$ (orbital period), $\gamma$
(center-of-mass velocity), $K_1$ (primary velocity semiamplitude), $e$
and $\omega_1$ (eccentricity and argument of periastron for the
primary), and $T_0$ (a reference time of periastron passage, for
eccentric orbits, or a time of maximum primary velocity, for circular
orbits). Orbits were assumed to be circular when initial solutions
solving for $e$ gave a value that was not statistically
significant. Relative weights for the observations were taken to be
inversely proportional to $\sigma_{\rm RV}^2$, where $\sigma_{\rm RV}$
is the formal error of the measurement. As those uncertainties can
sometimes be underestimated (or overestimated), we solved
simultaneously for a multiplicative scale factor $F$ for the
uncertainties in order to achieve reduced $\chi^2$ values near
unity. The orbital elements are collected in Table~\ref{tab:sb1}, and
include the derived quantities $f(M)$ (the standard mass function),
the coefficient of $M_2 \sin i$ (the minimum secondary mass), and $a_1
\sin i$ (the projected linear semimajor axis of the primary). The
number of observations, the standard deviation of the residuals, and
the error scaling factor are listed as well, and non-members are
flagged at the end of each line.

\setlength{\tabcolsep}{4pt}
\begin{deluxetable*}{r@{\hskip 1ex}lccccccccccc}
\tablecaption{Orbital Elements for the Single-lined Binaries in the Sample.\label{tab:sb1}}
\tablehead{
\colhead{} &
\colhead{Name} &
\colhead{$P$} &
\colhead{$\gamma$} &
\colhead{$K_1$} &
\colhead{$e$} &
\colhead{$\omega_1$} &
\colhead{$T_0$ (HJD)} &
\colhead{$f(M)$} &
\colhead{$M_2 \sin i$} &
\colhead{$a_1 \sin i$} &
\colhead{$\sigma$ (\kms)} &
\colhead{$F$}
\\
\colhead{} &
\colhead{} &
\colhead{(day)} &
\colhead{(\kms)} &
\colhead{(\kms)} &
\colhead{} &
\colhead{(degree)} &
\colhead{(2,400,000+)} &
\colhead{($M_{\sun}$)} &
\colhead{($M_{\sun}$)} &
\colhead{($10^6$ km)} &
\colhead{$N_{\rm obs}$} &
\colhead{}
}
\startdata
15  & PELS~7      &        7322. &        2.898 &        2.168 &        0.506 &        203.4 &       57996. &      0.00496 &       0.1705 &        188.2          &  0.099 & 0.786 \\ 
    &             &         109. &        0.048 &        0.044 &        0.013 &          2.1 &          23. &      0.00036 &       0.0041 &          5.2          &     33 &  NM   \\ [1.5ex]
         
26  & AK~II-346   &        6123. &         4.65 &         4.40 &        0.030 &         114. &       54024. &        0.054 &        0.378 &         371.          &  1.073 & 1.219 \\ 
    &             &         123. &         0.28 &         0.44 &        0.083 &         160. &        2754. &        0.016 &        0.038 &          37.          &     63 &       \\ [1.5ex]
         
44  & AK~III-664  &     14.07286 &         4.06 &        15.64 &        0.313 &        105.4 &    53383.133 &      0.00478 &       0.1685 &        2.875          &  0.512 & 1.349 \\ 
    &             &      0.00095 &         0.13 &         0.17 &        0.011 &          2.2 &        0.067 &      0.00016 &       0.0019 &        0.033          &     24 &       \\ [1.5ex]
         
74  & HII~120     &        2940. &        6.818 &        2.426 &        0.303 &        238.6 &       56530. &      0.00377 &       0.1556 &         93.5          &  0.228 & 0.935 \\ 
    &             &          12. &        0.041 &        0.057 &        0.026 &          4.4 &          37. &      0.00024 &       0.0033 &          2.1          &     53 &       \\ [1.5ex]
         
93  & HII~233     &       1241.5 &        5.321 &        1.978 &        0.242 &        248.9 &       51856. &     0.000910 &       0.0969 &        32.77          &  0.277 & 0.991 \\ 
    &             &          1.4 &        0.039 &        0.060 &        0.025 &          7.7 &          22. &     0.000080 &       0.0029 &         0.97          &     64 &       \\ [1.5ex]
         
98  & HII~250     &       971.59 &        4.206 &        3.180 &       0.6613 &        243.5 &      54910.3 &     0.001366 &       0.1110 &        31.87          &  0.175 & 1.014 \\ 
    &             &         0.67 &        0.029 &        0.046 &       0.0099 &          1.6 &          2.3 &     0.000050 &       0.0014 &         0.39          &     51 &       \\ [1.5ex]
         
133 & HII~522     &     23.83746 &        6.597 &         4.98 &        0.109 &          30. &     48736.23 &     0.000300 &       0.0669 &        1.623          &  0.451 & 1.079 \\ 
    &             &      0.00049 &        0.082 &         0.12 &        0.023 &          13. &         0.83 &     0.000022 &       0.0016 &        0.039          &     42 &       \\ [1.5ex]
         
144 & HII~571     &    15.872245 &        5.829 &       26.476 &       0.3281 &        88.48 &    52084.074 &      0.02573 &      0.29521 &        5.459          &  0.322 & 1.608 \\ 
    &             &     0.000032 &        0.060 &        0.088 &       0.0031 &         0.45 &        0.020 &      0.00021 &      0.00080 &        0.015          &     89 &       \\ [1.5ex]
         
158 & HII~727     &        7271. &         6.08 &          8.6 &        0.832 &        146.2 &       51330. &        0.081 &        0.433 &         476.          &  0.848 & 0.988 \\ 
    &             &         153. &         0.20 &          1.7 &        0.039 &          3.2 &         159. &        0.024 &        0.043 &          49.          &     65 &       \\ [1.5ex]
         
160 & HII~745     &       1541.4 &         5.28 &         6.80 &        0.120 &           6. &       56509. &       0.0492 &       0.3664 &        143.1          &  0.688 & 0.730 \\ 
    &             &          9.8 &         0.10 &         0.13 &        0.025 &          12. &          55. &       0.0030 &       0.0074 &          3.0          &     62 &       \\ [1.5ex]
         
201 & TRU~S93     &       1059.7 &        4.092 &         3.18 &        0.392 &        156.5 &       56204. &      0.00276 &       0.1402 &         42.7          &  0.390 & 1.032 \\ 
    &             &          3.8 &        0.071 &         0.12 &        0.037 &          4.7 &          14. &      0.00027 &       0.0046 &          1.4          &     56 &       \\ [1.5ex]
         
233 & HII~1407    &       953.08 &        5.285 &         7.76 &        0.322 &        262.2 &      55419.4 &       0.0392 &       0.3398 &         96.3          &  0.376 & 0.657 \\ 
    &             &         0.90 &        0.088 &         0.11 &        0.011 &          2.2 &          4.4 &       0.0018 &       0.0053 &          1.5          &     55 &       \\ [1.5ex]
         
235 & HII~1397    &     7.345274 &        6.903 &       13.214 &            0 &   \nodata    &    49630.041 &     0.001756 &      0.12065 &       1.3347          &  0.416 & 1.223 \\ 
    &             &     0.000018 &        0.069 &        0.097 &   \nodata    &   \nodata    &        0.015 &     0.000039 &      0.00088 &       0.0098          &     47 &       \\ [1.5ex]
         
251 & HII~1653    &        548.2 &         7.39 &         6.20 &            0 &   \nodata    &      55674.1 &       0.0135 &        0.238 &         46.7          &  1.667 & 0.781 \\ 
    &             &          1.1 &         0.52 &         0.51 &   \nodata    &   \nodata    &          9.8 &       0.0033 &        0.020 &          3.8          &     27 &       \\ [1.5ex]
         
254 & HII~1762    &        4017. &         6.29 &        10.48 &        0.468 &         350. &       58398. &        0.331 &        0.692 &         512.          &  1.485 & 0.634 \\ 
    &             &         155. &         0.38 &         0.49 &        0.039 &          10. &          65. &        0.042 &        0.029 &          28.          &     33 &       \\ [1.5ex]
         
277 & HII~2172    &     30.21296 &        5.967 &       14.667 &       0.3294 &       121.54 &    49574.579 &     0.008314 &      0.20258 &        5.753          &  0.332 & 1.162 \\ 
    &             &      0.00011 &        0.046 &        0.059 &       0.0037 &         0.90 &        0.062 &     0.000096 &      0.00078 &        0.022          &     75 &       \\ [1.5ex]
         
282 & HII~2284    &       807.39 &        5.705 &        3.617 &       0.4111 &         54.6 &      56747.8 &     0.003000 &      0.14422 &        36.61          &  0.079 & 0.492 \\ 
    &             &         0.45 &        0.015 &        0.024 &       0.0061 &          1.2 &          2.5 &     0.000060 &      0.00096 &         0.24          &     38 &       \\ [1.5ex]
         
291 & HII~2407    &    7.0504772 &        6.195 &       19.672 &            0 &   \nodata    &   52656.3988 &     0.005561 &      0.17717 &       1.9072          &  0.413 & 1.185 \\ 
    &             &    0.0000090 &        0.051 &        0.057 &   \nodata    &   \nodata    &       0.0064 &     0.000048 &      0.00051 &       0.0055          &     74 &       \\ [1.5ex]
         
297 & HII~2500    &        2391. &        5.628 &         6.32 &            0 &   \nodata    &       56290. &       0.0626 &       0.3970 &        207.8          &  0.586 & 1.010 \\ 
    &             &          17. &        0.094 &         0.13 &   \nodata    &   \nodata    &          11. &       0.0040 &       0.0084 &          4.8          &     48 &       \\ [1.5ex]
         
328 & HII~3104    &       1312.5 &        7.036 &         2.72 &        0.207 &         235. &       57554. &      0.00256 &       0.1368 &         48.0          &  0.295 & 0.799 \\ 
    &             &          4.5 &        0.093 &         0.14 &        0.051 &          21. &          74. &      0.00038 &       0.0067 &          2.4          &     22 &       \\ [1.5ex]
         
330 & HII~3097    &       780.38 &        5.549 &        4.537 &        0.777 &         29.8 &      51415.4 &      0.00188 &       0.1235 &        30.65          &  0.310 & 0.972 \\ 
    &             &         0.14 &        0.042 &        0.082 &        0.010 &          1.3 &          1.3 &      0.00011 &       0.0024 &         0.60          &     69 &       \\ [1.5ex]
         
342 & PELS~69     &       1327.5 &         7.66 &         4.96 &        0.436 &        232.3 &      55577.8 &       0.0122 &       0.2302 &         81.4          &  0.517 & 0.958 \\ 
    &             &          1.3 &         0.10 &         0.13 &        0.029 &          3.5 &          8.6 &       0.0010 &       0.0063 &          2.2          &     39 &  NM   \\ [1.5ex]
         
348 & AK~IV-287   &        1808. &         4.59 &         4.75 &        0.229 &          20. &       57765. &       0.0185 &        0.265 &         115.          &  1.585 & 1.621 \\ 
    &             &          26. &         0.29 &         0.43 &        0.064 &          31. &         151. &       0.0051 &        0.024 &          11.          &     42 &       
\enddata

\tablecomments{Uncertainties for the orbital elements and derived
quantities are given in the second line for each system. The symbol
$T_0$ represents a reference time of periastron passage for eccentric orbits,
and a time of maximum primary velocity for circular orbits.
$M_2 \sin i$ is the coefficient of the minimum secondary mass
multiplying the factor $(M_1+M_2)^{2/3}$. The first line of the last
column ($F$) represents the scale factor applied to the internal RV
uncertainties to produce a reduced $\chi^2$ of unity. Non-members are
indicated with ``NM'' in the second line of the last column. For
AK~II-346 and HII~1762, the orbits reported are for the secondary
component; the primary is not seen in our spectra.}
\end{deluxetable*}
\setlength{\tabcolsep}{6pt}  

For double-lined binaries we proceeded in a similar fashion, solving
for the additional parameter $K_2$ representing the velocity
semiamplitude of the secondary. As the secondary velocities are
typically poorer than those of the primary because the secondaries are
fainter, we allowed separate scaling factors for the two stars.  The
orbital elements are presented in Table~\ref{tab:sb2.1}. Derived
properties are listed separately in Table~\ref{tab:sb2.2}. They
include the minimum masses $M_1 \sin^3 i$ and $M_2 \sin^3 i$, the mass
ratio $q \equiv M_2/M_1$, the individual projected semimajor axes $a_1
\sin i$ and $a_2 \sin i$, and the total projected semimajor axis $a
\sin i$ in units of the solar radius.  In some cases there can be a
systematic offset between the primary and secondary velocities caused,
e.g., by a mismatch between the properties of the real stars and the
templates used for the components. If not accounted for, this can bias
the velocity semiamplitudes, or possibly the center-of-mass velocity
of the binary. Most commonly this mismatch will occur for the
secondary, as we are sometimes unable to determine its temperature or
rotational velocity independently from our spectra, and had to adopt
educated guesses based on models or other observables. Therefore, we
have added as a free parameter a velocity offset $\Delta RV$ that we
solved for simultaneously with the other elements. These offsets are
listed in the table as well, when they are different from zero by more
than twice their uncertainty.

\begin{figure}
\epsscale{1.15}
\plotone{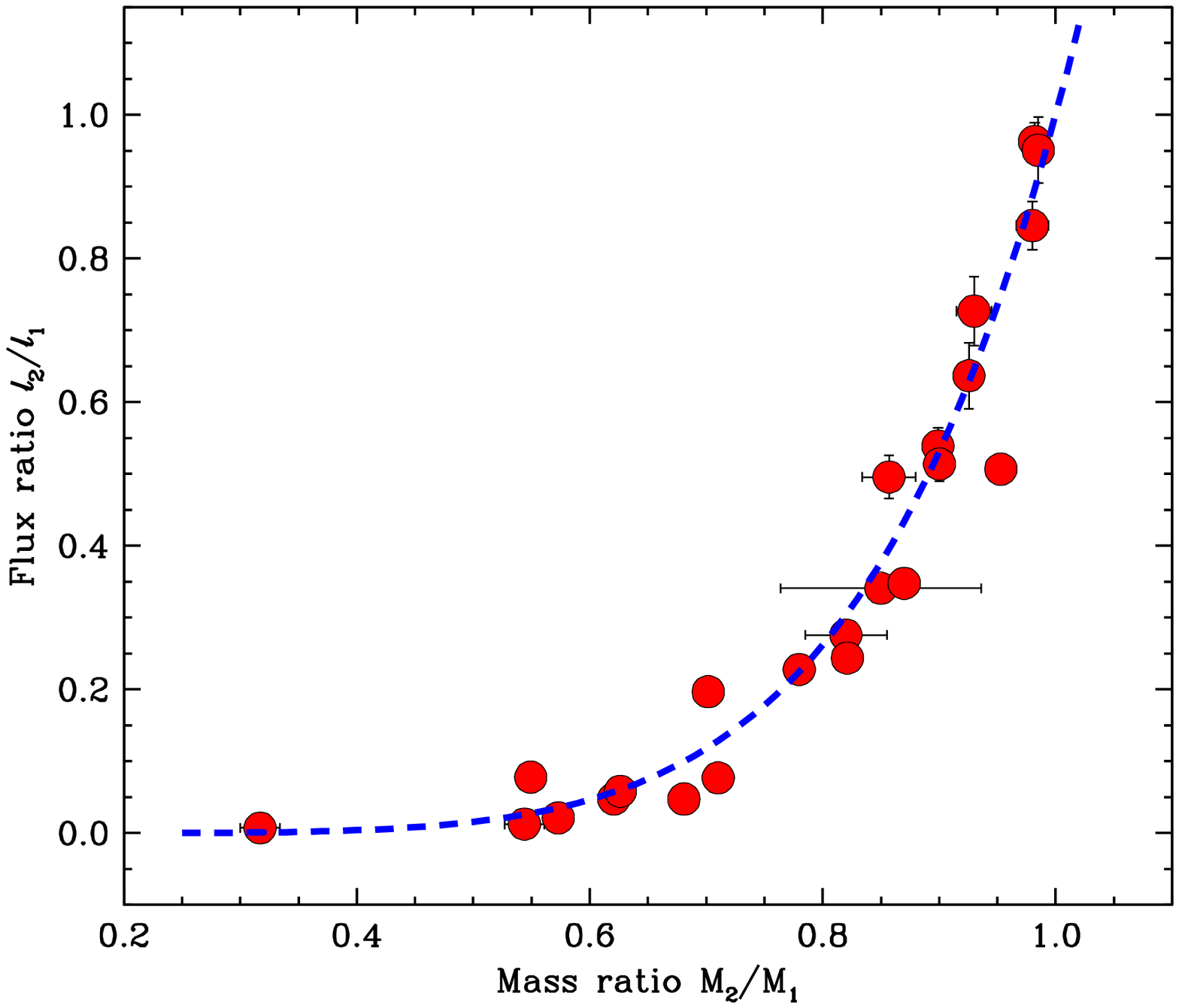}
\figcaption{Flux ratios as a function of the mass ratios for the
  double-lined binaries in the sample. The dashed line corresponds to
  a power law with an exponent of $\alpha = 6$, represented here on
  the linear scale of the observations.\label{fig:qlrat}}
\end{figure}

Table~\ref{tab:sb2.2} reports also the flux ratios determined with
{\tt TODCOR}, which correspond to the mean wavelength of our
observations ($\approx$5187~\AA). We expect a strong correlation
between the flux ratios and the mass ratios for the double-lined
binaries, resulting from the theoretical mass-luminosity relation.
This trend is illustrated in Figure~\ref{fig:qlrat}. The bolometric
mass-luminosity relation on the main sequence is often approximated as
a power law ($L \propto M^{\alpha}$), although the exponent depends on
the mass range ($\alpha \sim3$--4).  The value of $\alpha$ has been
shown to be larger when considering fluxes over a discrete bandpass
\citep[e.g.,][]{Goldberg:2002}, and indeed the PARSEC~1.2S isochrone
invoked earlier for the Pleiades predicts a slope for the
mass-luminosity relation that is $\sim$6 in a bandpass near our
spectral window, for masses between 0.5 and 1.5~$M_{\sun}$.  This
$\alpha = 6$ relation converted to the linear scale of the
measurements is shown as a dashed line in the figure, and is seen to
match the empirical relation well.

\setlength{\tabcolsep}{4pt}
\begin{deluxetable*}{r@{\hskip 1ex}lcccccccccc}
\tabletypesize{\scriptsize}
\tablecaption{Orbital Elements for the Double-lined Binaries in the Sample.\label{tab:sb2.1}}
\tablehead{
\colhead{} &
\colhead{Name} &
\colhead{$P$} &
\colhead{$\gamma$} &
\colhead{$K_1$} &
\colhead{$K_2$} &
\colhead{$e$} &
\colhead{$\omega_1$} &
\colhead{$T_0$ (HJD)} &
\colhead{$\Delta_{\rm CfA}$} &
\colhead{$\Delta_{\rm COR}$} &
\colhead{}
\\
\colhead{} &
\colhead{} &
\colhead{(day)} &
\colhead{(\kms)} &
\colhead{(\kms)} &
\colhead{(\kms)} &
\colhead{} &
\colhead{(degree)} &
\colhead{(2,400,000+)} &
\colhead{(\kms)} &
\colhead{(\kms)} &
\colhead{}
}
\startdata
1   & AK~III-31   &    5.0043210 &        5.511 &       54.774 &        87.45 &            0 &   \nodata    &   53131.5598 &   \nodata    &   \nodata      &  NM  \\ 
    &             &    0.0000090 &        0.056 &        0.065 &         0.73 &   \nodata    &   \nodata    &       0.0019 &   \nodata    &   \nodata      &      \\ [1.5ex]
                                                                                                                                                                  
21  & AK~III-419  &    34.321476 &        3.811 &        38.87 &        43.18 &       0.6493 &       345.18 &    53996.096 &   \nodata    &   \nodata      &      \\ 
    &             &     0.000067 &        0.033 &         0.20 &         0.22 &       0.0022 &         0.16 &        0.010 &   \nodata    &   \nodata      &      \\ [1.5ex]
                                                                                                                                                                  
81  & HII~164     &      268.704 &        4.968 &        9.830 &         30.9 &       0.2505 &         73.7 &      52933.9 &          2.9 &   \nodata      &      \\ 
    &             &        0.056 &        0.086 &        0.070 &          1.7 &       0.0072 &          2.8 &          1.6 &          1.4 &   \nodata      &      \\ [1.5ex]
                                                                                                                                                                  
86  & HII~173     &       481.25 &        5.216 &        15.99 &        16.79 &       0.0986 &        214.0 &      48399.4 &   \nodata    &   \nodata      &      \\ 
    &             &         0.10 &        0.076 &         0.12 &         0.15 &       0.0067 &          3.9 &          5.4 &   \nodata    &   \nodata      &      \\ [1.5ex]
                                                                                                                                                                  
89  & HII~177     &       2278.3 &       10.724 &        9.073 &       10.092 &       0.4900 &       331.45 &      54258.7 &   \nodata    &   \nodata      &  NM  \\ 
    &             &          1.2 &        0.030 &        0.066 &        0.059 &       0.0036 &         0.57 &          2.9 &   \nodata    &   \nodata      &      \\ [1.5ex]
                                                                                                                                                                  
112 & HII~320     &       757.01 &        5.759 &       12.248 &        14.30 &       0.3064 &        284.8 &      52047.4 &   \nodata    &   \nodata      &      \\ 
    &             &         0.22 &        0.059 &        0.065 &         0.38 &       0.0049 &          1.1 &          2.2 &   \nodata    &   \nodata      &      \\ [1.5ex]
                                                                                                                                                                  
119 & PELS~38     &    17.285878 &        6.910 &       19.078 &        33.30 &       0.0811 &        132.7 &    55314.946 &   \nodata    &   \nodata      &      \\ 
    &             &     0.000069 &        0.026 &        0.025 &         0.69 &       0.0018 &          1.5 &        0.068 &   \nodata    &   \nodata      &      \\ [1.5ex]
                                                                                                                                                                  
145 & HII~605     &     20.79762 &         5.08 &        42.27 &        68.09 &       0.4318 &       287.62 &    52405.746 &  $-$1.38\phs &   \nodata      &      \\ 
    &             &      0.00012 &         0.12 &         0.12 &         0.72 &       0.0026 &         0.51 &        0.020 &         0.60 &   \nodata      &      \\ [1.5ex]
                                                                                                                                                                  
165 & HII~761     &    3.3072926 &        5.349 &       51.477 &        75.60 &            0 &   \nodata    &   52392.7804 &   \nodata    &   \nodata      &      \\ 
    &             &    0.0000025 &        0.073 &        0.080 &         0.77 &   \nodata    &   \nodata    &       0.0015 &   \nodata    &   \nodata      &      \\ [1.5ex]
                                                                                                                                                                  
180 & AK~I-2-288  &     17.46668 &        4.568 &       18.548 &        21.32 &       0.2010 &       325.91 &    55197.296 &   \nodata    &   \nodata      &      \\ 
    &             &      0.00013 &        0.040 &        0.069 &         0.10 &       0.0030 &         0.91 &        0.044 &   \nodata    &   \nodata      &      \\ [1.5ex]
                                                                                                                                                                  
204 & HII~1117    &     26.02712 &        7.424 &        13.30 &        13.57 &       0.5745 &       136.95 &    49112.755 &         0.42 &  $-$0.08\phs   &      \\ 
    &             &      0.00010 &        0.099 &         0.15 &         0.15 &       0.0062 &         0.90 &        0.030 &         0.20 &         0.16   &      \\ [1.5ex]
                                                                                                                                                                  
227 & HII~1338    &     7.757171 &         3.14 &        58.45 &        63.15 &       0.0344 &        297.1 &     53459.92 &   \nodata    &   \nodata      &      \\ 
    &             &     0.000072 &         0.14 &         0.22 &         0.25 &       0.0032 &          5.8 &         0.12 &   \nodata    &   \nodata      &      \\ [1.5ex]
                                                                                                                                                                  
228 & HII~1348    &       94.805 &        6.367 &       20.163 &        25.85 &       0.5543 &        82.20 &    56452.796 &   \nodata    &   \nodata      &      \\ 
    &             &        0.012 &        0.022 &        0.054 &         0.33 &       0.0017 &         0.32 &        0.068 &   \nodata    &   \nodata      &      \\ [1.5ex]

234 & HII~1392    &       767.04 &         5.78 &        23.65 &        25.42 &       0.8206 &        221.7 &     57386.28 &         1.09 &   \nodata      &      \\ 
    &             &         0.25 &         0.24 &         0.32 &         0.28 &       0.0075 &          1.0 &         0.24 &         0.37 &   \nodata      &      \\ [1.5ex]
                                                                                                                                                                  
237 & HII~1431    &     2.461127 &         6.29 &       100.07 &       142.60 &            0 &   \nodata    &  52969.43596 &         1.74 &   \nodata      &      \\ 
    &             &     0.000015 &         0.20 &         0.23 &         0.28 &   \nodata    &   \nodata    &      0.00092 &         0.31 &   \nodata      &      \\ [1.5ex]
                                                                                                                                                                  
290 & HII~2406    &    33.006290 &        6.065 &       25.801 &         47.4 &       0.5109 &       218.29 &   57815.0878 &   \nodata    &   \nodata      &      \\ 
    &             &     0.000049 &        0.036 &        0.033 &          1.5 &       0.0011 &         0.21 &       0.0096 &   \nodata    &   \nodata      &      \\ [1.5ex]

299 & HII~2507    &    16.726227 &        6.428 &       39.358 &       71.651 &            0 &   \nodata    &   54799.3046 &        0.428 &   \nodata      &      \\ 
    &             &     0.000040 &        0.020 &        0.023 &        0.079 &   \nodata    &   \nodata    &       0.0037 &        0.075 &   \nodata      &      \\ [1.5ex]
                                                                                                                                                                  
300 & HCG~384     &       542.11 &         5.75 &        15.06 &        18.36 &       0.6624 &        167.3 &     57624.31 &   \nodata    &   \nodata      &      \\ 
    &             &         0.27 &         0.11 &         0.23 &         0.78 &       0.0080 &          1.2 &         0.79 &   \nodata    &   \nodata      &      \\ [1.5ex]
                                                                                                                                                                  
351 & DH~794      &     5.694369 &        7.426 &       40.200 &        56.60 &       0.0119 &        201.4 &     58470.71 &   \nodata    &   \nodata      &      \\ 
    &             &     0.000042 &        0.057 &        0.084 &         0.67 &       0.0021 &          7.9 &         0.12 &   \nodata    &   \nodata      &      \\ [1.5ex]
                                                                                                                                                                  
356 & HCG~489     &     3.108737 &        6.619 &        76.29 &        77.44 &            0 &   \nodata    &  57190.40052 &   \nodata    &   \nodata      &      \\ 
    &             &     0.000012 &        0.086 &         0.16 &         0.22 &   \nodata    &   \nodata    &      0.00075 &   \nodata    &   \nodata      &      \\ [1.5ex]
                                                                                                                                                                  
358 & HCG~495     &      8.57662 &         7.61 &        53.06 &        54.01 &       0.1368 &        272.4 &    57190.015 &         2.90 &   \nodata      &      \\ 
    &             &      0.00053 &         0.31 &         0.42 &         0.52 &       0.0071 &          2.6 &        0.060 &         0.51 &   \nodata      &      \\ [1.5ex]
                                                                                                                                                                  
359 & AK~V-151    &        3079. &        14.97 &         10.1 &        11.83 &        0.301 &        247.1 &       54181. &   \nodata    &   \nodata      &  NM  \\ 
    &             &          25. &         0.21 &          1.0 &         0.23 &        0.019 &          3.9 &          32. &   \nodata    &   \nodata      &      \\ [1.5ex]
         
366 & AK~V-198    &      176.364 &        7.858 &       14.599 &       17.779 &       0.2182 &        329.0 &     54750.00 &  $-$0.62\phs &   \nodata      &      \\ 
    &             &        0.011 &        0.068 &        0.043 &        0.081 &       0.0054 &          1.4 &         0.54 &         0.15 &   \nodata      &     
\enddata

\tablecomments{Uncertainties for the orbital elements are given in the
second line for each system. The symbol $T_0$ represents a reference time of
periastron passage for eccentric orbits, and a time of maximum
primary velocity for circular orbits. Primary/secondary velocity
offsets for CfA and the CORAVEL are listed under $\Delta_{\rm CfA}$
and $\Delta_{\rm COR}$. Non-members are indicated with ``NM'' in the
last column.}
\end{deluxetable*}
\setlength{\tabcolsep}{6pt}  

\setlength{\tabcolsep}{4pt}
\begin{deluxetable*}{r@{\hskip 1ex}lccccccccccc}
\tabletypesize{\scriptsize}
\tablecaption{Derived Properties for the Double-lined Binaries in the Sample.\label{tab:sb2.2}}
\tablehead{
\colhead{} &
\colhead{Name} &
\colhead{$M_1 \sin^3 i$} &
\colhead{$M_2 \sin^3 i$} &
\colhead{$a_1 \sin i$} &
\colhead{$a_2 \sin i$} &
\colhead{$a_{\rm tot} \sin i$} &
\colhead{$q$} &
\colhead{$\sigma_1$ (\kms)} &
\colhead{$\sigma_2$ (\kms)} &
\colhead{$F$} &
\colhead{$\ell_2/\ell_1$} &
\colhead{}
\\
\colhead{} &
\colhead{} &
\colhead{($M_{\sun}$)} &
\colhead{($M_{\sun}$)} &
\colhead{($10^6$ km)} &
\colhead{($10^6$ km)} &
\colhead{($R_{\sun}$)} &
\colhead{} &
\colhead{$N_1$} &
\colhead{$N_2$} &
\colhead{} &
\colhead{} &
\colhead{}
}
\startdata
1   & AK~III-31   &        0.917 &       0.5745 &       3.7692 &        6.018 &       14.068 &       0.6263        &  0.431 &  5.095 &  1.044 &   0.0583 &   NM  \\          
    &             &        0.017 &       0.0060 &       0.0045 &        0.050 &        0.072 &       0.0053        &     66 &     66 &  1.005 &   0.0089 &       \\ [1.5ex]  
                                                                                                                                                                             
21  & AK~III-419  &       0.4546 &       0.4093 &       13.951 &       15.497 &        42.33 &       0.9003        &  0.396 &  0.387 &  1.016 &   0.514  &       \\          
    &             &       0.0039 &       0.0035 &        0.047 &        0.050 &         0.12 &       0.0031        &     77 &     77 &  1.070 &   0.024  &       \\ [1.5ex]  
                                                                                                                                                                             
81  & HII~164     &         1.30 &        0.413 &        35.16 &        110.6 &        209.6 &        0.318        &  0.369 &  6.951 &  1.048 &   0.0076 &       \\          
    &             &         0.18 &        0.034 &         0.26 &          5.9 &          8.5 &        0.017        &     55 &     32 &  1.081 &   0.0040 &       \\ [1.5ex]  
                                                                                                                                                                             
86  & HII~173     &        0.886 &        0.844 &       105.31 &        110.6 &        310.3 &        0.953        &  0.813 &  1.041 &  1.210 &   0.507  &       \\          
    &             &        0.018 &        0.015 &         0.78 &          1.0 &          1.9 &        0.011        &     80 &     74 &  1.192 &   0.013  &       \\ [1.5ex]  
                                                                                                                                                                             
89  & HII~177     &       0.5796 &       0.5211 &        247.8 &        275.6 &        752.3 &       0.8990        &  0.405 &  0.362 &  1.030 &   0.539  &   NM  \\          
    &             &       0.0087 &       0.0086 &          1.8 &          1.7 &          3.9 &       0.0081        &     90 &     90 &  1.002 &   0.025  &       \\ [1.5ex]  
                                                                                                                                                                             
112 & HII~320     &        0.682 &        0.584 &       121.36 &        141.7 &        378.1 &        0.857        &  0.243 &  1.207 &  1.065 &   0.496  &       \\          
    &             &        0.040 &        0.020 &         0.77 &          3.8 &          5.9 &        0.023        &     46 &     23 &  1.137 &   0.030  &       \\ [1.5ex]  
                                                                                                                                                                             
119 & PELS~38     &       0.1620 &       0.0928 &       4.5198 &         7.89 &        17.84 &        0.573        &  0.092 &  2.939 &  0.991 &   0.0213 &       \\          
    &             &       0.0076 &       0.0025 &       0.0059 &         0.16 &         0.24 &        0.012        &     36 &     31 &  1.027 &   0.0066 &       \\ [1.5ex]  
                                                                                                                                                                             
145 & HII~605     &        1.311 &        0.814 &       10.902 &        17.57 &        40.92 &       0.6207        &  0.553 &  3.343 &  1.058 &   0.0478 &       \\          
    &             &        0.032 &        0.012 &        0.035 &         0.19 &         0.28 &       0.0068        &     42 &     42 &  1.057 &   0.0039 &       \\ [1.5ex]  
                                                                                                                                                                             
165 & HII~761     &       0.4183 &       0.2849 &       2.3411 &        3.438 &        8.307 &       0.6809        &  0.469 &  4.554 &  1.029 &   0.0475 &       \\          
    &             &       0.0093 &       0.0035 &       0.0037 &        0.035 &        0.051 &       0.0070        &     48 &     48 &  1.028 &   0.0055 &       \\ [1.5ex]  
                                                                                                                                                                             
180 & AK~I-2-288  &      0.05764 &      0.05015 &        4.364 &        5.016 &       13.483 &       0.8700        &  0.257 &  0.367 &  1.453 &   0.348  &       \\          
    &             &      0.00060 &      0.00044 &        0.016 &        0.024 &        0.042 &       0.0053        &     32 &     30 &  1.131 &   0.011  &       \\ [1.5ex]  
                                                                                                                                                                             
204 & HII~1117    &      0.01449 &      0.01420 &        3.897 &        3.975 &       11.315 &        0.980        &  0.836 &  0.838 &  1.305 &   0.846  &       \\          
    &             &      0.00034 &      0.00034 &        0.041 &        0.041 &        0.085 &        0.014        &     84 &     85 &  1.188 &   0.034  &       \\ [1.5ex]  
                                                                                                                                                                             
227 & HII~1338    &       0.7491 &       0.6933 &        6.231 &        6.732 &       18.632 &       0.9256        &  1.538 &  1.741 &  1.026 &   0.637  &       \\          
    &             &       0.0067 &       0.0060 &        0.023 &        0.026 &        0.052 &       0.0049        &     72 &     72 &  1.027 &   0.046  &       \\ [1.5ex]  
                                                                                                                                                                             
228 & HII~1348    &       0.3101 &       0.2418 &       21.878 &        28.05 &        71.77 &       0.7799        &  0.090 &  0.902 &  1.093 &   0.228  &       \\          
    &             &       0.0083 &       0.0036 &        0.046 &         0.35 &         0.51 &       0.0098        &     22 &     22 &  1.196 &   0.015  &       \\ [1.5ex]  
                                                                                                                                                                             
234 & HII~1392    &        0.908 &        0.845 &        142.6 &        153.3 &        425.2 &        0.930        &  1.025 &  0.872 &  1.045 &   0.727  &       \\          
    &             &        0.059 &        0.056 &          3.4 &          3.4 &          9.2 &        0.015        &     30 &     30 &  0.980 &   0.048  &       \\ [1.5ex]  
                                                                                                                                                                             
237 & HII~1431    &       2.1414 &       1.5028 &       3.3867 &       4.8260 &       11.805 &       0.7018        &  1.308 &  1.569 &  1.034 &   0.197  &       \\          
    &             &       0.0099 &       0.0072 &       0.0078 &       0.0094 &        0.018 &       0.0021        &     49 &     49 &  1.034 &   0.010  &       \\ [1.5ex]  
                                                                                                                                                                             
290 & HII~2406    &        0.551 &        0.300 &       10.067 &        18.49 &        41.04 &        0.544        &  0.194 &  7.402 &  1.120 &   0.0126 &       \\         
    &             &        0.040 &        0.012 &        0.015 &         0.59 &         0.85 &        0.017        &     68 &     38 &  1.053 &   0.0028 &       \\ [1.5ex] 

299 & HII~2507    &       1.5302 &       0.8406 &       9.0525 &       16.480 &       36.700 &      0.54931        &  0.143 &  0.519 &  1.034 &   0.078  &       \\          
    &             &       0.0040 &       0.0015 &       0.0052 &        0.018 &        0.028 &      0.00067        &     59 &     59 &  0.991 &   0.003  &       \\ [1.5ex]  
                                                                                                                                                                             
300 & HCG~384     &        0.484 &        0.397 &        84.12 &        102.5 &        268.3 &        0.820        &  0.517 &  2.563 &  1.076 &   0.276  &       \\          
    &             &        0.042 &        0.020 &         0.90 &          4.2 &          6.2 &        0.035        &     29 &     29 &  1.078 &   0.018  &       \\ [1.5ex]  
                                                                                                                                                                             
351 & DH~794      &       0.3129 &       0.2222 &       3.1476 &        4.432 &       10.895 &       0.7103        &  0.164 &  1.952 &  1.129 &   0.077  &       \\          
    &             &       0.0081 &       0.0032 &       0.0066 &        0.053 &        0.077 &       0.0085        &     19 &     19 &  1.137 &   0.011  &       \\ [1.5ex]  
                                                                                                                                                                             
356 & HCG~489     &       0.5895 &       0.5807 &       3.2612 &       3.3105 &        9.446 &       0.9851        &  0.433 &  0.581 &  0.907 &   0.951  &       \\          
    &             &       0.0035 &       0.0030 &       0.0070 &       0.0092 &        0.017 &       0.0034        &     19 &     19 &  1.256 &   0.046  &       \\ [1.5ex]  
                                                                                                                                                                             
358 & HCG~495     &        0.535 &       0.5255 &        6.199 &        6.310 &        17.98 &        0.982        &  1.055 &  1.322 &  1.201 &   0.963  &       \\          
    &             &        0.011 &       0.0098 &        0.048 &        0.060 &         0.11 &        0.012        &     17 &     17 &  1.202 &   0.026  &       \\ [1.5ex]  
                                                                                                                                                                             
359 & AK~V-151    &         1.57 &         1.33 &         406. &        477.6 &        1270. &        0.850        &  1.574 &  0.797 &  1.799 &   0.341  &   NM  \\          
    &             &         0.16 &         0.26 &          41. &          8.7 &          62. &        0.086        &     11 &     44 &  1.126 &   0.020  &       \\ [1.5ex]  
                                                                                                                                                                             
366 & AK~V-198    &       0.3166 &       0.2600 &        34.55 &        42.08 &       110.15 &       0.8212        &  0.236 &  0.451 &  1.211 &   0.244  &       \\          
    &             &       0.0032 &       0.0021 &         0.10 &         0.19 &         0.33 &       0.0043        &     45 &     44 &  1.038 &   0.010  &   
\enddata

\tablecomments{Uncertainties for the derived
quantities are given in the second line for each system.  $M_2 \sin i$
is the coefficient of the minimum secondary mass multiplying the
factor $(M_1+M_2)^{2/3}$. The column with the $F$ heading contains
the scale factors applied to the internal RV uncertainties for the
primary and secondary in order to produce a reduced $\chi^2$ of unity.
Non-members are indicated with ``NM'' in the last column.}
\end{deluxetable*}
\setlength{\tabcolsep}{6pt}  

\begin{figure}
\epsscale{1.15}
\plotone{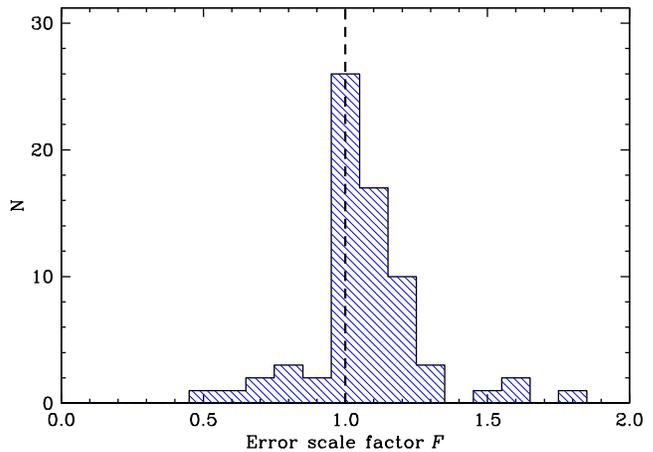}
\figcaption{Histogram of the error scaling factors for SB1s and SB2s
  (see the text), showing that the internal errors from CfA and the
  northern CORAVEL are realistic. \label{fig:scalefactors}}
\end{figure}

The error scaling factors $F$ from our orbital solutions for the SB1s
and SB2s are fairly close to unity in most cases, indicating that our
internal uncertainties (and those from the CORAVEL) are quite
realistic. Their distribution is shown in
Figure~\ref{fig:scalefactors}.  Graphical representations of the
orbital solutions for each SB1 and SB2 are presented at the end of the
paper in Figures~\ref{fig:sb.1}--\ref{fig:sb.4}, and notes for some of
the systems with features of interest are given in the Appendix.

\subsection{Spectroscopic Triple Systems}
\label{sec:triples}

Two of the objects with variable velocity in our sample are
hierarchical spectroscopic triple systems, discussed here. Several
others are spectroscopic binaries with outer astrometric companions.

\subsubsection{HII~1338}

This object was reported as a double-lined spectroscopic binary by
several authors on the basis of one or a few observations
\citep{Abt:1970, Liu:1991, Soderblom:1993}. A spectroscopic orbit with
a period of 7.75 days was first presented by \cite{Raboud:1998},
although as far as we are aware the original velocities were never
published (and do not appear in the catalog of
\citealt{Mermilliod:2009}).  \cite{Raboud:1998} stated only that they
obtained 25 and 21 measurements for the primary and secondary over a
period of about two years (presumably sometime between 1978 and 1995),
and noted that they saw hints of the presence of a third star at
certain phases.

Our observations with the Digital Speedometers also reveal signs of
the tertiary, and although we have been able to measure its velocity
using {\tt TRICOR}, the measurements are very poor due to the
faintness of that component. We list them nonetheless in
Table~\ref{tab:HIIx1338}, along with those of the primary and
secondary. The template used for the tertiary has a temperature of
5250~K and no rotational broadening. The primary and secondary have a
flux ratio of $\ell_2/\ell_1 = 0.637 \pm 0.046$, and between the
tertiary and the primary the ratio is $\ell_3/\ell_1 = 0.335 \pm
0.046$.

\setlength{\tabcolsep}{6pt}
\begin{deluxetable*}{cccccc}
\tablecaption{Radial Velocity Measurements for the Triple-lined System HII~1338.\label{tab:HIIx1338}}
\tablehead{
\colhead{HJD} &
\colhead{$RV_1$} &
\colhead{$RV_2$} &
\colhead{$RV_3$} &
\colhead{S/N} &
\colhead{Inst}
\\
\colhead{(2,400,000+)} &
\colhead{(\kms)} &
\colhead{(\kms)} &
\colhead{(\kms)} &
\colhead{} &
\colhead{}
}
\startdata
  51920.6419 &  $-36.18 \pm 2.08$\phs &   $51.17 \pm 2.36$\phn     &   $16.2 \pm 10.5$     &  27 & 1 \\
  51939.5019 &   $33.24 \pm 2.21$     &  $-26.69 \pm 2.50$\phn\phs &   \phn$1.5 \pm 11.1$  &  26 & 1 \\
  52508.8216 &   $15.74 \pm 1.62$     &  $-10.48 \pm 1.83$\phn\phs &    $4.5 \pm  8.1$     &  35 & 1 \\
  52567.9234 &   $35.50 \pm 1.59$     &  $-31.95 \pm 1.80$\phn\phs &   $21.9 \pm  8.0$\phn &  36 & 1 \\
  52603.7973 &  $-50.82 \pm 1.66$\phs &   $64.88 \pm 1.88$\phn     &    $2.7 \pm  8.4$     &  34 & 1 
\enddata
\tablecomments{The heliocentric radial velocities are on the reference
frame of the minor planets in the Solar
System \citep[see][]{Stefanik:1999}.  Signal-to-noise ratios in the
S/N column are per resolution element. Codes in the Inst column are 1
for the Digital Speedometers and 2 for TRES.  (This table is available
in its entirety in machine-readable form.)}
\end{deluxetable*}
\setlength{\tabcolsep}{6pt}  

The center-of-mass velocity of the SB2 is $\gamma = 3.14 \pm
0.14~\kms$ (see Table~\ref{tab:sb2.1} and Figure~\ref{fig:sb.2}),
which deviates significantly from the cluster mean. On the other hand,
the weighted mean of our 72 tertiary velocities, assuming they are
constant, is $7.15 \pm 0.94~\kms$, which is on the opposite side of
the cluster mean, consistent with membership in the Pleiades if the
tertiary and the SB2 are physically associated. Interestingly,
\cite{Raboud:1998} reported a rather different $\gamma$ velocity for
the SB2 of $6.3 \pm 0.4~\kms$, and also that the tertiary velocities
appeared to be close to the cluster mean, whereas our tertiary
velocities are somewhat higher. All of this would be consistent with a
hierarchical triple scenario in which the SB2's center of mass and the
tertiary were closer in velocity to each other and to the cluster mean
at the time of the CORAVEL observations, and farther apart when we
observed the object a decade or two later.  No significant trend is
detected in the residuals of the primary or secondary in our
observations, taken over a period of seven years, suggesting the outer
period is much longer than that.

A visual companion to HII\,1338 was detected by speckle interferometry
in 2005 at a separation of 0\farcs20 \citep{Mason:2009}, and again in
2010 by lunar occultations \citep{Loader:2012}. At the distance to the
Pleiades, the expected orbital period is of the order of a century.
This visual companion may be the tertiary seen in our spectra. The
previous data release (DR2) of the Gaia catalog listed a parallax
and proper motion for HII\,1338 that were inconsistent with membership
in the cluster, but had very large uncertainties indicating that the
astrometric solution was compromised. The latest data release (EDR3)
no longer reports either a parallax or the proper motion, again
suggesting difficulty with the astrometric solution, possibly caused
by the visual companion.

\subsubsection{HII~2027}

\cite{Mermilliod:1992a} recognized this as a triple system consisting
of a single-lined spectroscopic binary with a period of 48.6 days
(components Ba and Bb), accompanied by a brighter star (A) that was
also visible in their cross-correlation profiles. They detected a
drift in the center-of-mass velocity of the binary and a change in the
other direction for the third star, indicating motion in a wide orbit
supporting the physical association. Additional velocities for the two
visible stars A and Ba were reported by \cite{Mermilliod:2009}. The
total time span of the CORAVEL measurements is slightly over 19 years,
from 1978 to 1997.

The CfA observations of HII\,2027 were collected between 1982 and
early 2021, extending the time baseline enough to permit a solution
for the elements of the outer orbit for the first time.  Radial
velocities for components A and Ba were derived with {\tt TODCOR}, and
while experiments using {\tt TRICOR} revealed hints of the secondary
in the 48.6~day orbit (star Bb), we were not able to measure its
velocities reliably. Our measured velocities for A and Ba are listed
in Table~\ref{tab:HIIx2027.rvs}. The flux ratio between components Ba
and A is $0.737 \pm 0.030$ at a mean wavelength of 5187~\AA.

\setlength{\tabcolsep}{6pt}
\begin{deluxetable}{ccccc}
\tablecaption{Radial Velocity Measurements for the Double-lined Triple System HII~2027.\label{tab:HIIx2027.rvs}}
\tablehead{
\colhead{HJD} &
\colhead{$RV_{\rm A}$} &
\colhead{$RV_{\rm Ba}$} &
\colhead{S/N} &
\colhead{Inst.}
\\
\colhead{(2,400,000+)} &
\colhead{(\kms)} &
\colhead{(\kms)} &
\colhead{} &
\colhead{}
}
\startdata
  45302.8342  &  $6.50 \pm 1.28$  &   $6.48 \pm 1.09$     & 10 & 1  \\
  45339.8090  &  $6.96 \pm 1.04$  &  $22.57 \pm 0.88$\phn & 12 & 1  \\
  52907.8553  &  $6.86 \pm 1.32$  &  $-4.18 \pm 1.12$\phs &  9 & 1  \\
  52958.6826  &  $4.41 \pm 1.19$  &  $-3.15 \pm 1.01$\phs & 10 & 1  \\
  53001.6534  &  $3.86 \pm 1.14$  &  $-7.54 \pm 0.97$\phs & 11 & 1
\enddata

\tablecomments{The heliocentric radial velocities are on the reference
frame of the minor planets in the Solar
System \citep[see][]{Stefanik:1999}.  Signal-to-noise ratios in the
S/N column are per resolution element. Codes in the Inst column are 1
for the Digital Speedometers and 2 for TRES.  (This table is available
in its entirety in machine-readable form.)}

\end{deluxetable}
\setlength{\tabcolsep}{6pt}  

\begin{figure}
\epsscale{1.15}
\plotone{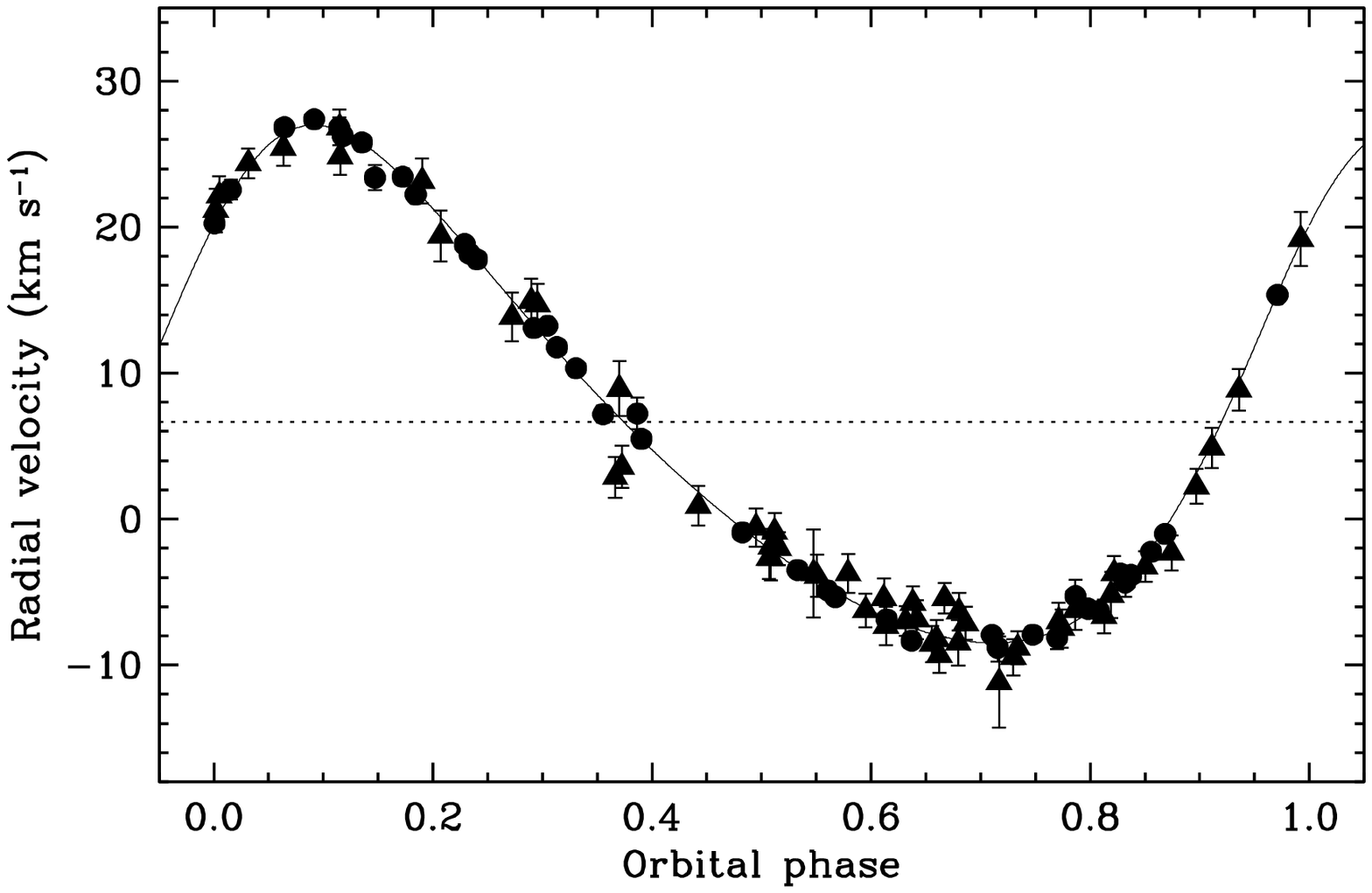}
\vskip 5pt
\plotone{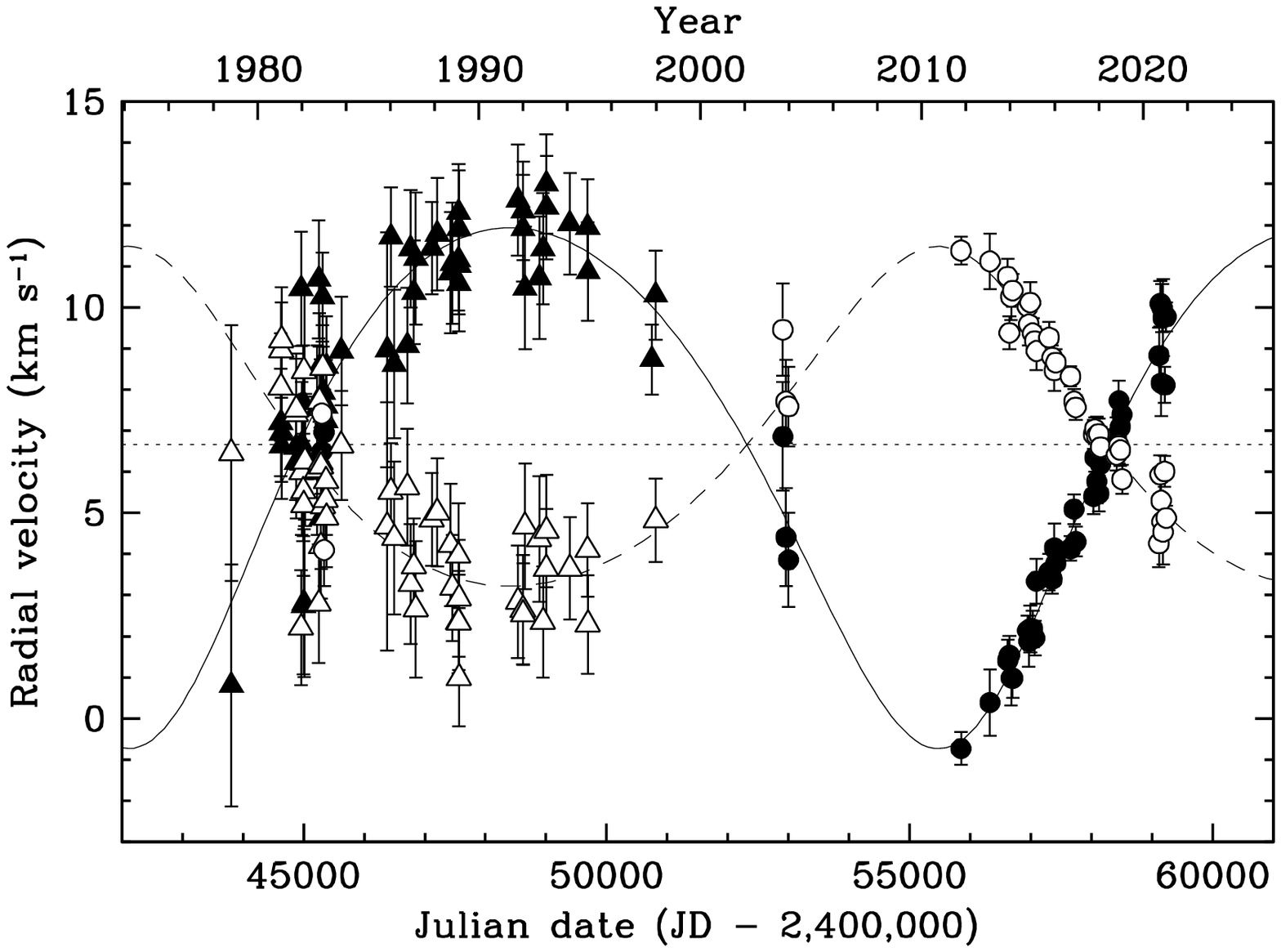}
\figcaption{\emph{Top}: Observations and inner 48.6~day orbit model of
  HII\,2027. Motion in the outer orbit has been subtracted from the
  measured velocities of star Ba. Triangles represent CORAVEL
  measurements, and circles are for RVs from the CfA. The dotted line
  represents the center-of-mass velocity of the triple system.
  \emph{Bottom}: Observations and model for the outer orbit, with
  motion in the inner orbit subtracted from the velocities of star Ba
  (open symbols). The point style is the same as in the top
  panel. \label{fig:HIIx2027}}
\end{figure}

We combined the CORAVEL measurements with ours to solve for the inner
and outer orbits simultaneously, assuming they are dynamically
independent. Internal errors were adjusted separately for the primary
and secondary to give reduced $\chi^2$ values near unity. The CfA
errors listed in Table~\ref{tab:HIIx2027.rvs} already include those
adjustments; the adjustment factors for the CORAVEL measurements were
2.1 and 1.8.  The elements are given in Table~\ref{tab:HIIx2027.elem},
and the orbits together with all observations are represented
graphically in Figure~\ref{fig:HIIx2027}. The outer orbit has a period
of 36.5 yr, the longest in our survey of the Pleiades.

\setlength{\tabcolsep}{6pt}
\begin{deluxetable}{lc}
\tablecaption{Orbital Elements for the Double-lined Triple System HII~2027. \label{tab:HIIx2027.elem}}
\tablehead{
\colhead{~~~~~~~~~Parameter~~~~~~~~~} &
\colhead{Value}
}
\startdata
 $P_{\rm AB}$ (days)              &  $13351 \pm 79$\phm{222} \\
 $\gamma$ (\kms)                  &  $6.615 \pm 0.053$ \\
 $K_{\rm A}$ (\kms)               &  $6.22 \pm 0.24$ \\
 $K_{\rm B}$ (\kms)               &  $4.19 \pm 0.19$ \\
 $e_{\rm AB}$                     &  $0.174 \pm 0.021$ \\
 $\omega_{\rm A}$ (degree)        &  $196.2 \pm 8.8$\phn\phn \\
 $T_{\rm AB}$ (HJD$-$2,400,000)   &  $42559 \pm 329$\phn\phn \\
 $P_{\rm B}$ (days)               &  $48.62538 \pm 0.00055$\phn \\
 $K_{\rm Ba}$ (\kms)              &  $17.736 \pm 0.094$\phn \\
 $e_{\rm B}$                      &  $0.2368 \pm 0.0051$ \\
 $\omega_{\rm Ba}$ (degree)       &  $307.8 \pm 1.4$\phn\phn \\
 $T_{\rm B}$ (HJD$-$2,400,000)    &  $45867.54 \pm 0.21$\phm{2222} \\ [1ex]
\noalign{\hrule} \\ [-1.5ex] 
\multicolumn{2}{c}{Derived quantities} \\ [1ex]
\noalign{\hrule} \\ [-1.5ex]
 $M_{\rm A} \sin^3 i_{\rm AB}$ ($M_{\sun}$)   &  $0.599 \pm 0.070$ \\
 $M_{\rm B} \sin^3 i_{\rm AB}$ ($M_{\sun}$)   &  $0.891 \pm 0.098$ \\
 $a_{\rm A} \sin i_{\rm AB}$ ($10^6$ km)      &  $1125 \pm 44$\phn\phn \\
 $a_{\rm B} \sin i_{\rm AB}$ ($10^6$ km)      &  $757 \pm 36$\phn \\
 $q \equiv M_{\rm B}/M_{\rm A}$               &  $1.487 \pm 0.060$ \\
 $f(M)$ ($M_{\sun}$) for inner pair           &  $0.02578 \pm 0.00040$ \\
 $M_{\rm Bb} \sin i_{\rm B}$ ($[M_{\rm Ba}+M_{\rm Bb}]^{2/3} M_{\sun}$)  & $0.2954 \pm 0.0015$ \\
 $a_{\rm Ba} \sin i_{\rm B}$ ($10^6$ km)      &  $11.522 \pm 0.060$\phn \\
 $N_{\rm A}$ (CfA / CORAVEL)                  &  40 / 53 \\
 $N_{\rm Ba}$ (CfA / CORAVEL)                 &  40 / 52 \\
 Cycles covered for outer orbit               &  1.15
\enddata
\tablecomments{Subscripts ``AB'' correspond to the outer pair (A+B). Star Ba is
the primary of the inner binary.}
\end{deluxetable}
\setlength{\tabcolsep}{6pt}

HII\,2027 has a visual companion found with adaptive optics imaging in
1996 \citep{Bouvier:1997}, with a current separation of 0\farcs2.  It
corresponds to the inner 48.6~day binary.

\subsection{Other Multiple Systems}
\label{sec:othermultiples}

Later in Section~\ref{sec:astrometric} we report known astrometric
companions to our targets, and in some cases these additional objects
have been found around systems that already have (closer)
spectroscopic companions.  Examples of such hierarchical triple
systems are HII~102, HII~120, HII~571, HII~717, HII~1348, and possibly
HII~745 and HCG~384, although the physical association of the
tertiaries in the last two cases has not been confirmed. Additionally,
HII~3197 is also a triple system in which the secondary and tertiary
were both detected astrometrically. The 16\arcsec\ pair HII~303 and
HII~302 is another case, provided they are bound, in which the
northern, brighter star (HII~303) has a closer 1\farcs8 visual
companion. More detailed information for all of these objects may be
found in the Appendix.

Even higher multiplicity systems may also be present in our sample.
Each component of the 5\farcs7 visual pair HII~1392 and HII~1397 is in
turn a spectroscopic binary (SB2 in the case of HII~1392), which would
make this a hierarchical quadruple system, provided the visual
components are physically bound. A more complex case is that of
HII~2507, HII~2503, and HII~2500. The latter two objects are 3\farcs3
and 10\farcs1 from the first, respectively, and HII~2507 itself is an
SB2, while HII~2500 is an SB1 and has another companion at 0\farcs3 that
is not the same as the spectroscopic one.  Therefore, this could well
be a sextuple system.

\section{Long-term Trends}
\label{sec:longperiod}

More than a dozen objects in our sample exhibit long-term trends in
their velocities indicating they are binaries, but the observations
are insufficient to determine a period unambiguously. Most of them
benefit greatly from the longer baseline that comes from including the
CORAVEL observations. The time histories of these objects are shown in
Figure~\ref{fig:long} at the end of the paper.

In several cases (AK~I-1-29, HII~727, HII~338, HII~717, HII~916) we have
supplemented the ${\rm CfA} + {\rm CORAVEL}$ observations with other
measurements from the literature that help cover gaps or support the
trends \citep{Liu:1991, Soderblom:1993, Barrado:1998, White:2007,
  Kunder:2017}.\footnote{The \cite{Soderblom:1993} and
  \cite{White:2007} measurements have been adjusted as indicated by
  \cite{Torres:2020b}, to place them on the same zero-point as the CfA
  observations.}  For HII~727, the observations cover a single
periastron passage, which causes some ambiguity in the period (see
figure).  Nevertheless, based on the distribution of the rest of our
velocities including archival ones, we have derived a tentative orbit
with a period of about 7200~days that is shown in
Figure~\ref{fig:sb.2} and listed in Table~\ref{tab:sb1}.

\section{Other Published Orbital Solutions in the Pleiades}
\label{sec:otherpublished}

Future studies of the binary population in the Pleiades may benefit
from having the collection of all known binaries with orbital
solutions in one place. With that in mind, we have supplemented our
own discoveries with a compilation of all other orbits from the
literature that we know of, presented in
Table~\ref{tab:publishedorbits}. All are considered members of the
cluster.

The first four systems in the table are part of our original sample,
but were published separately. They include three rapidly-rotating
early-type stars from \cite{Torres:2020b}, and HII\,2147
\citep{Torres:2020}, which was spatially resolved with the technique
of very long baseline interferometry \citep{Melis:2014} and yielded
absolute masses for the components. Next we include the
spectroscopic-interferometric binary Atlas \citep[27~Tau,
  HII\,2168;][]{Zwahlen:2004}, and the Be shell star Pleione
\citep[28~Tau, HII\,1180;][their Solution~3]{Nemravova:2010}, which
has a period of 218 days. A second, tentative 35~yr long periodicity
in the radial velocities for Pleione has been reported, but has not
been confirmed to be due to orbital motion, and may simply reflect a
periodicity in the shell episodes \citep[see, e.g.,][]{Luthardt:1994,
  Katahira:1996}. We do not include it in the table.

\setlength{\tabcolsep}{6pt}
\begin{deluxetable*}{lccccccccccc}
\tablecaption{Orbital Elements for Other Astrometric or Spectroscopic
  Pleiades Binaries Published Previously.\label{tab:publishedorbits}}
\tablehead{
\colhead{Name} &
\colhead{$P$} &
\colhead{$\gamma$} &
\colhead{$K_1$} &
\colhead{$K_2$} &
\colhead{$e$} &
\colhead{$\omega_1$} &
\colhead{$T_0$ (HJD)} &
\colhead{$a^{\prime\prime}$} &
\colhead{$i$} &
\colhead{$\Omega$} &
\colhead{Ref}
\\
\colhead{} &
\colhead{(day)} &
\colhead{(\kms)} &
\colhead{(\kms)} &
\colhead{(\kms)} &
\colhead{} &
\colhead{(degree)} &
\colhead{(2,400,000+)} &
\colhead{(mas)} &
\colhead{(degree)} &
\colhead{(degree)} &
\colhead{}
}
\startdata

HII 2147  &   6641         & 5.70    & \nodata  &  7.102     & 0.105  & 256.8         & 47284              & 62.32     &  75.78        & 141.80        & 1 \\ [0ex]
          & \phm{22}42     & 0.17    & \nodata  &  0.081     & 0.011  & \phm{22}4.6   & \phm{22}107        & \phn0.45  & \phn0.46      & \phm{22}0.19  &   \\ [1.5ex]

TRU S26   &     71.8198    & 3.94    & 21.1     &  \nodata   & 0.284  & 111           & 58403.4            &  \nodata  &   \nodata     &    \nodata    & 2 \\ [0ex]   
          &  \phn0.0084    & 0.78    & \phn1.3  &  \nodata   & 0.048  & \phn11        & \phm{2222}2.0      &  \nodata  &   \nodata     &    \nodata    &   \\ [1.5ex]

HII 563   &   3172.4       & 5.66    &  7.64    &  \nodata   & 0.391  & 350           & 37784              &  \nodata  &   \nodata     &    \nodata    & 2 \\ [0ex]   
          & \phm{222}9.9   & 0.34    &  0.88    &  \nodata   & 0.079  & \phn11        & \phm{222}82        &  \nodata  &   \nodata     &    \nodata    &   \\ [1.5ex]
 
TRU S194  &   3635         & 7.183   &  3.72    &  \nodata   & 0.528  & 207.8         & 49190              &  \nodata  &   \nodata     &    \nodata    & 2 \\ [0ex]   
          & \phm{22}19     & 0.068   &  0.16    &  \nodata   & 0.035  & \phm{22}3.4   & \phm{222}47        &  \nodata  &   \nodata     &    \nodata    &   \\ [1.5ex]

HII 2168  &    290.984     & \nodata & 26.55    & 36.89      & 0.2385 & 151.9         & 50583.0            & 13.08     & 107.87        & 154.0         & 3 \\ [0ex]
          & \phm{22}0.079  & \nodata & \phn1.41 & \phn0.22   & 0.0063 & \phm{22}2.2   & \phm{2222}1.9      & \phn0.12  & \phm{22}0.49  & \phm{22}0.7   &   \\ [1.5ex]

HII 1180  &    218.053     & \nodata &  6.39    &  \nodata   & 0.745  & 157.3         & 52039.73           &  \nodata  &   \nodata     &    \nodata    & 4 \\ [0ex]
          & \phm{22}0.053  & \nodata &  0.46    &  \nodata   & 0.026  & \phm{22}3.5   & \phm{2222}0.73     &  \nodata  &   \nodata     &    \nodata    &   \\ [1.5ex]

HCG 76    &     32.7470    & 5.31    & 26.75    & 29.19      & 0.1328 &  30.8         & 57068.748          &  \nodata  &  89.126       &    \nodata    & 5 \\ [0ex]
          &   \phn0.0013   & 0.17    & \phn0.30 & \phn0.29   & 0.0043 &  \phn3.2      & \phm{2222}0.001    &  \nodata  &  \phn0.029    &    \nodata    &   \\ [1.5ex]

MHO 9     &     42.80      & 4.6     & 16.4     & 39.3       & 0.406  & 132.0         & 57099.21943        &  \nodata  &  89.278       &    \nodata    & 5 \\ [0ex]
          &    \nodata     & 1.3     & \phn3.0  & \phn6.8    & 0.056  & \phm{22}9.8   & \phm{2222}0.00064  &  \nodata  &  \phn0.094    &    \nodata    &   \\ [1.5ex]

HII 3197  &  11106         & \nodata & \nodata  &  \nodata   & 0.5562 & 221.51        & 55909.4            & 80.84     &  47.67        &  22.77        & 6 \\ [0ex]
          & \phm{222}99    & \nodata & \nodata  &  \nodata   & 0.0018 & \phm{22}0.29  & \phm{2222}4.4      & \phn0.40  &  \phn0.34     &  \phn0.56     &   \\ [1.5ex] 

PPL 15    &  5.825         & 6.5     & \nodata  &  \nodata   & 0.42   &  62           & 50782.59           &  \nodata  &  \nodata      &  \nodata      & 7 \\ [0ex]
          &  0.3\phm{22}   & 2.0     & \nodata  &  \nodata   & 0.05   & \nodata       & \phm{2222}0.01     &  \nodata  &  \nodata      &  \nodata      &   
\enddata

\tablecomments{$T_0$ is a reference time of primary eclipse for
  HCG~76 and MHO~9, and a time of periastron passage for the other
  systems. The last three elements are the angular semimajor axis
  ($a^{\prime\prime}$), the inclination angle ($i$), and the position angle of the
  ascending node ($\Omega$). The second line for each system lists the
  1$\sigma$ uncertainties. References in the last column are: (1)
  \cite{Torres:2020}; (2) \cite{Torres:2020b}; (3)
  \cite{Zwahlen:2004}, (4) \cite{Nemravova:2010}; (5)
  \cite{David:2016}; (6) \cite{Schaefer:2014}; and (7)
  \cite{Basri:1999}.}

\end{deluxetable*}
\setlength{\tabcolsep}{6pt}  

Also listed are the elements for two low-mass eclipsing binaries in
the Pleiades uncovered by the Kepler/K2 mission \citep[HCG~76
  and MHO~9;][]{David:2016}, and for the 30.2~yr astrometric binary
HII\,3197 \citep{Schaefer:2014}, for which we have only a handful of
radial velocity measurements over a period of less than 100 days.

Finally, we mention the interesting case of PPL~15, a 5.8~day
double-lined spectroscopic binary in the Pleiades consisting of brown
dwarfs \citep{Basri:1999}. The Gaia EDR3 catalog confirms it is a
cluster member, although the parallax uncertainty is quite large.

\section{Radial Velocity Variability}
\label{sec:variability}

As stated earlier, for this work we rely on the merged ${\rm CfA}+{\rm
  CORAVEL}$ data sets, and we will assume that any scatter in the
velocities significantly in excess of the observational uncertainties
is caused by a binary companion, or perhaps several. Note that
because of the relatively young age of the Pleiades, some low-level
variability is also to be expected from stellar activity, which is
higher in the Pleiades than in the field, but is sometimes difficult
to distinguish from changes due to orbital motion.

As a quantitative measure of the variability we adopted the $e/i$
(external/internal) statistic, defined as the ratio of the standard
deviation of the RVs to the measurement precision. This is a commonly
used metric in many spectroscopic surveys
\citep[e.g.,][]{Mermilliod:1992a, Hole:2009, Geller:2015}. We
calculated it as described in the second of those references, making
use of the individual uncertainty of each measurement. The $e/i$
values for the 377 stars in our sample are listed in
Table~\ref{tab:meanRV}, where we also specify the number of CORAVEL
measurements for each target.

\begin{figure}
\epsscale{1.15}
\plotone{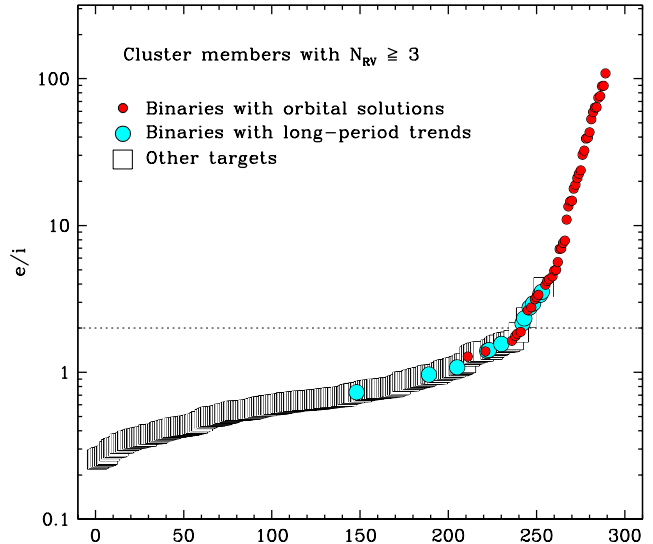}
\figcaption{Values of $e/i$ sorted in increasing order for Pleiades
  members in our sample, based on the merged CfA and CORAVEL data
  sets. Known binaries are marked as labeled, and the dotted line
  represents the adopted threshold for this work, above which stars
  are considered to be binaries.\label{fig:e/i}}
\end{figure}

In Figure~\ref{fig:e/i} we have sorted the $e/i$ values in increasing
order, and show them for all stars that are considered to be Pleiades
members, and that have three or more observations. We include the two
CORAVEL objects mentioned previously that were not observed at the CfA
(PELS~30 and PELS~39).  Targets that are known to be spectroscopic
binaries are distinguished with different symbols from stars not
previously flagged as binaries (Table~\ref{tab:meanRV}).  The
distribution shows a change in slope, or ``knee'', at a value of $e/i
\approx 2$.  The vast majority of objects with larger values are known
spectroscopic binaries reported in Sections~\ref{sec:orbits},
\ref{sec:longperiod}, and \ref{sec:otherpublished}, with solved orbits
or with long-term trends. We have therefore chosen to adopt this $e/i$ value
as our threshold for variability in the survey, and it is marked with
a dotted line in the figure. A few of the binaries with solved orbits
also appear below the line, along with several that have long-term
trends. The latter have been identified mostly by visual inspection
rather than by the scatter of the RVs.  Not surprisingly, the binaries
below the line tend to have small velocity amplitudes. We expect that
a few other cases with smaller $e/i$ values may also be long-period
and/or low-amplitude binaries, but those are undetectable given the
precision of our measurements.

Two objects not previously known to be binaries have $e/i$ values
larger than 2.0, and we therefore consider them to be velocity
variables: AK~I-2-121, and PELS~30. The first is rotating quite
rapidly ($v \sin i = 49~\kms$), and the second was observed only with
the CORAVEL, and has four measurements.

In addition to the $e/i$ statistic, we have also computed the $\chi^2$
value for each object along with its associated probability,
$P(\chi^2)$. We include those probabilities in Table~\ref{tab:meanRV}
as well, for readers who would like to use them to make their own
selection of variables. We have found, however, that $P(\chi^2)$ is
much more sensitive to outliers than $e/i$, and for any reasonable
probability threshold such as 0.01 or even 0.001, it leads to several
dozen objects with low $e/i$ values to be classified as variables,
whereas visual inspection suggests they are most likely just more
active. For this reason we have preferred to use the more conservative
$e/i$ statistic.

\section{Completeness}
\label{sec:completeness}

While our sample contains several dozen confirmed spectroscopic
binaries that have either computed orbits or obvious long-term trends,
or that are revealed by their excess radial velocity scatter ($e/i >
2$), we are likely to have missed others because of our time sampling
and the limited precision of our observations. This is surely the case
for systems with low-mass companions, or for longer-period binaries,
as not all our targets have been observed for the full duration of the
survey. To estimate our completeness, we simulated a large number of
binaries for each of our targets
and used the same $e/i$
criterion as above to decide when each synthetic binary would have
been recovered.  The number of recovered binaries divided by the
number of simulations for each object then represents our
completeness, or detection fraction.

We carried out these simulations for the 231 targets that have not
previously been flagged as spectroscopic binaries, and that have three
or more observations.  Synthetic binaries were generated using the
distributions of orbital elements for field stars, following the
general procedure described by \cite{Geller:2015}. We sampled the
orbital periods and mass ratios from the distributions of
\cite{Raghavan:2010} for solar-type binaries. For the eccentricities,
those authors proposed a distribution that is flat up to at least $e =
0.6$, with a deficit of higher values possibly caused either by
dynamical interactions or by a lack of measurements.
\cite{Geller:2012} showed that a good representation of the entire set
of eccentricities can be achieved with a Gaussian model having a mean
of 0.39 and a standard deviation of 0.31. We adopted the latter
function, and assumed circular orbits for periods shorter than the
tidal circularization period, which is $P_{\rm circ} = 7.2$ days in
the Pleiades (see Section~\ref{sec:circularization}).  We also
restricted the simulated binaries to configurations that are detached,
as determined from the masses and radii adopted for the two components
(see below).  Our simulations accounted for the correlation between
the mass ratios and periods that is illustrated in Figure~17 of
\cite{Raghavan:2010} \citep[see also][]{Geller:2015}.  Inclination
angles were assumed to be distributed isotropically, and arguments of
periastron and orbital phases were sampled from uniform distributions.
Approximate primary masses and radii sufficient for this purpose were
obtained from the PARSEC isochrone for the Pleiades shown in
Figure~\ref{fig:cmd}, interpolated at the $G_{\rm BP}-G_{\rm RP}$
color of each target as listed the Gaia EDR3 catalog. Secondary masses
then followed from the mass ratios.  Radii for the secondary
components, which are also needed to compute the size of the Roche
lobes and verify whether the configurations are detached, were assumed
to scale with their masses.  In generating binaries for any given
object, we randomly perturbed the primary masses using a Gaussian
distribution with a standard deviation of 50\%, to account for possible
systematic errors in the PARSEC model. With the primary and secondary
masses known, we then calculated the velocity semiamplitudes, and
predicted velocities at the actual times of observation for each
star. Finally, we added Gaussian noise corresponding to the measured
uncertainty at each epoch, and computed the $e/i$ values. We repeated
this for a population of $10^5$ synthetic binaries for each target.
Simulated systems resulting in $e/i > 2$ were considered to be
detectable, as with the real data.

\begin{figure*}
\epsscale{1.15}
\plotone{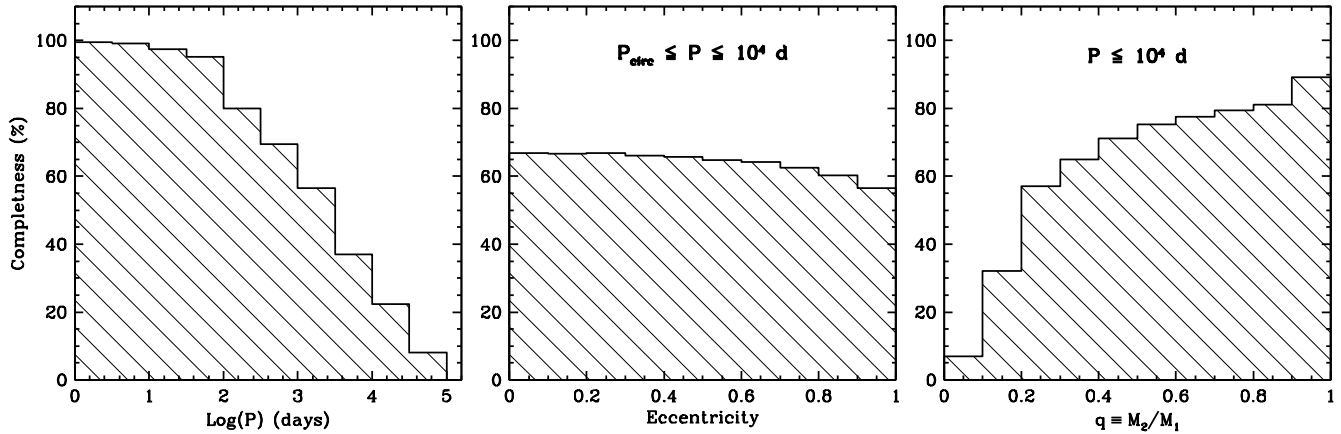}
\figcaption{Completeness curves for the Pleiades binaries from our
  Monte Carlo simulations. The panels show our sensitivity as a
  function of orbital period, eccentricity (for periods $P_{\rm circ}
  \leq P \leq 10^4$~days), and mass ratio ($P \leq 10^4$~days). See
  the text for an explanation of these
  ranges. \label{fig:completeness}}
\end{figure*}

We find that our survey is highly sensitive to binaries with orbital
periods up to 100 days (97\% completeness), and that we detect about
84\% of the binaries with periods less than 1000 days, and 67\% of the
ones up to $10^4$~days. We are increasingly less sensitive to longer
periods, as shown in the left panel of Figure~\ref{fig:completeness}.
Two of our targets have known orbital periods beyond $10^4$~days.  One
is HII~3197, with $P \approx 11,\!100$~days, but we do not include it
in the statistical analysis below because its orbit was determined
astrometrically, rather than from our RV measurements, which show
little variation. The other long-period system is the hierarchical
triple HII~2027, with $P \approx 13,\!400$~days for its outer orbit.
As a precaution we will exclude it as well, as the orbital elements
may have suffered changes due to internal dynamics over the lifetime
of the Pleiades, particularly for the inner pair.  All other binaries
have much shorter periods than this. Therefore, for the analysis of
the distribution of orbital parameters below, we have chosen to
restrict the study to binaries with periods less than $10^4$~days.

Our completeness as a function of eccentricity out to periods of
$10^4$~days, shown also in Figure~\ref{fig:completeness}, is fairly
flat and drops only slightly at the higher values. On the other hand,
while we have good sensitivity to binaries with equal mass components
(right panel), we miss many systems with low-mass secondaries,
particularly below $q = 0.2$. Note that the completeness curves in
each of these three parameters are not independent of each other. The
orbital period has the largest impact on detectability, because
shorter periods will typically lead to higher velocity amplitudes. The
overall lower completeness in both eccentricity and mass ratio is
caused in part by the longer-period binaries that are more difficult
to detect, and that occur at all mass ratios and eccentricities.

\section{Distribution of Orbital Elements}
\label{sec:distribution}

Armed with estimates of our ability to detect binaries with different
properties, we now investigate the distributions of their orbital
elements using the sample of 25 single-lined binaries in the Pleiades
with known orbits, and 20 double-lined systems. As illustrated below,
the majority of these binaries have primary components quite similar
in mass to the Sun. Given the relatively small size of the sample, we
refrain from dividing it up by mass, and will consider it as a single
set of ``solar-type'' binaries for comparison with other populations.

\subsection{Period Distribution}

\begin{figure}
\epsscale{1.15}
\plotone{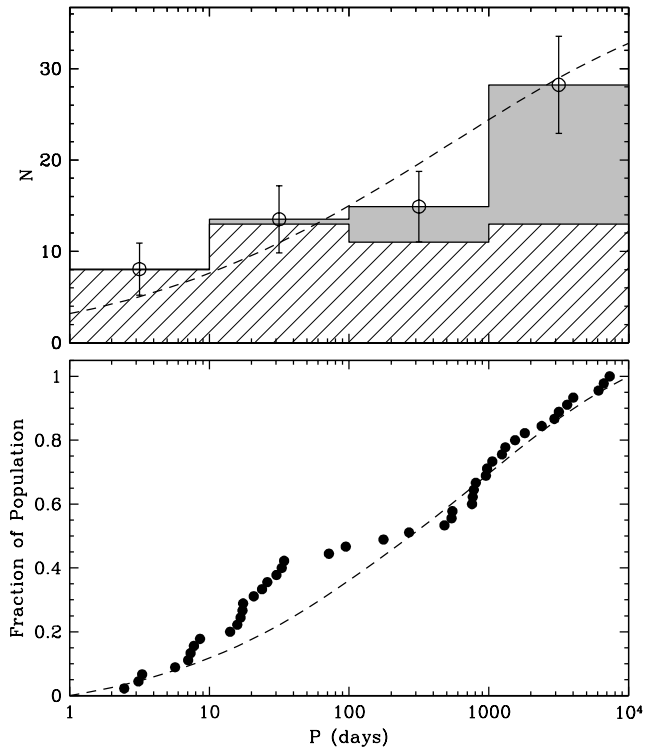}
\figcaption{Observed period distribution for the Pleiades binaries in
  our survey with periods up to $10^4$ days, shown as a hatched
  histogram in the top panel, and in cumulative distribution form in
  the bottom panel (filled circles). In the top panel, the solid gray
  histogram includes the corrections for incompleteness derived in
  Section~\ref{sec:completeness}. Error bars on the corrected
  distribution are from counting statistics. For reference, in both
  panels we show also the log-normal period distribution for field
  stars from \cite{Raghavan:2010}, normalized in the top panel so that
  the integral under the curve equals the number of binaries in the
  sample, after corrections for incompleteness. \label{fig:distribp}}
\end{figure}

In Figure~\ref{fig:distribp} (top) we show the distribution of orbital
periods for the 45 systems in our Pleiades sample with $P \leq
10^4$~days (i.e., excluding HII~3197 and HII~2027; see above). The
hatched histogram is the observed distribution, and the solid gray
histogram includes the corrections for incompleteness described in the
previous section. For reference, we show also with a dashed line the
log-normal period distribution from \cite{Raghavan:2010} for
solar-type field stars, which we have normalized to the same number of
binaries found in our sample up to $10^4$ days. The field distribution
is characterized by a mean period of $\log P = 5.03$ and a standard
deviation of $\sigma_{\log P} = 2.28$, with $P$ in units of days.

The cumulative distribution of the observed periods is shown in the
bottom panel, along with the corresponding distribution from
\cite{Raghavan:2010} transformed so that it reflects the same level of
incompleteness as the observed distribution. We obtained this curve by
multiplying the Gaussian model by our corresponding incompleteness
function from Figure~\ref{fig:completeness}, and then integrating.  An
Anderson-Darling test on the two distributions gives a $p$ value of
0.033, implying the shape of the cumulative distribution function of
binary periods in the Pleiades up to $10^4$ days is not statistically
distinguishable from that of solar-type binaries in the field.

\subsection{Eccentricity Distribution}

\begin{figure}
\epsscale{1.15}
\plotone{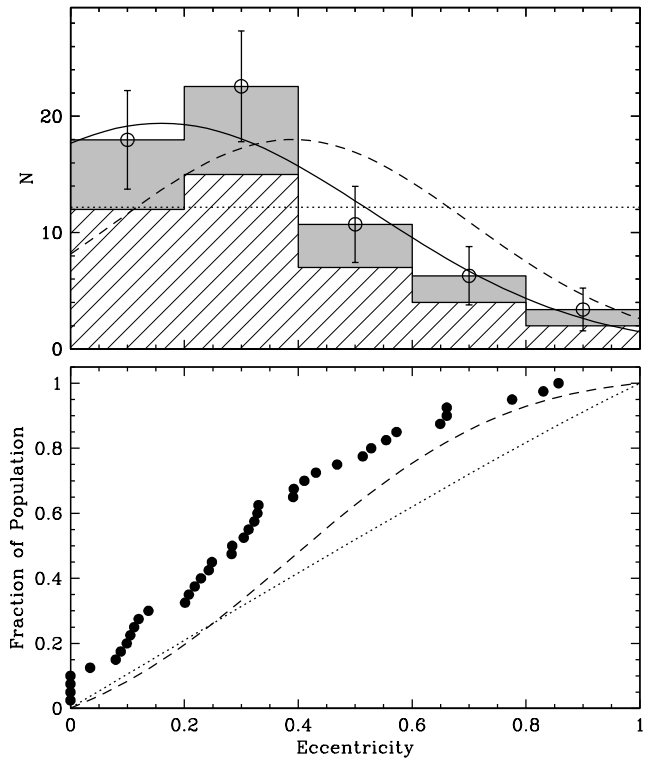}

\figcaption{Observed eccentricity distribution for the Pleiades
  binaries with $P_{\rm circ} \leq P \leq 10^4$~days, shown in the
  same way as the periods in Figure~\ref{fig:distribp}. The dashed
  line in the top panel is the Gaussian model proposed by
  \cite{Geller:2012} for field binaries of solar type, and the dotted
  line represents the uniform distribution of \cite{Raghavan:2010},
  both normalized to the same number of binaries in our sample.  The
  solid line corresponds to a Gaussian fit to the Pleiades
  distribution, with a mean of $e = 0.16$ and $\sigma_e = 0.37$.  The
  bottom panel shows the corresponding cumulative distribution
  functions, where the two models have our observational
  incompleteness applied so that they can be compared directly with
  the observations.\label{fig:distribe}}
\end{figure}

The observed eccentricity distribution for the 40 binaries with
periods between the circularization period (7.2~days) and $10^4$~days
is presented in Figure~\ref{fig:distribe} (top). Also shown in gray is
the corresponding histogram corrected for incompleteness, based on the
results in Section~\ref{sec:completeness}. We indicate with a dashed
line the eccentricity distribution proposed by \cite{Geller:2012} that
we adopted for our simulations in Section~\ref{sec:completeness},
which is a Gaussian with a mean of $e = 0.39$ and a standard deviation
of $\sigma_e = 0.31$. The dotted line is for the flat distribution
proposed originally by \cite{Raghavan:2010}. The corresponding
cumulative distribution functions are shown in the bottom panel, where
the filled circles represent the observed distribution. The two models
have been modified in the same way as explained above for the period
distribution, to include the same incompleteness as the observations.
Anderson-Darling tests indicate the observed distribution is
statistically distinct from both the Gaussian model and the flat
model, with $p$ values under 0.001 in both cases.\footnote{Strictly
  speaking, the flat model was proposed by \cite{Raghavan:2010} to be
  a reasonable representation for field binaries only up to $e = 0.6$,
  beyond which they noted a deficit of eccentric systems. Even over
  this restricted range, the Anderson-Darling test indicates a rather
  significant difference with the observed distribution in the
  Pleiades ($p$ = 0.0044).}  While qualitatively similar eccentricity
distributions as in the Pleiades are found in some older clusters such
as M67 \citep[4~Gyr;][]{Geller:2021} and NGC~188
\citep[7~Gyr;][]{Geller:2012}, in the sense of peaking at small values
and tailing off at high values, there may be differences in detail, as
we find in the field. The Pleiades distribution is well represented by
a Gaussian model with a mean of $e = 0.16$ \citep[smaller than found
  by][]{Geller:2012} and $\sigma_e = 0.37$. This is shown with a solid
line in the top panel of Figure~\ref{fig:distribe}.

\subsection{Mass Ratio Distribution}
\label{sec:massratio}

The distribution of mass ratios in a population of binaries provides
valuable constraints on their formation mechanisms
\citep[e.g.,][]{Bodenheimer:1993, Bate:1997, Clarke:2001,
  Halbwachs:2003}, and on the dynamical evolution of star clusters to
which the binaries belong \citep{Hut:1992, Benacquista:2013}.  A
typical sample of binaries with known orbits will usually consist of
both single-lined and double-lined systems.  The ones that are
double-lined provide a direct measure of $q \equiv M_2/M_1$ from the
ratio of the velocity semiamplitudes, $K_1/K_2$. For single-lined
binaries, on the other hand, the information is more limited and comes
only in the form of the mass function, $f(M)$, which requires both the
primary mass and the orbital inclination to be known in order to
calculate the mass ratio.

It is therefore common to apply statistical inversion techniques in
order to infer the shape of the mass-ratio distribution for SB1s from
the distribution of $f(M)$, or some function of $f(M)$, and to then
add in the SB2s. Many such methods with varying degrees of
sophistication have been developed \citep[see, e.g.,][]{Lucy:1979,
  Mazeh:1992, Heacox:1995, Halbwachs:2003, Carquillat:2007,
  Boffin:2010, Cure:2015, Shahaf:2017}. A few of them are parametric
in the sense of having to propose some a priori form for the
mass-ratio distribution of SB1s, while others are not. A common
assumption in all of these methods is that there is no other
information on the masses except perhaps for a rough estimate of
$M_1$, e.g., from a spectral type.

However, in a cluster such as the Pleiades with a known age and
metallicity, this is not necessarily the case. Brightness measurements
for the combined light of a binary contain useful information on the
individual masses that can be extracted if the parallax is known, and
if there is no contaminating flux from other sources. The only other
ingredient needed is a model isochrone for the cluster, such as the
one shown in Figure~\ref{fig:cmd}. With these constraints and the
dynamical information from the orbits, it is possible to estimate the
mass ratio directly for each SB1, obviating the need for a statistical
procedure. This is the approach we have chosen to follow here.  For a somewhat
similar application of this idea, see \cite{Goldberg:1994}.

In addition to using the brightness measurements for each binary in
the three Gaia bandpasses ($G$, $G_{\rm BP}$, $G_{\rm RP}$), we also
extracted the near-infrared $JHK_{\rm S}$ magnitudes from the 2MASS
catalog \citep{Cutri:2003}, which should be more sensitive to the
light contribution from the secondaries because they are of later
spectral type than the primaries. We corrected the magnitudes for
extinction following \cite{Cardelli:1989}, assuming an average
reddening for the cluster of $E(B-V) = 0.04$~mag, and adjusted the
magnitudes for the distance modulus of each SB1 using its parallax as
listed by Gaia EDR3. For the comparison of these magnitudes with the
predictions from the isochrone, we found it more convenient to use as
a constraint the minimum secondary mass, $M_2 \sin i = (2\pi G)^{-1/3}
P^{1/3} \sqrt{1-e^2} K_1 (M_1+M_2)^{2/3}$, as listed in
Table~\ref{tab:sb1}, rather than the mass function. For each binary we
then explored a range of $M_1$ and $\sin i$ values, and for each trial
pair $\{M_1, \sin i\}$ we solved for $M_2$ using the equation above,
requiring the mass ratio to be no larger than unity. If this last
condition was not met during the iterations, we simply reversed the
masses, as this does not affect the photometry for the combined
light. This was continued until the $\chi^2$ value between the six
observed magnitudes and the model fluxes at $\{M_1, M_2\}$ was
minimized.  We adopted the 2MASS uncertainties as given, but to be
conservative and allow for errors in the model fluxes, we inflated the
uncertainties for the Gaia magnitudes to 0.02~mag, as they are
typically much smaller (often less than 0.001~mag).

\begin{figure}[!b]
\epsscale{1.15}
\plotone{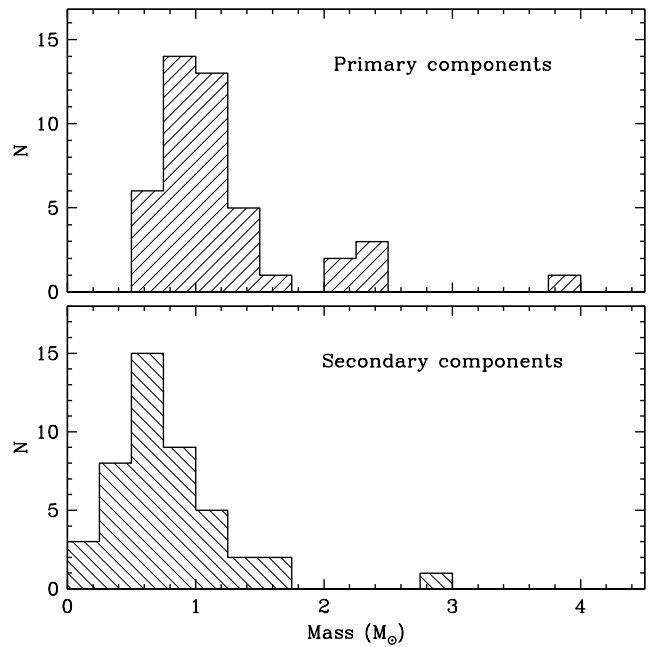}
\figcaption{Distribution of the primary and secondary masses of our
  SB1 and SB2 binaries, derived from the dynamical information provided
  by our orbital solutions combined with photometric information from
  Gaia and 2MASS (see the text).\label{fig:masshist}}
\end{figure}

The composition of our binary sample by mass is illustrated in
Figure~\ref{fig:masshist}. The panels show the nominal primary and
secondary values from the procedure described above, to which we have
added the corresponding values for the SB2s from a similar exercise
using the known mass ratios instead of the minimum secondary masses.
The primary masses for the ensemble of binaries are seen to be
concentrated around one solar mass.

To propagate uncertainties for the SB1s, we used a Monte Carlo
procedure whereby we repeated the fits to the photometry 1000 times
for each binary, perturbing the values of all measurements in each
simulation. This was done by adding Gaussian noise to $M_2 \sin i$
with a standard deviation equal to its uncertainty, and similarly for
the magnitudes. We also added Gaussian noise to the reddening in the
amount of 0.02~mag, to account for the fact that reddening is not
uniform across the cluster \citep[see][]{Breger:1986, Taylor:2008}.
Finally, we broadened the individual mass distributions by perturbing
each estimate by 20\% of its value, to allow for further systematic
errors in the model masses and to provide a degree of smoothing given
the relatively small size of the SB1 sample.  We then merged the
resulting individual mass-ratio distributions for all binaries to
construct the distribution for the ensemble.

Out of our sample of 25 SB1s, a search in the literature revealed that
one of them, HII~2500, has a visual companion currently separated by
0\farcs3 that is 1.7~mag fainter in the $K$ band \citep{Bouvier:1997},
and which is not the secondary in the binary (see
Section~\ref{sec:othermultiples}). As this third star is likely
affecting the photometry and may therefore bias the mass estimates, we
have chosen to remove it. Similarly, HII~563 has a companion about
1.6~mag fainter detected multiple times by the lunar occultation
technique, which appears also to not be the secondary in the SB1
system \citep[see][]{Torres:2020b}. We have excluded it as well,
leaving 23 objects in the sample with periods up to $10^4$~days.

\begin{figure}
\epsscale{1.15}
\plotone{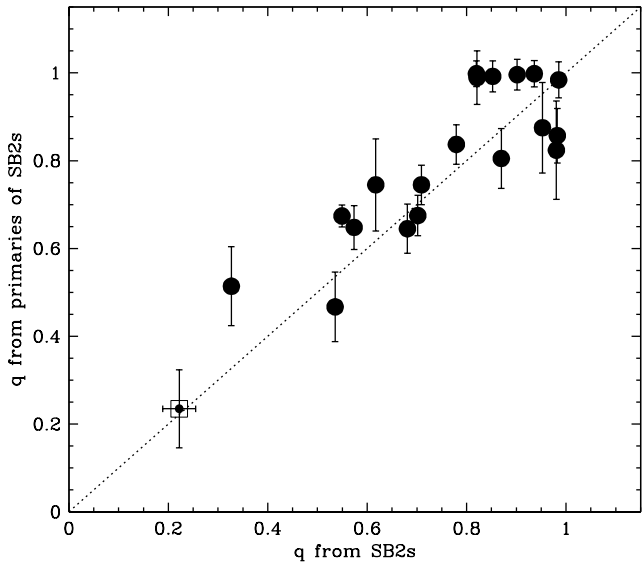}
\figcaption{Comparison between inferred and measured mass ratios for
  19 SB2s in the Pleiades, with HII~1338 excluded. The inferred values
  were derived by using the model isochrone shown in
  Figure~\ref{fig:cmd}, along with brightness measurements for the
  combined light from Gaia and 2MASS, and the measured minimum
  secondary masses calculated as if the system were single-lined. The
  dotted line represents the one-to-one relation. The smaller dot with
  a square corresponds to HII~2407, a single-lined eclipsing system
  used as a check (see the text). \label{fig:compareq}}
\end{figure}

Given that these mass ratios are strictly model-dependent, one may ask
to what extent systematic errors in the models might cause distortions
in the true shape of the mass-ratio distribution. To investigate this,
we used the sample of SB2s, and calculated the minimum secondary mass
from the orbital elements as if the secondaries had not been detected,
i.e., using only information about the primaries. We then followed the
same procedure explained earlier for the SB1s to infer the mass
ratios, and compared them with the $q$ values directly measured for
these systems. For this test we did not consider the triple-lined
system HII~1338 because the photometry is contaminated by the
light from the third star. Only one other SB2 (HII~1348) has a known
close visual companion at about 1\farcs1, but it is more than five
magnitudes fainter in the $K$ band, so its effect will be negligible.
The result of the comparison is shown in Figure~\ref{fig:compareq},
along with a line representing the one-to-one relation. The average
mass ratio difference in the sense $q_{\rm model} - q_{\rm obs}$ is
0.04, with a scatter of 0.11. The median absolute deviation from the
one-to-one relation is 0.076, which corresponds to about a 10\% error
for a typical mass ratio of $\sim$0.8 for this sample.

An additional check is available from the fact that one of our SB1s,
HII~2407, is an eclipsing binary discovered by \cite{David:2016} based
on Kepler/K2 photometry. The individual masses we derive using our
$M_2 \sin i$ value, the Gaia and 2MASS magnitudes, and the isochrone,
are $M_1 = 0.80\pm 0.04~M_{\sun}$ and $M_2 = 0.19 \pm 0.07~M_{\sun}$,
giving a ratio of $q = 0.24 \pm 0.09$.  These agree well with the
values $M_1 = 0.81 \pm 0.08~M_{\sun}$, $M_2 = 0.18 \pm 0.02~M_{\sun}$,
and $q = 0.22 \pm 0.03$ reported by \cite{David:2016}, supporting the
accuracy of our procedures at the low end of the distribution. We
represent our mass ratio for this system along with theirs as a small
dot and a square in Figure~\ref{fig:compareq}.  Another consistency
check is provided by HII~2147, for which \cite{Torres:2020} determined
individual masses from a combination of astrometric observations from
VLBI and radial velocities for the secondary only. They reported
primary and secondary masses of $0.978 \pm 0.024~M_{\sun}$ and $0.897
\pm 0.022~M_{\sun}$, giving $q = 0.917 \pm 0.004$. We obtained $0.946
\pm 0.049~M_{\sun}$, $0.918 \pm 0.049~M_{\sun}$, and $0.969 \pm
0.073$, respectively, which are within 3\% for the masses and within
6\% for $q$.

\begin{figure}
\epsscale{1.15}
\plotone{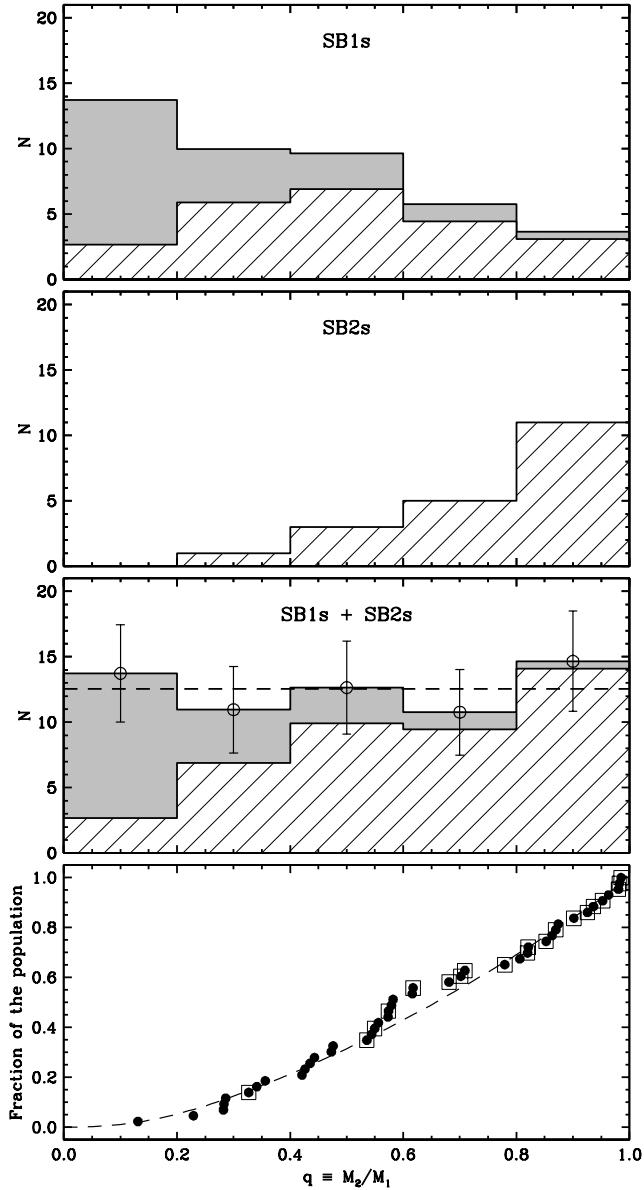}
\figcaption{Mass ratio distribution for the SB1s and SB2s (top two
  panels) with periods $P < 10^4$~days, along with the merged samples
  (third panel). Hatched histograms correspond to the observed
  distributions, and the gray histograms add the incompleteness
  corrections from Section~\ref{sec:completeness}. For the SB1s the
  observed distribution is based on estimates of the individual mass
  ratios using brightness measurements from Gaia and 2MASS, the
  measured minimum secondary masses, and a model isochrone for the
  Pleiades (see the text). Note that the smallest bin for the SB1s may
  be unreliable as the incompleteness corrections are four times
  larger than the number of observed binaries.  Error bars in the third
  panel are based on counting statistics, and the dashed line
  represents a flat distribution. The bottom panel shows the
  cumulative distribution function for the SB1s (dots) and SB2s (dots
  with squares), along with the curve corresponding to the flat
  distribution modified to have the same incompleteness as the
  observations. \label{fig:qdistrib}}
\end{figure}

These tests suggest that, while not perfect, our inferred mass ratios
for the SB1s in the sample are sufficiently accurate for our
statistical purposes. The distribution of the $q$ values is displayed
in histogram form in the top panel of Figure~\ref{fig:qdistrib}.
Because of the relatively small number of SB1s (23), we have chosen a
bin size of 0.2, equivalent to about twice the scatter in $q_{\rm
  model} - q_{\rm obs}$ mentioned earlier.  The gray histogram
includes the corrections for incompleteness from
Section~\ref{sec:completeness}.  As expected for SB1s, the
distribution rises toward smaller values, although the leftmost bin
should be interpreted with caution as the incompleteness corrections
are a factor of four larger than the actual number of binaries in that
range. A somewhat surprising feature of this distribution, however, is
that the rightmost bin is not empty, as would be expected for SB1s,
but in fact contains three binaries with mass ratios larger than 0.8,
which are more typical of SB2s. The three systems are AK~II-346,
HII~1762, and HII~2147.  In all three cases (and no others) there is
evidence that the star whose velocities we measured is the secondary
rather than the primary. This was confirmed for HII~2147 by
\cite{Torres:2020}. For the other two, the evidence is based on the
appearance of the cross-correlation profiles, which display broad
wings suggesting the presence of a much more rapidly rotating star
that we presume to be the (more massive) primary, in addition to the
narrow(er) peak that we are able to measure. The fact that we detect
both stars is then an indication of similar brightness, and therefore
of similar mass.

The second panel of Figure~\ref{fig:qdistrib} shows the distribution
of measured mass ratios for the 20 SB2s. In this case it is not
obvious that corrections for incompleteness should be applied, because
double-lined systems are often detectable from a single exposure, and
are then immediately placed on higher observing priority than stars
that only show single lines. We therefore present the distribution
without any corrections. The preference for higher mass ratios is
typical of SB2s in other populations, and responds to the fact that
companions are more easily detected when they are similar to the
primary in mass, and therefore in brightness.

The sum of the SB1 and SB2 histograms is shown in the third panel of
Figure~\ref{fig:qdistrib}, with error bars added to indicate formal
uncertainties from counting statistics. The distribution is consistent
with being flat, as indicated by the dashed line, with the caveat
mentioned above about the smallest bin. The cumulative distribution of
mass ratios, measured directly for the SB2s and inferred as explained
above for the SB1s, is shown in the bottom panel of the same figure.
The dashed line corresponds to the flat distribution modified as we
described previously for the period and eccentricity distributions, to
reflect the same level of incompleteness as the observations. There is
no distinction between the two based on the Anderson-Darling test.

An earlier study of the mass-ratio distribution in the Pleiades by
\cite{Goldberg:1994} was based on a sample of only 9 spectroscopic
binaries, more than four times smaller than ours, but also found it to
be essentially flat, or perhaps rising slightly toward $q = 1$.  A
comparison between the mass-ratio distribution in the Pleiades and in
other populations suggests there may be some differences. In NGC~188,
for example, \cite{Geller:2012} found a distribution that rises toward
lower mass ratios, along with a slight excess of binaries with $q >
0.9$.  The distribution for M67 was also found to be rising toward
small mass ratios \citep{Geller:2021}, but without an excess of
equal-mass binaries.  For solar-type binaries in the field,
\cite{Raghavan:2010} reported a roughly flat distribution in the range
$0.2 < q < 0.95$, with some evidence for an excess of equal-mass
pairs. A study of a sample of high proper motion field stars by
\cite{Goldberg:2003} found a rise down to $q \sim 0.2$, a drop below
that, and a peak near $q = 0.8$. A similar sample with a higher
fraction of SB2s was investigated by \cite{Mazeh:2003}, and their
results indicated the mass ratios follow a uniform distribution
between 0.3 and 1.0, with an apparent rise toward lower values down to
0.1. It is unclear to what extent these differences may depend on the
analysis methodologies, or on the details of the sample selection,
including the mass and period ranges.

\section{The Spectroscopic Binary Frequency in the Pleiades}
\label{sec:binaryfrequency}

As indicated earlier, only two of the objects in our sample with known
orbits have periods approaching the duration of our survey, which is
$\sim$15,000 days, or $\sim$40~yr: HII~3197 ($P \approx
11,\!100$~days), and the triple system HII~2027 ($P \approx
13,\!400$~days for the outer orbit). We do not consider the former
because its binary nature was discovered astrometrically rather than
spectroscopically \citep[see][]{Schaefer:2014}. All other systems have
periods roughly half that of the outer orbit of HII~2027, or
shorter. As for our studies of the distribution of orbital properties,
we therefore chose here to adopt a cutoff period of $10^4$~days for
computing the multiplicity frequency, which we define as the fraction of targets
in our sample that are in multiple systems, including triples. We
refer to it in the following simply as the binary frequency. Our
detection completeness in this period range is 67\%
(Section~\ref{sec:completeness}).

Of the 289 objects in our survey that we consider to be Pleiades
members and have three or more observations, we have identified 25
SB1s and 21 SB2s (we include here the inner binary of the triple
system HII~2027). Two additional objects have variable radial
velocities ($e/i > 2$), but we lack sufficient observations or
coverage to determine a period; we will assume here that both periods
are shorter than $10^4$ days. The raw binary frequency is then 17\%
(48/289), which after correction for incompleteness becomes $25 \pm
3$\%.  While this binary fraction includes stars of all spectral types
(from B to M), restricting the sample to just the ``solar-type'' stars
of spectral type FGK does not change the result.

\setlength{\tabcolsep}{8pt}
\begin{deluxetable}{lccc}
\tablecaption{Binary Frequency in the Pleiades Compared with Other Populations.\label{tab:binfreq}}
\tablehead{
\colhead{Population} &
\colhead{Age (Gyr)} &
\colhead{Frequency (\%)} &
\colhead{Source}
}
\startdata
Blanco~1  &  0.1      &  $20 \pm 6$    &     1        \\
Pleiades  &  0.125    &  $25 \pm 3$    &  This paper  \\
M35       &  0.15     &  $24 \pm 3$    &     2        \\
NGC~7789  &  1.6      &  $31 \pm 4$    &     3        \\
NGC~6819  &  2.5      &  $22 \pm 3$    &     4        \\
M67       &  4        &  $34 \pm 3$    &     5        \\
NGC~188   &  7        &  $29 \pm 3$    &     6        \\
Field     &  \nodata  &  $14 \pm 2$    &     7        \\
Halo      &  $\sim$10 &  $15 \pm 2$    &     8
\enddata

\tablecomments{Binary frequencies for solar-type stars in different populations up to orbital periods of $10^4$~days, corrected for incompleteness. Sources are: 
(1) \cite{Mermilliod:2008};
(2) \cite{Leiner:2015};
(3) \cite{Nine:2020};
(4) \cite{Milliman:2014};
(5) \cite{Geller:2021};
(6) \cite{Geller:2012}
(7) \cite{Raghavan:2010};
(8) \cite{Latham:2002}.
The estimate for Blanco~1 did not specify an upper limit to the
period, but is presumed to be similar to the
others \citep[see][]{Geller:2012}, and is not corrected for
incompleteness. Similarly for the halo sample. For Blanco~1 we have
added an uncertainty based on the number of binaries discovered in
that survey.}

\end{deluxetable}
\setlength{\tabcolsep}{6pt}  

Table~\ref{tab:binfreq} lists determinations of the binary fraction up
to the same orbital period for solar-type (FGK) stars in six other
open clusters, and in two field samples in the solar neighborhood
(disk stars in the field, and halo stars).  Interestingly, we find
that within the uncertainties, our estimate for the Pleiades is quite
consistent with what has been found in a variety of other populations
spanning nearly two orders of magnitude in age, except perhaps for the
field and halo samples. We note that the latter has not been corrected
for incompleteness, so it is only a lower limit. The binary frequency
for solar-type field stars appears decidedly smaller than in several
of the clusters, particularly NGC~7789, M67, and
NGC~188. \cite{Geller:2012} had already pointed this out for NCG~188,
and suggested it may be a dynamical signature, as $N$-body simulations
predict single stars are preferentially lost from a cluster through
evaporation because they are generally lighter than binaries.

Even though our sensitivity to binaries drops for longer periods, the
fact that we have uncovered about a dozen systems with long-term
velocity trends shows that we can still detect binaries with periods
that are longer than the duration of the survey.  These systems are
valuable because they begin to bridge the gap between spectroscopic
and astrometric binaries, and provide useful information on the
multiplicity fraction out to much larger binary separations.  While
their orbital periods are not known, they are likely to be a few times
longer than the span of our observations, perhaps of order a century
or two. Circumstantial evidence for this is given by the case of
HII~717, which shows a monotonically decreasing velocity drift
(Figure~\ref{fig:long}) and has a visual companion at 0\farcs213
\citep{Mason:1993} that is almost certainly the same as the companion
we detect spectroscopically. At the distance to the Pleiades, this
separation corresponds to a period of roughly a century, or perhaps
slightly longer. A few other objects with long-term RV drifts also
have visual companions, but they are much wider and are unlikely to be
the ones responsible for the RV variations.

If we then adopt a new period cutoff of $10^5$~days, corresponding to
about 270~yr, we find that our completeness is reduced to 49\%.
Including now in the tally the 12 systems with long-term drifts, the
raw multiplicity frequency out to this period is 60/289 = 21\%, or $42
\pm 4$\% once the correction for undetected binaries is applied. Based
on the study of \cite{Raghavan:2010}, solar-type binary systems in the
field occur with a frequency of about $22 \pm 3$\% up to the same
period of $10^5$~days, which is significantly lower than in the
Pleiades. These wider binaries are more susceptible to ejection from a
cluster, so it is possible that this difference is in fact a signature
of ongoing evaporation in the Pleiades, as proposed above for NGC~188
by \cite{Geller:2012}.

If the period distribution in the Pleiades is assumed to be consistent
with that of solar-type binaries in the field, an extrapolation of our
binary frequency from a cutoff of $10^4$~days to orbits of any period
results in a total fraction of $76 \pm 5$\%, again significantly
higher than the rate in the field \citep[44\%;][]{Raghavan:2010}.
Studies by others of the total binary frequency in the Pleiades using
the distribution of stars in the color-magnitude diagram have produced
similarly high estimates: \cite{Kahler:1999} reported 60--70\%, and
\cite{Converse:2008} obtained 68--76\%. We note, however, that the
extrapolation mentioned above may not be valid, as \cite{Deacon:2020}
have concluded that the wide binary fraction in the Pleiades (and also
in other clusters) is low: approximately 2\% between 300--3000~au,
corresponding to a range of $\log P$ in days of about 6.3--7.8.

\section{Astrometric Binaries}
\label{sec:astrometric}

Much wider binaries than we can detect spectroscopically in the
Pleiades have been found by adaptive optics, speckle interferometry,
and lunar occultation techniques. The most extensive and systematic
surveys (excluding those for substellar companions) have been carried
out by \cite{Bouvier:1997}, \cite{Richichi:2012}, and
\cite{Hillenbrand:2018}, and resulted in the discovery of several
dozen visual companions to our targets. Many others have been found
serendipitously by other investigators.  Most of these companions are
within a few arc seconds of the primaries, and are likely physically
associated with them.

Cross-matching our target list against the Gaia EDR3 catalog revealed
companions to another dozen or so objects, some much wider and fainter
than reported by others. Many of these have been announced previously
by \cite{Deacon:2020}. In several cases the companions are also on our
target list, and therefore have measured velocities. For many of them
Gaia also lists the proper motion and parallax, so it is possible to
confirm the physical association, or at least membership in the
cluster.  We have ignored any Gaia companions failing this check.

\setlength{\tabcolsep}{6pt}
\begin{deluxetable*}{llclcl}
\tablecaption{Astrometric Companions to our Target Stars.\label{tab:astrometry}}
\tablehead{
\colhead{} &
\colhead{Name} &
\colhead{WDS ID} &
\colhead{Discov.} &
\colhead{$\rho$ (\arcsec)} &
\colhead{Mag.\ diff.\ (mag)}
}
\startdata
\phn7   &  AK III-153 &  03313+2515 &  A 1825              &  2.9  &  4.86($G$)              \\
56*     &  HCG 65     &   \nodata   &  \nodata             &  1.3  &  0.08($G$)              \\
69*     &  HII 102    &  03434+2314 &  BOV   2    AB       &  3.6  &  3.12($K$), 5.34($G$)   \\
70      &  HII 97     &  03434+2500 &  BOV   1             &  0.7  &  1.62($K$)              \\
72*     &  TRU S45    &  03435+2244 &  OCC  97    Aa,Ab    &  0.1  &  0.0                    \\  
72*     &  TRU S45    &  03435+2244 &  STF 438 AB          &  1.7   &  1.0($V$), 0.88($G$)   
\enddata
\tablecomments{
Companions are mostly from the Washington Double Star Catalog (WDS),
and are listed only for cluster members. Columns after the target name
give the WDS identifier, the discoverer code and component
designations as listed in the WDS, the angular separation from the
most recent measurement available, and magnitude differences from the
WDS or other sources, including Gaia. Bandpasses are indicated in
parentheses, when available. An asterisk after the running number
calls attention to notes in the Appendix. (This table is available in
its entirety in machine-readable form.)}
\end{deluxetable*}
\setlength{\tabcolsep}{6pt}  

Table~\ref{tab:astrometry} is a compilation of 75 of our targets with
companions listed in the Washington Double Star Catalog
\citep[WDS;][]{Hartkopf:2001} separated by less than 20\arcsec,
supplemented with other discoveries from the literature. We restrict
the list to members of the cluster, and omit companions that have been
shown to be background stars. Targets with notes of interest in the
Appendix are flagged in the last column of the table. Three of the
companions are reported to be substellar: those around HII~1348
\citep{Geissler:2012}, AK~I-2-199 \citep{Konishi:2016}, and HII~1132
\citep{Rodriguez:2012}.

There are 58 objects in our sample that have one or more visual
companions that are \emph{not} also spectroscopic binaries, i.e., that
represent new multiple systems not considered previously. This list is
likely incomplete, however, in part because of the intrinsic
sensitivities of the various astrometric surveys. For example, the
Gaia EDR3 catalog is known to be missing many close pairs within about
1\farcs5 due to instrumental limitations, and many more below 0\farcs7
\citep{Fabricius:2020}.  Furthermore, not all our targets have been
examined for close companions.

If the total binary fraction in the Pleiades is as high as our
estimate of 76\% in the previous section, which assumes the same
period distribution as in the field, it would mean there should be a
total of 220 binary or multiple systems in our sample. The number of
spectroscopic binaries uncovered by our survey up to periods of $10^5$
days is 60, which after correction for incompleteness becomes 122
(60/49\%).  This period cutoff corresponds to an angular separation of
$\sim$0\farcs3 at the distance to the Pleiades, and most of the
astrometric binaries in Table~\ref{tab:astrometry} (43 out of the 58
that represent additional multiples) are indeed wider than this, so
they complement the spectroscopic observations well by sampling the
regime to which we have no sensitivity. We would therefore expect a
total of $220 - 122 = 98$ binaries wider than about 0\farcs3, whereas
we find only 43.\footnote{Note that the majority of the separations
  under 0\farcs3 are from lunar occultations, which are only lower
  limits because they represent the angular separation projected in
  the direction of the motion of the Moon during the event. Their true
  separations are unknown.} This would imply a total binary frequency
of $(122 + 43)/289 = 57$\%, which is independent of any assumption on
the shape of the period distribution, yet is still higher than in the
field. It is only a lower limit, however, as we have excluded lunar
occultation pairs that may actually be wider than 0\farcs3, and some
number of astrometric companions to our targets have likely gone
undetected.

\section{Mean Cluster Velocity and Internal Velocity Dispersion}
\label{sec:dispersion}

\begin{figure}
\epsscale{1.15}
\plotone{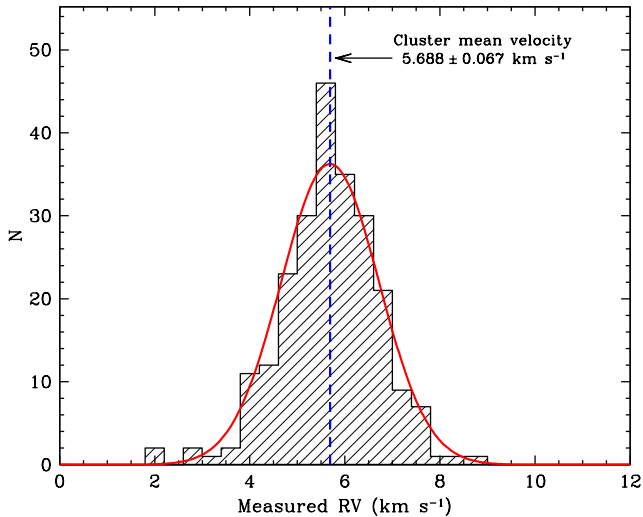}
\figcaption{Histogram of the weighted mean radial velocities for 234
  members of the Pleiades cluster. Binaries for which we only see
  long-term trends are excluded, as are the early-type stars reported
  by \cite{Torres:2020b} (see the text). For binaries with orbits, we
  used their center-of-mass velocities.\label{fig:meanRV}}
\end{figure}

A histogram of the weighted mean velocities for 234 cluster members is
shown in Figure~\ref{fig:meanRV}. It includes all objects with at
least three RV observations, but excludes the early-type stars
published earlier \citep{Torres:2020b} because their velocities were
measured in a different way that makes it difficult to place them on
the same reference system as the ones in this paper. Objects with
long-term trends have also been excluded, along with a few with very
large velocity uncertainties that make them unreliable. For binaries
with orbital solutions, we used the center-of-mass velocity rather
than the mean RV. We left out six of the binaries with orbital fits
(HII~164, HII~605, HCG~495, AK~V-198, HII~1392, and HII~1431) because our
solutions in Table~\ref{tab:sb2.1} show a significant primary/secondary
velocity offset of more than 0.5~\kms, which can potentially bias the
center-of-mass velocity (see Section~\ref{sec:orbits}).

The mean radial velocity, $5.688 \pm 0.067$~\kms, is very close to the
result from the Gaia mission (DR2), which is $5.65 \pm
0.09$~\kms\ \citep{Gaia:2018b}. The standard deviation of the 234
measurements is 1.03~\kms, although this does not represent the true
velocity dispersion of the cluster in the radial direction.
Observational errors will tend to inflate the scatter, as will the
presence of unrecognized binaries and any unrecognized non-members
there may still be in the sample.  Furthermore, because of the large
angular extent of the Pleiades on the plane of the sky, there is a
radial velocity gradient of several \kms\ across the cluster due to
changes in the projection of the mean space velocity along the line of
sight. Additionally, the measured radial velocities do not reflect the
true motion of a star's center of mass, as they are affected by the
gravitational redshift and convective blueshift, which depend on
spectral type. This can cause extra scatter in the measured
velocities.

The projection effect dominates the scatter of the raw velocities. We
removed this bias by subtracting from each star the radial velocity
predicted using the object's coordinates and the space velocity vector
of the center of the cluster, determined on the basis of the proper
motions and parallaxes of member stars \citep[see,
  e.g.,][eq.\ A.13]{Gaia:2017}. These predicted velocities, which
range between 3.5 and 8.2~\kms, are referred to as astrometric radial
velocities.  The components of the velocity vector in the equatorial
system, adopted from the latest determination from Gaia DR2
\citep{Gaia:2018b}, are:
\begin{eqnarray*}
V_x &=& \,\,\,-1.311 \pm 0.070~\kms \\
V_y &=& +21.390 \pm 0.105~\kms \\
V_z &=& -24.457 \pm 0.057~\kms.
\end{eqnarray*}
The mean residual velocity for the 234 stars (observed minus
predicted) is $\Delta = +0.015 \pm 0.050$~\kms, indicating excellent
agreement between the spectroscopic and astrometric velocities, on
average. The scatter around this value is reduced to 0.77~\kms.

\begin{figure}
\epsscale{1.15}
\plotone{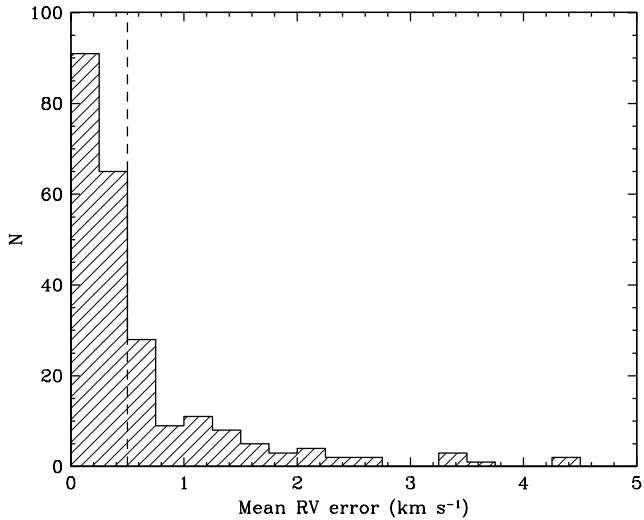}
\figcaption{Histogram of the errors of the weighted mean radial
  velocities for 234 members of the Pleiades cluster (see the text).
  Only objects with errors smaller than 0.5~\kms\ (dotted line) have
  been used for the calculation of the true velocity dispersion in the
  cluster.\label{fig:histerr}}
\end{figure}

Observational errors in our sample depend quite strongly on the
projected rotational velocity, among other factors. A histogram of the
errors of the mean velocities is seen in Figure~\ref{fig:histerr}.  To
ensure that we use the most precise measurements for the estimate of
the velocity dispersion in the Pleiades, we restricted the analysis to
the 157 objects with errors smaller than 0.5~\kms. The standard
deviation of the mean velocities for these objects, with the
projection effect removed, is 0.62~\kms. After correction for
observational errors following \cite{McNamara:1986}, we obtained a
preliminary estimate of the dispersion within the cluster of $0.56 \pm
0.04$~\kms.

This still contains the effect of undetected binaries, mostly with
very long periods. These systems will show constant velocities over
the duration of our survey, but their average RV may be slightly
offset from the expected radial motion in the cluster.  To account for
this bias as much as possible, we removed from the above set of 157
targets the 21 stars that have astrometric companions reported in
Section~\ref{sec:astrometric}, leaving 136 objects. The dispersion is
reduced to $0.50 \pm 0.04$~\kms, showing that astrometric binaries do,
in fact, contribute to the scatter.

Closer examination of the RVs with the projection effect removed
revealed a monotonically decreasing trend with effective temperature.
Dividing the sample into three bins (4000--5000~K, 5000--6000~K,
6000--7000~K), we obtained average differences between the
spectroscopic and astrometric velocities of $\Delta_1 = +0.440 \pm
0.073$~\kms, $\Delta_2 = +0.113 \pm 0.060$~\kms, and $\Delta_3 =
-0.175 \pm 0.090$~\kms, respectively. The difference between the first
and last bins is significant at the 5.3$\sigma$ level.

\begin{figure}
\epsscale{1.15}
\plotone{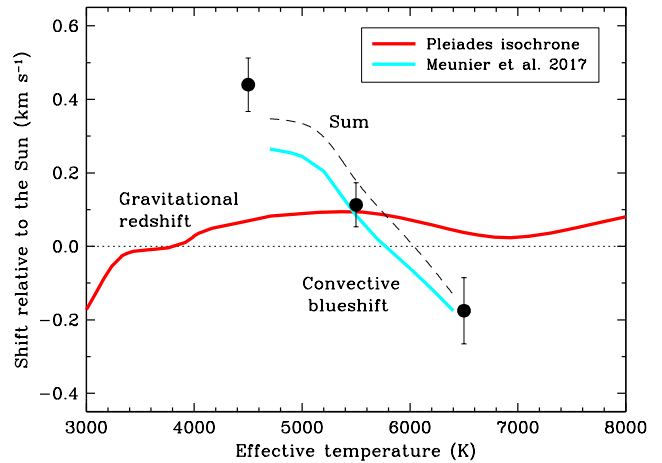}
\figcaption{Astrophysical components of the measured RVs, relative to
  the effects in the Sun. The differential gravitational redshift
  curve is based on masses and radii from the Pleiades isochrone shown
  in Figure~\ref{fig:cmd}. The convective blueshift curve, with the
  value for the Sun subtracted out, is taken from Figure~3 of
  \cite{Meunier:2017}. Adding these two effects together results in
  the curve represented with a dashed line. Dots with error bars
  correspond to the measured differences ($\Delta_1$, $\Delta_2$,
  $\Delta_3$) in three temperature bins between the spectroscopic and
  astrometric velocities for stars with mean RV errors smaller than
  0.5~\kms, and shows rather good agreement with the dashed
  curve. \label{fig:gravred}}
\end{figure}

The astrometric RVs measure the true motion of the center of mass of
each star, while the spectroscopic RVs do not, as mentioned earlier.
By construction, our RVs have the gravitational redshift of the Sun
taken out (0.63~\kms), as well as its convective blueshift, because
our velocity zero-point is based on observations of minor planets in
the solar system, which represent reflected sunlight.  Stars with
different properties than the Sun will have their velocities affected
differently by those two effects, and will therefore be biased to some
degree.  We illustrate the expected magnitude of these effects in
Figure~\ref{fig:gravred}. The curve for the differential gravitational
redshift is based on the masses and radii tabulated in the same
Pleiades isochrone we have used throughout this paper, and is seen to
vary relatively little over most of the temperature range shown. On
the other hand, the convective blueshift has a steeper dependence on
temperature. The curve shown was derived from Figure~3 of
\cite{Meunier:2017}, which is based on measurements in a sample of
F7--K4 stars, after subtracting out the value for the Sun from the
same figure ($-0.34$~\kms). The dashed line in our
Figure~\ref{fig:gravred} corresponds to the sum of the two effects,
and is dominated by the convective blueshift.  Our measurements of the
difference between the spectroscopic and astrometric velocities as a
function of temperature ($\Delta_1$, $\Delta_2$, $\Delta_3$) also
capture both effects, and are represented in the figure by the dots
with error bars.  The observations appear quite consistent with
expectations, supporting the accuracy of our radial velocities, and
showing they are precise enough to enable us to detect the dominant
effects of varying convective blueshifts among solar-type stars in the
Pleiades.

These systematic differences with spectral type make our earlier
estimate of the cluster's velocity dispersion slightly larger than it
should be. To remove this bias, we performed a simple linear fit to
the differences between the measured and predicted velocities as a
function of temperature, and subtracted it from those differences. The
resulting scatter is reduced slightly to $0.48 \pm 0.04$~\kms. This
represents our best estimate of the internal dispersion in the Pleiades
cluster in the radial direction.

Earlier estimates of the one-dimensional (1D) dispersion in the
cluster have varied significantly. \cite{Jones:1970} reported
0.46~\kms\ based on proper motions from photographic plates, while
\cite{Makarov:2001} obtained 0.69~\kms\ and 0.22~\kms\ for a sample of
weak and strong X-ray sources, respectively, using proper motions from
the Tycho-2 catalog \citep{Hog:2000}.  We have rescaled these
three determinations here from their originally reported values in
order to account for the difference between the adopted distance to
the cluster in those studies and the current estimate of 136~pc from
Gaia \citep{Gaia:2018b}. An additional estimate by \cite{Galli:2017}
using proper motions from \cite{Buoy:2015} and Tycho-2 gave a
1D dispersion of 0.8~\kms. \cite{Rosvick:1992} used radial velocity
observations for 34 stars with the CORAVEL instrument, and reported an
estimate of 0.68~\kms, which they noted was only an upper limit due to
undetected long-period binaries. \cite{Raboud:1998} also used CORAVEL,
and obtained 0.36~\kms\ from a sample of 67 stars.

The Gaia mission now presents an opportunity to obtain high-precision
estimates of the velocity dispersion in two directions orthogonal to
our own determination. 
To perform this calculation, we
used the membership list from \cite{Gaia:2018b}, and filtered it using
the quality criteria recommended by \cite{Arenou:2018} (their eqs.\ 1
and 2) to retain only stars with good astrometric solutions. This
still left many stars with statistically significant excess
astrometric noise (Gaia parameter ${\tt
  astrometric\_excess\_noise\_sig} > 2$), which we chose to also
exclude as in some cases this may be caused by unrecognized
binarity. We then removed projection and distance effects following
\cite{Gaia:2017} (eq.\ A.13), using the same space velocity vector as
above. Because the predicted p.m.\ components subtracted from the
measurements to remove projection effects are directly proportional to
the parallax, we applied the further condition that the relative
errors in the parallaxes should be 2\% or less. With this, and an
adopted distance to the cluster of 136~pc, we obtained identical
\mbox{1D} velocity dispersions in the right ascension and declination
directions of $0.48 \pm 0.02$~\kms\ (corrected for observational
errors), based on 292 stars. Some contamination by non-members and
remaining long-period binaries affecting the Gaia proper motions is to
be expected, so this may represent an upper limit.  In any case, these
determinations happen to agree with our value in the radial direction,
strongly suggesting the velocity field is isotropic. Our best estimate
of the 3D velocity dispersion in the Pleiades is then $0.83 \pm
0.03$~\kms.

Our determination of the internal dispersion enables an estimate of
the total mass of the Pleiades through the virial theorem, on the
assumption that the cluster is in dynamical equilibrium. Following
\cite{Geller:2015}, the total mass can be expressed as $M_{\rm tot} =
10 r_{\rm hp} \sigma_r^2 / G$, where $r_{\rm hp}$ is the projected
half-mass radius, $\sigma_r^2$ is the velocity dispersion in the
radial direction, and $G$ is the gravitational constant. This relation
assumes the velocity dispersion is isotropic, a conclusion our results
above seem to support. Adopting a projected half-mass radius of $53
\pm 10$~arc~min from \cite{Raboud:1998}, which at the 136~pc distance
to the cluster corresponds to $2.1 \pm 0.4$~pc, we obtain $M_{\rm tot}
= 840 \pm 200~M_{\sun}$. This value is consistent with estimates by
others done in different ways \citep[e.g.,][]{Raboud:1998,
  Pinfield:1998, Adams:2001, Converse:2008, Danilov:2020}.

\section{Tidal Circularization}
\label{sec:circularization}

The eccentricity versus log period diagram is a valuable diagnostic
tool to investigate the effectiveness of tidal forces in binary
systems belonging to coeval populations. It relies on the fact that
tidal forces tend to circularize the orbits through energy
dissipation, although the precise mechanisms are still not well
understood. As the strength of tidal forces is very sensitive to the
orbital separation \citep[e.g.,][]{Zahn:1975}, coeval binaries of
similar mass are observed to have circular orbits up to a certain
orbital period, beyond which the orbits display a range of
eccentricities. For older and older populations, the effects of tidal
forces affect wider and wider binaries and the critical period
separating circular from eccentric orbits increases.

The precise definition of this transition period has varied among
different studies \citep[see, e.g.][]{Mazeh:2008}. Nevertheless,
estimates are now available in a variety of populations of different
ages and different masses, including G-type stars in the solar
neighborhood \citep{Duquennoy:1991, Raghavan:2010}, A-type stars
\citep{Matthews:1992}, M67 \citep{Latham:1992b, Geller:2021}, NGC~188
\citep{Mathieu:2004, Geller:2012}, M35 \citep{Meibom:2005,
  Leiner:2015}, NGC~7789 \citep{Nine:2020}, NGC~6819
\citep{Milliman:2014}, Hyades/Praesepe \citep{Meibom:2005}, the
Pleiades \citep{Mermilliod:1992a}, and even among giants
\citep{Mermilliod:1992b, Verbunt:1995}, halo stars
\citep{Latham:2002}, and pre-main-sequence stars \citep{Melo:2001}.
These studies have confirmed that the transition period, which we
refer to here as the tidal circularization period $P_{\rm circ}$,
following \cite{Meibom:2005}, varies as a function of age, and it has
even been proposed as a clock to date coeval populations
\citep{Mathieu:1988}. The trend with age has been shown most recently
by \cite{Nine:2020}, compared against several theoretical predictions.

The latest study of tidal circularization in the Pleiades is that of
\cite{Meibom:2005}. These authors defined a procedure to determine the
circularization period involving the fitting of a function dependent
on several parameters that control its shape. They calibrated these
parameters with binaries in several clusters as well as with numerical
simulations. They then used a sample of 12 solar-type binaries in the
Pleiades known at the time from the work of \cite{Mermilliod:1992a,
  Mermilliod:1997}, and reported a circularization period of $7.2 \pm
1.9$~days.

\begin{figure}
\epsscale{1.15}
\plotone{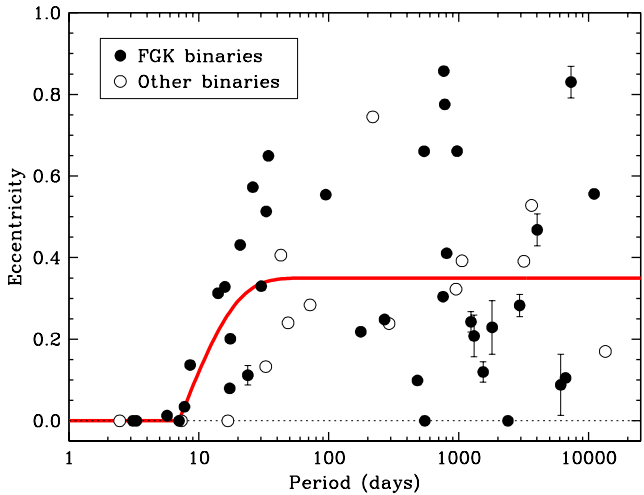}
\figcaption{Eccentricity versus log period diagram for the 52 binaries
  in the Pleiades cluster with orbital solutions. The curve represents
  the circularization function of \cite{Meibom:2005} fitted to the 38
  binaries of spectral type FGK (filled symbols), resulting in a
  circularization period $P_{\rm circ} = 7.2 \pm 1.0$~days. Open
  symbols represent binaries of spectral type B, A, and
  M. \label{fig:elogp}}
\end{figure}

There are now a total of 52 main-sequence binaries or triples in the
cluster with well characterized orbital solutions, including several
new systems near that critical orbital period. They range in spectral
type from B to M, and they are shown in a diagram of eccentricity
versus orbital period in Figure~\ref{fig:elogp}. The circularization
period may depend on stellar mass to some degree \citep[see,
  e.g.,][]{Mathieu:1988}, and most studies of the $e$-$\log P$ diagram
have focused on stars more or less similar to the Sun. We have
therefore done the same, and restricted the sample to primaries of
approximately solar type (filled circles), broadly defined here as
those with spectral type FGK. As a precaution we have left out the
triple system HII~2027 (with two entries), which may have suffered
eccentricity changes due to internal dynamics between the inner binary
and the outer perturber. We are left with 38 systems. Our fit of the
circularization function of \cite{Meibom:2005} gives a circularization
period of $P_{\rm circ} = 7.2 \pm 1.0$~days, which is identical to
that of those authors, but about twice as precise.\footnote{As a
  check, we note that one of the fixed parameters of this function,
  the average eccentricity of binaries with periods longer than 50
  days, was set by \cite{Meibom:2005} to a value of 0.35 based on the
  mean eccentricity found for binaries in the Pleiades, M35,
  Hyades/Praesepe, M67, and NGC 188. We have verified for our expanded
  sample of 38 solar-type systems in the Pleiades that the average for
  $P > 50$~days is indeed 0.35, with a formal uncertainty for the mean
  of 0.06.}  Including the remaining twelve B, A, and M-type systems
gives essentially the same value, $P_{\rm circ} = 7.3 \pm 1.0$~days.

We note that DH~794 has an orbital period of 5.69~days that is shorter
than $P_{\rm circ}$, yet the orbit appears to be slightly eccentric
($e = 0.0126 \pm 0.0022$). A possible explanation is the presence of a
tertiary component, given that dynamical interactions with the inner
pair can pump up the eccentricity in a short-period system that
normally would have already been circularized \citep{Mazeh:1990}.
While no such companions have been directly imaged around DH~794, the
Gaia EDR3 catalog indicates excess astrometric noise that is
statistically significant, and which could be due to an unresolved
companion, although there may be other causes.  It is also possible
that DH~794 was formed with a very high eccentricity, which would delay
circularization.

\section{Prospects for Absolute Mass Determinations from Gaia}
\label{sec:masses}

With its long observational history and close distance to the Earth,
the Pleiades cluster would seem like an ideal place to test models of
stellar evolution using well measured properties of stars. The most
fundamental of these is the mass. As of this writing, however,
model-independent absolute mass determinations have only been made for
four binary systems in the Pleiades. One is the B-type
spectroscopic-interferometric binary Atlas
\citep[27~Tau;][]{Zwahlen:2004}, which has formal fractional errors
for the masses of 5--7\%, making the measurements only marginally
useful.  Another is the A-type eclipsing system HII~1431
\citep[HD~23642;][]{Munari:2004, Southworth:2005, Groenewegen:2007,
  David:2016}, which is in our sample. A third system is HCG~76
\citep[V612~Tau;][]{David:2016}, also an eclipsing system with
low-mass components of spectral type M. The last is HII~2147, a
spatially resolved G-type binary on our target list that was studied
separately by \cite{Torres:2020}.

The Gaia mission promises to provide many more examples. It will
detect the orbital motion of virtually all binaries in the Pleiades
astrometrically, even for periods longer than the duration of the
observations. However, with the instrument's angular resolution of
$\sim$0\farcs1, as currently stated in the mission's online
documentation\footnote{\url{https://www.cosmos.esa.int/web/gaia/science-performance}}
\citep[see also][]{Gaia:2016}, the individual components will remain
spatially unresolved at the 136~pc distance to the cluster for orbital
periods shorter than 35~yr, if the total mass is 2~$M_{\sun}$, or
50~yr, if it is 1~$M_{\sun}$. Only the motion of the photocenter will
be detected. Modeling this motion will provide in many cases the
inclination angle of the orbit, along with the angular size of the
photocentric ellipse ($a^{\prime\prime}_{\rm phot}$) and other shape
and orientation elements, but not the true semimajor axis
($a^{\prime\prime}$), without which the absolute masses cannot be
determined. For this to be possible, the binaries must have SB2
orbits. The absolute masses of the components can then be calculated
from the spectroscopic minimum masses $M_{1,2} \sin^3 i$ and the
inclination angle $i$.

For the most stringent tests of models, the fractional mass errors
(and therefore also those of the minimum masses from spectroscopy)
should be smaller than about 3\% \citep[see, e.g.,][]{Andersen:1991,
  Torres:2010}. Not counting the eclipsing system HII~1431 mentioned
above, there are 14 SB2s in Table~\ref{tab:sb2.2} that satisfy this
requirement, and should therefore yield high-quality masses at the
conclusion of the Gaia mission. This will more than quadruple the
number of current systems with absolute mass determinations.

In addition to the masses, it will also be possible to derive the
individual brightness (absolute magnitude) of each component in the
Gaia $G$ band, even if the pair is not resolved.  This can be done by
combining the Gaia parallax and inclination angle with the total
projected semimajor axis in linear units from spectroscopy ($a_{\rm
  tot} \sin i$; Table~\ref{tab:sb2.2}) to give $a^{\prime\prime}$, and
then using the photocenter semimajor axis $a^{\prime\prime}_{\rm
  phot}$, also from Gaia.  These last two quantities are related by
the classical expression $a^{\prime\prime}_{\rm phot} =
a^{\prime\prime} (B-\beta)$ \citep[e.g.,][]{VanDeKamp:1981}. Here $B
\equiv M_2/(M_1+M_2) = q/(1+q)$ is the mass fraction, and $\beta
\equiv \ell_2/(\ell_1+\ell_2) = 1/(1+10^{0.4 \Delta G})$ is the
fractional luminosity, with $\Delta G$ being the magnitude difference
in the Gaia bandpass. As the mass ratio $q$ is known for SB2s, the
brightness difference in the Gaia bandpass immediately
follows. Finally, the measured combined-light $G$-band magnitude along
with $\Delta G$ and the parallax provide the absolute magnitude of
each star. In this way it becomes possible to compare the measurements
with the predicted $G$-band model fluxes at each mass. Our flux ratios
from {\tt TODCOR} near a wavelength of
5187~\AA\ (Table~\ref{tab:sb2.2}) can be easily converted to $V$,
providing another constraint on the model fluxes for the same masses.

Alternatively, if the binaries are only single-lined and are
unresolved by Gaia, individual component masses can still be derived
if the systems can be spatially resolved from the ground at least once
using other techniques. Most of our SB1s are beyond the sensitivity
limits of current long-baseline interferometers such as CHARA, but the
ones with the longer periods should be reachable with adaptive optics
observations on large telescopes.

\section{Summary and Conclusions}
\label{sec:summary}

In this paper we have derived key properties of the binary population
in the Pleiades cluster based on more than 6100 radial velocity
measurements for 377 stars, obtained over 39 years using four
different telescope/spectrometer combinations at the CfA. We augmented
this material with some 1150 other velocity measurements of similar
precision obtained by the CORAVEL team, overlapping substantially in
time with ours. The total duration of the observations is more than 43
years.

Nearly three dozen new spectroscopic binary and multiple systems have
been identified, and their orbital solutions are presented here. The
periods range from 3~days to more than 36~yr. The current tally of
spectroscopic systems (binaries and triples) with orbital solutions in
the Pleiades stands at 52, including systems reported by others. An
additional 10 objects show long-term drifts in their radial
velocities, implying periods longer than the duration of the survey.

With the enlarged sample of binaries, we have revisited the
determination of the tidal circularization period and obtained $P_{\rm
  circ} = 7.2 \pm 1.0$ days for FGK stars, identical to the estimate
of \cite{Meibom:2005}, but with an uncertainty reduced by half. The
binaries that carry the most weight for this determination are those
with periods near $P_{\rm circ}$.  Given this, and because of the high
degree of completeness of our binary survey for short periods (97\%
for periods up to 100 days), our $P_{\rm circ}$ result for FGK binaries seems
unlikely to change much in the future.

An investigation of the binary orbital properties of the sample shows
the distribution of orbital periods (up to $10^4$ days) to be similar
in shape to that of solar-type binaries in the field
\citep{Raghavan:2010}, after corrections for incompleteness. The
eccentricity distribution for $P > P_{\rm circ}$, on the other hand,
is different \citep[see][]{Raghavan:2010, Geller:2012} and is well
represented by a Gaussian with a mean of $e = 0.16$ and $\sigma_e =
0.37$. We find the distribution of mass ratios to be flat, again
similar to that in the field, except that we do not find an excess of
equal-mass pairs as \cite{Raghavan:2010} proposed.

The fraction of binaries in the Pleiades up to periods of $10^4$ days
is $25 \pm 3$\%, which is nearly double that of solar-type binaries in
the field and in the halo.  This excess of binaries is similar to what
has been found in several other open clusters (Blanco~1, M35,
NGC~7789, NGC~6819, M67, NGC~188) and has been proposed as evidence
that single stars have been preferentially lost to the clusters
through dynamical interactions, consistent with results from $N$-body
simulations.  Accounting for the several dozen astrometric binaries in
the cluster discovered in adaptive optics and lunar occultation
surveys, as well as wide pairs identified by Gaia, we estimate the
total binary fraction in the Pleiades to be at least 57\%. This lower
limit is again higher than the multiplicity frequency of solar-type
stars in the field \citep[44\%;][]{Raghavan:2010}.  Estimates of the
total binary frequency by others are even higher \citep{Kahler:1999,
  Converse:2008}.

Taking the average radial velocity of 234 member stars, we derive a
mean RV for the cluster of $5.688 \pm 0.067$~\kms\ on the velocity
system defined by minor planets in the solar system, in excellent
agreement with the result from Gaia DR2. The internal velocity
dispersion in the radial direction is $0.48 \pm 0.04$~\kms, after
accounting for perspective effects, observational errors, and
long-period (astrometric) binaries that we are not sensitive to. The
precision of our velocities is high enough that we are able to detect
and remove from the dispersion calculation a systematic change in the
average velocity as a function of effective temperature over the range
4000--7000~K. We ascribe this effect mostly to changes in the strength
of convective blueshifts affecting the Doppler measurements
\citep{Meunier:2017}, and to a lesser degree to variations in the
gravitational redshift. The magnitude of the changes we measure agrees
very well with expectations for these astrophysical effects. Proper
motion measurements from Gaia result in new estimates of the velocity
dispersions in the right ascension and declination directions of $0.48
\pm 0.02$~\kms\ each, which are identical to our measurement in the
radial direction, and support the conclusion that the velocity field
is isotropic.

Finally, with our measure of the dispersion in the radial direction,
we have used the virial theorem to estimate the total mass of the
Pleiades. Our result, $M_{\rm tot} = 840 \pm 200~M_{\sun}$, is in good
agreement with earlier estimates using different methodologies.

\begin{acknowledgements}

This paper is dedicated to the memory of John R.\ Stauffer
(1952--2021), who contributed so much to our knowledge of the Pleiades
cluster.  A large number of observers, students, and colleagues at the
CfA have contributed to gathering the spectroscopic observations for
this survey over the past four decades. We thank them all. We are also
grateful to R.\ Davis and J.\ Mink for maintaining the CfA Digital
Speedometer and TRES databases over the years. Helpful comments and
suggestions by the anonymous referee and the statistics editor are
appreciated.  We acknowledge the long-term support from the
Smithsonian Astrophysical Observatory that enabled this study, in the
form of telescope time and instrumentation.  The research has made
extensive use of the SIMBAD and VizieR databases, operated at the CDS,
Strasbourg, France, and of NASA's Astrophysics Data System Abstract
Service.  We also acknowledge the use of the Fourth Catalog of
Interferometric Measurements of Binary Stars \citep[][and online
  updates]{Hartkopf:2001}, and of the Washington Double Star Catalog,
maintained at the U.S. Naval Observatory. The work has used data from
the European Space Agency (ESA) mission {\it Gaia}
(\url{https://www.cosmos.esa.int/gaia}), processed by the {\it Gaia}
Data Processing and Analysis Consortium (DPAC,
\url{https://www.cosmos.esa.int/web/gaia/dpac/consortium}). Funding
for the DPAC has been provided by national institutions, in particular
the institutions participating in the {\it Gaia} Multilateral
Agreement.

\end{acknowledgements}

\appendix
\label{sec:appendix}
\section{1. Notes}

Included below are notes of interest for many of the objects in our
sample, concerning spectral peculiarities or issues with the radial
velocity measurements, multiplicity, or comparison with RV measures by
others in a few cases.  To facilitate reference to the tables in the
main text, we include the object number from Table~\ref{tab:sample}
(in order of increasing right ascension) along with the SIMBAD name.
\vskip 5pt

\vni{1.~AK~III-31.} \cite{Mermilliod:1997} found this star to be a
5~day SB1, and presented an orbit.  A very faint secondary is seen in
our spectra, making it an SB2.  Independent estimates of the secondary
temperature and rotational velocity were not possible from our
observations. A template for that star for the {\tt TODCOR}
measurements was selected based on the primary temperature with the
help of properties predicted from a model isochrone for the cluster.
Although the center-of-mass velocity agrees with the cluster mean,
Gaia EDR3 places the object in the foreground of the Pleiades at half
the distance, so it is not a member. This explains its location far
above the main-sequence.

\vni{5.~AK~III-79.} Our mean RV and the Gaia p.m.\ indicate this is a
non-member, even though the parallax agrees with the cluster mean.

\begin{figure*}[b!]
\epsscale{1.10}
\plotone{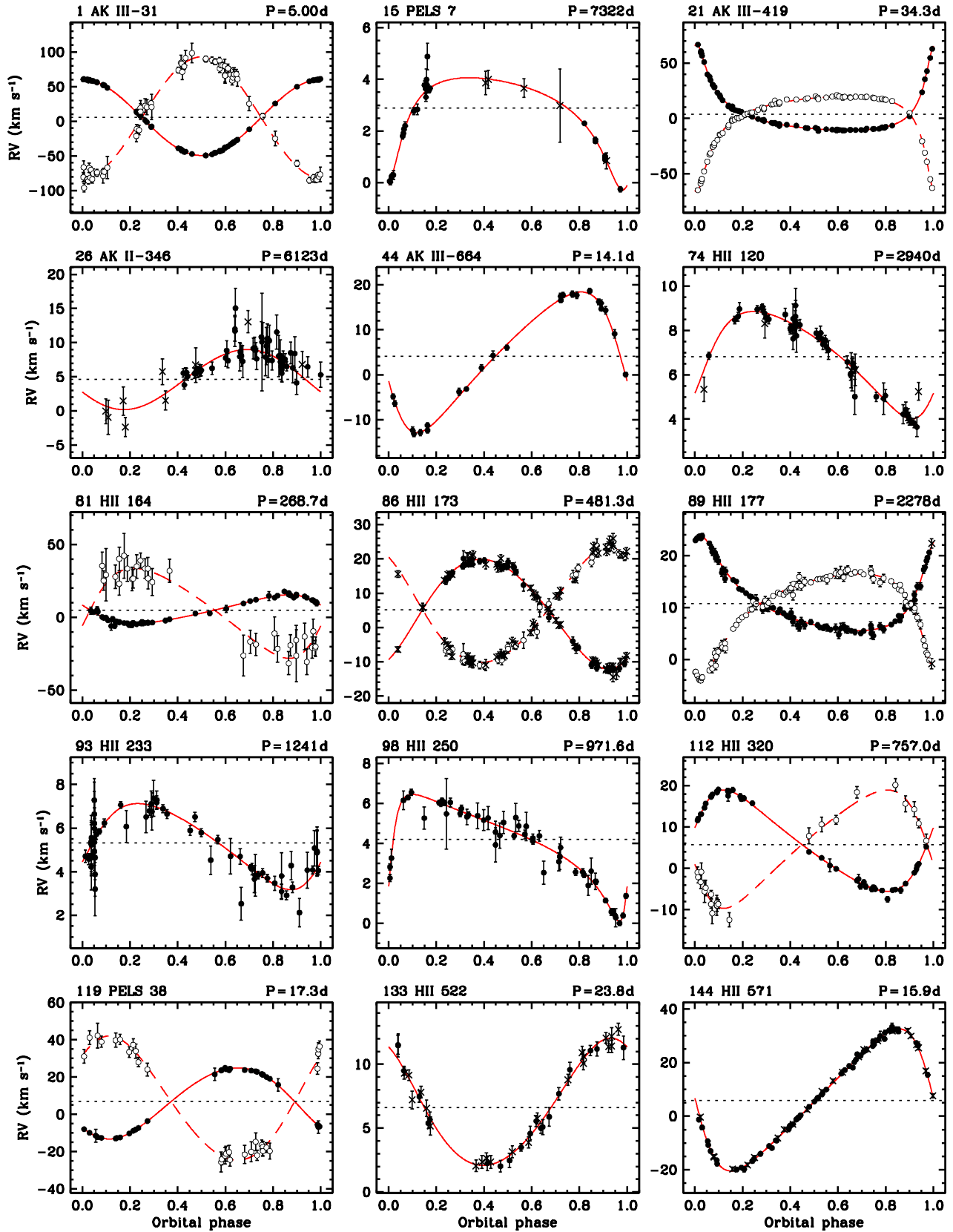} \figcaption{Plots
  of orbital solutions. Solid symbols are used for the primary
  velocities, open symbols for the secondary, and crosses for
  velocities from CORAVEL or other sources. The dotted line represents
  the center-of-mass velocity.\label{fig:sb.1}}
\end{figure*}

\vni{15.~PELS~7.} Our measurements show the velocities to be clearly
variable. A minimum occurred in September of 2017, and a handful of
CORAVEL observations include another low point in October of 1995. An
SB1 orbital solution with a period of 20~yr is presented in
Table~\ref{tab:sb1} and shown in Figure~\ref{fig:sb.1}, but is only
tentative due to poor phase coverage. The center-of-mass velocity is
inconsistent with cluster membership, and the Gaia catalog indicates
the parallax and p.m.\ are as well.

\vni{21.~AK~III-419.} Our observations reveal this to be a 34~day SB2.

\vni{22.~AK~III-416.} The Gaia EDR3 parallax and p.m., as well as its
velocity, indicate this is a non-member. \cite{Mermilliod:1997} found
it to be a single-lined binary, and published an orbit with a 10.9~day
period.

\vni{26.~AK~II-346.} There is spectroscopic evidence for the presence
of a narrow-lined star and a broad-lined star from the shape of the
cross-correlation functions (CCFs), which show
asymmetries. \cite{Queloz:1998} also reported it to be an SB2. While
we are not able to reliably detect the broad-lined star due to strong
blending, the sharp-lined star (which we assume is the secondary)
shows clear signs of long-term variability. However, our observations
alone are not sufficient to determine an orbit unambiguously, due to a
gap in the data between our Digital Speedometer and TRES
measurements. The object was also observed by the CORAVEL group
\citep{Rosvick:1992, Mermilliod:2009}, and fortunately those ten
measurements fill the gap and point to a period of about 6100~days.
Our orbital fit results in a large coefficient for the minimum mass of
the companion of 0.38~$M_{\sun}$, consistent with our presumption that
it corresponds to the more massive primary star, and which we do not
detect directly because of its rapid rotation.  On one of their
epochs, \cite{Mermilliod:2009} claim to have resolved the companion
and report for it a velocity measurement of +39.01~\kms\ that seems
much too large, if the orbit determined here is approximately correct,
because it would imply unreasonable minimum masses for the components
of 27 and 4 solar masses.  We have chosen to disregard that
measurement here, and use only their ``primary'' velocity at that
epoch, which fits our orbit well. We consider our solution in
Table~\ref{tab:sb1} to be somewhat preliminary at this time, although
we do note that an old Mt.\ Wilson velocity from 1921 reported by
\cite{Abt:1970} is consistent with our model, and so is the median RV
reported by Gaia (gathered over a timespan much shorter than the
period of the orbit).  The orbital period we determine corresponds to
a semimajor axis of about 67~mas at the distance to the Pleiades. We
are not aware of any visual companions reported for this object.

\vni{32--33.~TRU~S25~A, TRU~S25.} The similar SIMBAD names suggest an
association, but the stars are nearly 4\arcdeg\ apart on the sky.
TRU~S25~A is a non-member according to Gaia and our RV measurements. TRU~S25
is listed in the WDS as a visual binary, but the accompanying note
indicates this is dubious. Its velocities have been reported by
\cite{Torres:2020b}.

\vni{35.~TRU~S26.} RVs together with a 71~day SB1 orbit for this
rapidly-rotating A star were reported by \cite{Torres:2020b}. A
previously published 4.7~day orbit by \cite{Pearce:1975} is incorrect.

\vni{43.~PELS~25.} The mean RV is lower than the cluster mean, but the
entry in the Gaia EDR3 catalog shows the parallax and p.m.\ to be
consistent with membership.

\vni{44.~AK~III-664.} Our observations show this to be a 14~day SB1.

\vni{56.~HCG~65.} A 1\farcs3 astrometric companion is listed in the
Gaia EDR3 catalog.

\vni{69.~HII~102.} This was observed with the CORAVEL by
\cite{Mermilliod:1992a}, but variability was not noticed. Our own
velocities also show little change, but are 2--3~\kms\ higher. A time
history of all the observations (Figure~\ref{fig:long}) indicates this
is a very long period binary. Additional velocities by
\cite{Soderblom:1993} and \cite{White:2007} fit the trend, as does the
median value listed in the Gaia EDR3 catalog. A faint visual
companion at 3\farcs6 is listed in the WDS and has an entry in the
Gaia catalog, but seems too wide to be the spectroscopic secondary.
The system may therefore be a hierarchical triple.

\vni{72.~TRU~S45.} A close 0\farcs1 astrometric companion
discovered by lunar occultations has been reported only once in 1930,
and is unconfirmed.  A wider, 1\farcs7 companion to the south is
confirmed by Gaia to be physically associated. Our spectroscopic
observations do not resolve it, but signs that this is a close binary
are seen from the broad wings of the CCF and the impression that the
spectral lines are diluted compared to those of other stars of similar
spectral type. This suggests the presence of a very rapidly rotating
star, in addition to the one we are able to measure velocities for. We
presume the star we measure is the fainter secondary of the 1\farcs7
pair, from the fact that its spectroscopically determined temperature
is consistent with the $G_{\rm BP}-G_{\rm RP}$ color from Gaia,
whereas the Gaia color of the primary star is much bluer.

\vni{74.~HII~120.} RV variability was first noticed by
\cite{Raboud:1998} based on 5 measurements. An 8~yr orbit is reported
for the first time in the present paper. \cite{Bouvier:1997} detected
a close astrometric companion, but considered the target to be a
non-member; Gaia confirms membership. No parallax or p.m.\ are listed
in Gaia EDR3 for the 3\farcs4 astrometric companion listed in the WDS
(Table~\ref{tab:astrometry}). If the visual companion is physically
associated, it would make the system a triple.

\vni{80.~MT~41.} The parallax and p.m.\ components are somewhat
different than the cluster mean, but the very high value of ${\rm {\tt
    RUWE}} = 10.146$ indicates the astrometry may be biased. The
rotation period from \cite{Rebull:2016a, Rebull:2016b} is consistent
with membership. We retain it as a possible member.

\vni{81.~HII~164.} The secondary is very faint, and is only seen in
our TRES spectra. The best parameters for the corresponding template
could not be determined independently, and are merely educated
guesses. As different choices change the secondary velocities to some
degree, we consider the orbit for that component (specifically, $K_2$)
to be preliminary.

\vni{86.~HII~173.} This SB2 was found by \cite{Mermilliod:1992a}, who
published the first orbit with a period of 1.3~yr. We have combined
all CORAVEL observations \citep[including others
  by][]{Mermilliod:2009} with ours to update the orbital solution.

\vni{89.~HII~177.} \cite{Queloz:1998} reported this as an SB2, but no
orbit was presented. A single CORAVEL observation by
\cite{Mermilliod:2009} that resolves the components agrees with the
2280~day orbit we report here. This is a non-member based on the
center-of-mass velocity, as well as information from the Gaia mission.

\vni{93.~HII~233.} \cite{Mermilliod:1992a} reported a drift in their
CORAVEL velocities suggesting a long orbital period. Our observations
reveal it to be an SB1 with a period of 3.5~yr.

\vni{98.~HII~250.} We find this to be a 970~day SB1.

\vni{100.~HII~263.} RV variability was first noticed by
\cite{Raboud:1998}.  Merging the CORAVEL velocities with ours indeed
shows this to be a binary with a period longer than the duration of
our survey. No visual companions have been reported.
\cite{Prosser:1993} reported an anomalous dip in brightness in late
1992, and speculated it might be an eclipse. However, the velocities
now show that event to have occurred near the bottom of the RV curve,
so an eclipse by this companion is ruled out. Spottedness on the
primary seems like a natural explanation.

\vni{103--104.~HII~298, HII~299.} This is a 5\farcs7 pair with
membership confirmed by Gaia EDR3. The brighter star is HII~299,
which is slightly hotter than the companion, as expected.  The
velocities are similar.

\vni{106--107.~HII~303, HII~302.} This is a 16\arcsec\ pair showing
similar RVs. The northern (brighter) component, HII~303, is in turn a close
1\farcs8 visual binary that we have observed separately only once. We
refer to it in this paper as HII~303~B. The CCFs for all other
observations of HII~303 show variable widths, which may be caused by
variable, seeing-dependent contamination from the close companion.

\vni{112.~HII~320.} This is an SB2. \cite{Mermilliod:1992a} reported
it as single-lined, and presented the first orbit with a period of
2~yr, but noted their suspicion that the companion is a
rapidly-rotating star, which they could not measure. We confirm this,
and are able to measure the secondary RVs in our TRES
spectra. Application of {\tt TODCOR\/} to our Digital Speedometer
spectra shows hints of the secondary, but the velocities for that star
are too poor to be useful. Nevertheless, the use of {\tt TODCOR\/} has
the benefit of avoiding ``peak-pulling'' for the primary
velocities\footnote{``Peak-pulling'' refers to a bias in the measured
  velocities for a star caused by blending with an unresolved
  companion. The 1D cross-correlation peak will have a (sometimes
  subtle) shoulder on the side of the companion, resulting in a RV for
  the visible star shifted in the direction of the companion, i.e., in
  the direction of the center-of-mass velocity of the binary.
  Depending on the brightness of the companion, this will generally
  lead to a velocity semi-amplitude for the visible star that is too
  small.}. We find those primary velocities to be consistent with
those from TRES, so they have been incorporated into our final
solution. We have chosen not to use the CORAVEL velocities in our
updated orbital solution because the semiamplitude of the primary from
those velocities is significantly smaller than ours, compared to the
errors, suggesting the CORAVEL measurements may be affected by
peak-pulling. A $v \sin i$ measurement for the primary star from
CORAVEL \citep[$10.8 \pm 0.5$~\kms;][]{Queloz:1998} agrees with our
estimate of $10 \pm 2$~\kms. For the secondary we measure $v \sin i =
35 \pm 4$~\kms, and a cooler temperature than the primary.

\vni{113.~HII~338.} Our RVs show long-term variability, possibly
associated with a 3\farcs8 visual companion listed in the WDS.

\vni{119.~PELS~38.} \cite{Rosvick:1992} reported it as a spectroscopic
binary, on the basis of three observations. We find it to be a 17~day
SB2 with a faint secondary, and present the first orbit. The {\tt
  TODCOR\/} template for the secondary is an educated guess.

\vni{123.~AK~I-2-199.} A 0\farcs49 visual companion was reported by
\cite{Konishi:2016}, and is substellar.

\vni{127.~HII~468.} This is Electra (17~Tau). The two astrometric
companions discovered by lunar occultations may be one and the same
\citep[see][]{Richichi:1996}. A 100~day spectroscopic orbit reported by
\cite{Abt:1965} is spurious \citep[see][]{Torres:2020b}. RV
measurements were reported separately in the latter work.

\vni{129.~HII~476.} \cite{Raboud:1998} claimed their RVs
indicated a drift. We do not see that in our measurements, and our
$e/i$ metric classifies the object as non-variable.

\vni{133.~HII~522.} An SB1 orbit with a 23.8~day period was presented
by \cite{Mermilliod:1992a}, which is consistent with ours. Those
CORAVEL observations have been included with our own in an updated
solution.

\vni{134.~HII~514.} A 0\farcs78 astrometric companion was reported by
\cite{Makarov:2001} from a special reduction of the Tycho-2
observations.

\vni{138.~HII~541.} An astrometric companion has been discovered by
lunar occultations. See also \cite{Torres:2020b}.

\vni{140.~HII~563.} This is Taygeta (19~Tau). RVs and a tentative
8.7~yr spectroscopic orbit for this object were reported by
\cite{Torres:2020b}, who presented new velocities that are
inconsistent with an earlier orbit \citep{Abt:1965}. An astrometric
companion is known from lunar occultations. The spectroscopic and
astrometric companions may be the same.

\vni{144.~HII~571.} A 15.9~day SB1 orbit was published by
\cite{Mermilliod:1992a} that is consistent with ours. We have combined
the observations to update the solution. A 3\farcs8 visual companion
is listed in the WDS, and has a separate entry in Gaia. This would
make it a triple system.

\begin{figure*}
\epsscale{1.15}
\plotone{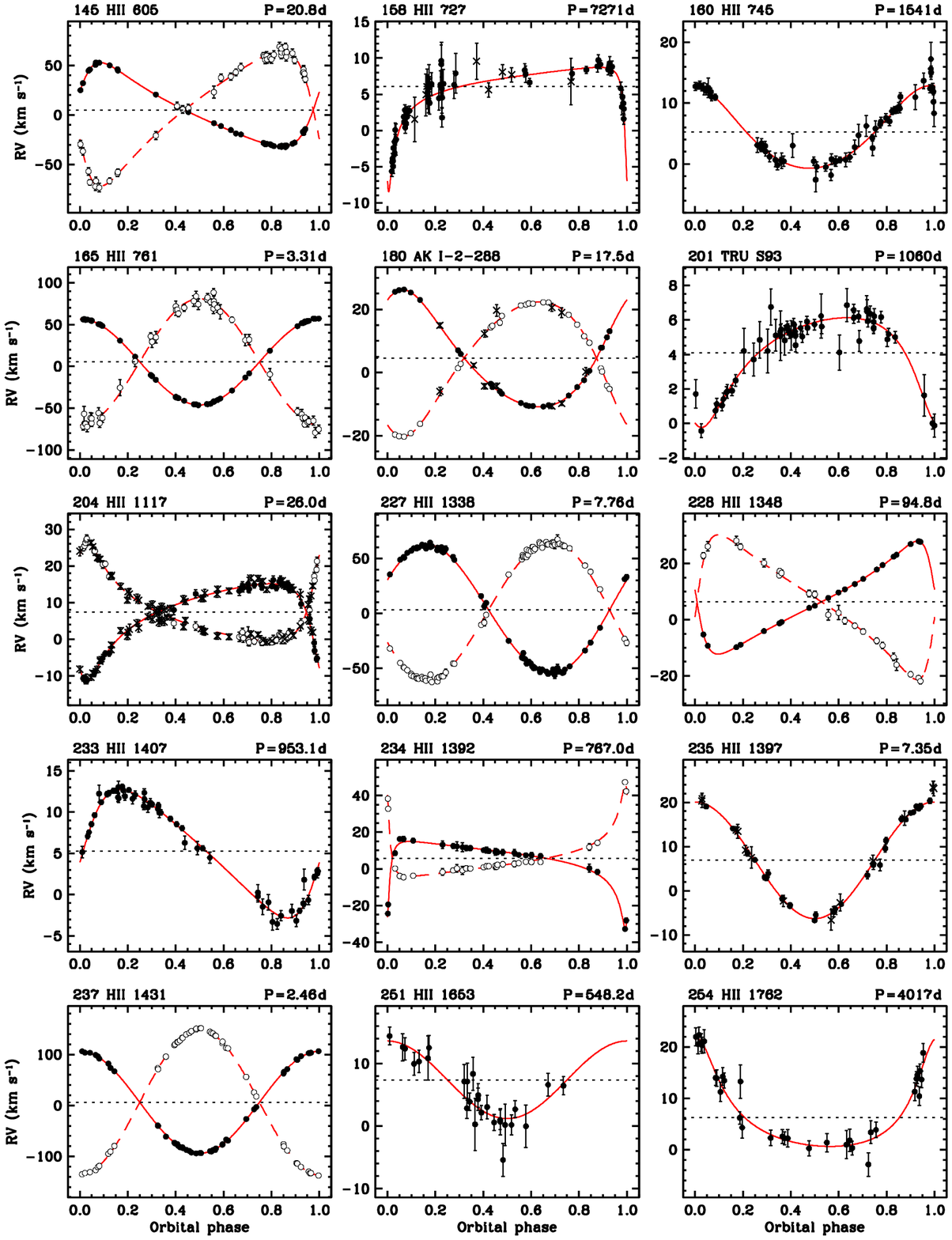}
\figcaption{Plots of orbital solutions (continued).\label{fig:sb.2}}
\end{figure*}

\vni{145.~HII~605.} \cite{Mermilliod:1992a} reported not being able to
measure this object with CORAVEL because of its rapid rotation. Our
observations show instead that it is double-lined, with components of
relatively sharp features. The secondary is too faint to determine its
rotational velocity independently. We adopt for it $v \sin i =
0$~\kms. Our SB2 orbital solution with a period of 21 days is
presented in Tables~\ref{tab:sb2.1}--\ref{tab:sb2.2} and shown in
Figure~\ref{fig:sb.2}.

\vni{156.~HII~717.} A close visual companion is known at a
separation of 0\farcs2, corresponding to a period of roughly a
century. The RVs show a slow downward drift consistent with this (see
Figure~\ref{fig:long}). RVs by \cite{Liu:1991} and
\cite{Soderblom:1993} agree with the trend.  A wider, very faint, and
physically associated 5\farcs3 companion is listed in the Gaia EDR3
catalog, which would make the system a triple.

\vni{158.~HII~727} This is a long-period binary with incomplete phase
coverage from our own observations, showing a single periastron
passage in late 2019. By adding six CORAVEL observations from
\cite{Mermilliod:2009}, two from \cite{Liu:1991}, and one from
\cite{Soderblom:1993}, we are able to find a satisfactory orbital
solution with a period of about 20~yr, and a high eccentricity of $e =
0.83$. The minimum secondary mass is fairly large, but we do not
detect the companion in our spectra.

\vni{160.~HII~745.} We find this to be a 4.2~yr SB1. Although the
minimum secondary mass is quite large, we are not able to detect the
secondary with confidence.  Two measurements of the primary star by
\cite{Liu:1991} fit our orbit well. The Gaia catalog lists a very wide
astrometric companion at 15\farcs8 that shares the parallax and
p.m.\ of the primary. The system may thus be triple.

\vni{162.~HII~739.} An astrometric companion was discovered by lunar
occultations.

\vni{165.~HII~761.} A 3.3~day SB1 orbit was reported by
\cite{Mermilliod:1992a}. We have found it to be double-lined and
measured the secondary's velocities. However, the secondary is too
faint to determine its temperature independently, so we have relied
for that on an estimate from a model isochrone for the cluster.

\vni{169.~HII~785.} This is Maia (20 Tau). An astrometric
companion was discovered by lunar occultations.

\vni{172.~HII~817.} This is Asterope (20 Tau). An astrometric companion
is reported in the Gaia EDR3 catalog.

\vni{175--176.~HII~879, HII~883.} The WDS lists this as a wide 17\farcs5
visual pair, the brighter primary being HII~879. Gaia EDR3 gives
nearly identical parallaxes and p.m.

\vni{177.~HII~890.} A visual companion discovered by
\cite{Bouvier:1997}, currently at 1\farcs2, is also listed in the
Gaia EDR3 catalog, but the entry has no parallax or p.m.

\vni{178.~HII~885.} A 0\farcs9 visual companion discovered by
\cite{Bouvier:1997} is also listed in the Gaia EDR3 catalog, but
the entry has no parallax or p.m.

\vni{180.~AK~I-2-288.} \cite{Mermilliod:1997} first reported this object
as an SB2, but lacked sufficient observations to solve for the
orbit. Their seven measurements are combined with ours for the new SB2
orbital solution presented in Tables~\ref{tab:sb2.1}--\ref{tab:sb2.2}.

\vni{182.~HII~916.} RV variability was first noticed by
\cite{Raboud:1998}. The combination of the CORAVEL measurements with
our own does seem to support a long-term trend (see
Figure~\ref{fig:long}).

\vni{185.~HII~956.} Visual binary with a preliminary orbit by
\cite{Popovic:1995} giving an angular semimajor axis of
$a^{\prime\prime} = 1\farcs9$ and a period of 1400~yr. Another
determination by \cite{Malkov:2012} reports $a^{\prime\prime} =
1\farcs2$ and $P = 900$~yr. The star whose velocities we are able to
measure has a line broadening corresponding to $v \sin i = 55$~\kms,
but the lines appear shallower than expected for its temperature,
suggesting dilution by another star. Estimates in the literature of
the rotational velocity of HII~956 are much higher than we measure:
\cite{Smith:1944} reported $v \sin i = 150$~\kms, \cite{Morse:1991}
gave 200~\kms, and \cite{Kounkel:2019} estimated 174~\kms. We
speculate these determinations correspond to the primary star of the
pair, which would explain the dilution we see. The velocities we
measure would therefore be for the secondary.

\vni{199.~HII~1084.} An astrometric companion was discovered by lunar
occultations.

\vni{201.~TRU~S93.} Our observations show this to be a single-lined
binary with a period of 2.9~yr.

\vni{203.~HII~1100.} The Gaia EDR3 catalog reports that the 0\farcs8
visual companion listed in the WDS has a slightly different p.m.\ than
the target, although the Gaia quality flags indicate the astrometric
solutions were problematic (large {\tt RUWE} values for both
components).

\vni{204.~HII~1117.} An SB2 orbit with a period of 26~days was
published by \cite{Mermilliod:1992a}. Additional velocities were
reported by \cite{Mermilliod:2009}, though some of them have the
primary and secondary interchanged. All of these CORAVEL observations
have been incorporated into our updated orbital solution.

\vni{206.~HII~1132.} A 2\farcs6 astrometric companion reported by
\cite{Geissler:2012} and \cite{Yamamoto:2013} is substellar.

\vni{218.~HII~1284.} RVs have been reported by \cite{Torres:2020b}.
Our new effective temperature estimate is $7740 \pm 200$~K.

\vni{219.~HII~1298.} The closer of two astrometric companions reported
in the WDS (0\farcs6) was discovered by lunar occultations. It may be
the same as the 1\farcs2 companion.

\vni{221.~HII~1306.} An astrometric companion was discovered by lunar
occultations.

\vni{227.~HII~1338.} See Section~\ref{sec:triples}.

\vni{228.~HII~1348.} \cite{Queloz:1998} reported this to be an SB2
based on three spectra from the ELODIE instrument, although no
velocities were published. Our velocities yield the 95~day SB2 orbit
presented in Tables~\ref{tab:sb2.1}--\ref{tab:sb2.2}. A visual
companion at 1\farcs1 is known, which is substellar
\citep[see][]{Geissler:2012, Yamamoto:2013}. This would therefore be
an interesting triple system with a circumbinary brown dwarf.

\vni{232.~HII~1375.} The astrometric companion listed in the WDS was
discovered by lunar occultations.

\vni{233.~HII~1407.} Our observations show this to be a single-lined
binary with a period of 2.6~yr.

\vni{234--235.~HII~1392, HII~1397.} This pair is separated by 5\farcs7.
The brighter star, HII~1397, is a metallic-line star, and an SB1 with a
7~day orbit that was first reported by \cite{Conti:1968}. Our
independent solution is consistent with theirs. The elements reported
in Table~\ref{tab:sb1} combine all observations together. We find
HII~1392 to be an SB2 with a 2~yr period and a large eccentricity ($e =
0.82$), which would make this a quadruple system.  \cite{Raboud:1998}
reported four measurements of HII~1392 with CORAVEL, but they are all
near times of conjunction so the double nature of the source was not
detected.  \cite{Queloz:1998} measured a projected rotational velocity
of $15.7 \pm 1.6$~\kms\ that is larger than our values for the two
components (10--12~\kms); this is likely the result of line blending.

\vni{237.~HII~1431.} This is a 2.5~day eclipsing binary (HD~23642)
discovered independently by \cite{Miles:1999} and \cite{Torres:2003}
using Hipparcos photometry. It has been used to address the
controversy over the distance to the Pleiades, following the Hipparcos
determination of a value smaller than the canonical one
\cite[see][]{Munari:2004, Southworth:2005, Groenewegen:2007}. The
secondary of this SB2 is a metallic-line A star. Consequently, we used
a synthetic template with a metallicity of ${\rm [Fe/H]} = +0.5$ for
that star, which provided better results.  Nevertheless, a
statistically significant primary/secondary velocity offset remains in
our orbital solution, which we attribute to template mismatch stemming
from the anomalous chemical composition of the secondary.

\vni{245.~TRU~S115.} Our RVs for this rapid rotator show a hint of a
downward trend (about 2~\kms\ over 6000 days), which may or may not be
significant. The $e/i$ value is smaller than our threshold for
variability.

\vni{249.~HII~1645.} This is a non-member, and appears to have
variable RV.

\vni{251.~HII~1653.} Our RVs show this to be a single-lined binary with
a period of 1.5~yr.

\vni{254.~HII~1762.} The 10.7~yr spectroscopic orbit reported
here has a very large minimum secondary mass. The RVs measured are
believed to be for the secondary component. This is based on the broad
wings seen in the CCF, and the unexpectedly shallow spectral lines for
a star of near-solar temperature as determined from our spectra,
corresponding to a spectral type of G1. This suggests dilution of the
lines from the presence of a much more rapidly rotating star. This
interpretation is supported by the large $v \sin i$ of
180~\kms\ reported for HII~1762 by \cite{Uesugi:1970}, and the combined
color index from Gaia EDR3, which, after correction for reddening,
corresponds to an early F star (consistent with the \ion{A9}{5}
classification given in SIMBAD). The pair was resolved once by speckle
interferometry, and found unresolved on two other occasions. The
companion may be the same one detected spectroscopically.
\cite{Liu:1991} reported double lines in one of their two
observations, but the velocity difference they measured is far too
large to correspond to the 10.7~yr period.

\vni{261.~HII~1823.} The astrometric companion listed in the WDS was
discovered by lunar occultations.

\vni{264.~HII~1876.} The astrometric companion listed in the WDS was
discovered by lunar occultations. A new effective temperature estimate
from this paper gives $9580 \pm 200$~K. RVs have been published by
\cite{Torres:2020b}.

\vni{266.~HII~1912.} This target has no parallax or p.m.\ from Gaia
EDR3, and is listed as having a highly significant excess astrometric
noise. A poorly determined p.m.\ from the Gaia DR2 catalog does appear
consistent with membership, while the (also poor) parallax is smaller
than expected. It has been considered a cluster member by some authors
\citep[e.g.,][]{Schilbach:1995, Belikov:1998, Sampedro:2017}, and more
doubtful by others \citep{Olivares:2018}. A rotation period of 3.17
days by \cite{Oelkers:2018} seems to agree with the rotational
sequence for the Pleiades, arguing for membership.  A close (0\farcs2)
visual companion is listed in the WDS, and a wider one at 0\farcs85
was reported by \cite{Makarov:2001} from a special reduction of
Tycho-2 observations. The latter measurement is uncertain, and it is
unclear whether it corresponds to a different companion. In any case,
one or both of these companions probably explain the difficulty with
Gaia's astrometric solution. Our spectra suggest a blend of broad and
narrow lines. We report velocities for the narrow-lined star only,
which are constant and agree with the cluster mean.

\vni{267.~TRU~S127.} Our velocities indicate this very rapid rotator ($v
\sin i = 175$~\kms) is a long-period binary (see
Figure~\ref{fig:long}), but the orbit is as yet undetermined. As the
center-of-mass is not yet known, we have retained it as a possible
Pleiades member even though the Gaia parallax is formally different
from the cluster mean (but the p.m.\ is consistent with membership).
The Gaia EDR3 quality flags indicate the astrometric solution was
problematic, which may have affected the parallax.

\vni{269.~HII~2027.} See Section~\ref{sec:triples}.

\vni{274.~HII~2147.} Single-lined binary with an 18~yr orbit and
absolute mass determinations from \cite{Torres:2020}, from a
combination of astrometric and spectroscopic observations. The star
seen spectroscopically is the secondary.

\vni{277.~HII~2172.} A 30.2~day SB1 orbit was presented by
\cite{Mermilliod:1992a}. We have combined their observations with ours
to update the solution. We see no sign of the secondary in our
spectra.

\vni{278.~HII~2195.} The astrometric companion listed in the WDS was
discovered by lunar occultations.

\vni{282.~HII~2284.} \cite{Raboud:1998} reported this object as a
spectroscopic binary based on eight CORAVEL measurements, but did not
have enough velocities for an orbit. We confirm it to be a
single-lined binary with a period of 807 days.

\vni{283.~HII~2278.} A 0\farcs4 visual companion discovered by
\cite{Bouvier:1997} is also listed in the Gaia EDR3 catalog, but
has no parallax or p.m.

\vni{290.~HII~2406.} A 33~day single-lined orbit was reported by
\cite{Mermilliod:1992a}. We have detected the secondary in our TRES
spectra, and updated the solution using all measurements.

\vni{291.~HII~2407.} \cite{Mermilliod:1992a} reported the first orbit
for this SB1 with a 7~day period. It was recently discovered by
\cite{David:2015} to be an eclipsing binary based on observations from
NASA's K2 mission. We combine the original CORAVEL velocities
\citep[as transformed to the IAU system by][]{Mermilliod:2009} with
our own, more numerous measurements, to provide an improved
solution. The secondary is very faint and is not seen in our spectra.

\vni{297--299.~HII~2500, HII~2503, HII~2507.} These three objects may
well form a multiple system. The brighter star, HII~2507, is an SB2
($P = 16.7$~days) first announced as an SB1 by \cite{Abt:1965} and
later by \cite{Pearce:1975}.  The secondary has been detected here for
the first time. The two astrometric companions to HII~2507 reported in
the WDS correspond to HII~2503 (3\farcs3) and HII~2500 (10\farcs1). The
latter is itself a close visual binary (0\farcs3), and also an SB1
with a period of about 6.5~yr, although these two companions cannot be
the same, making the object at least a triple. The variability of
HII~2500 was first noticed by \cite{Raboud:1998}, who did not have
enough observations for an orbit. There is no indication of a change
in the width of the CCF for HII~2500, as might be expected from the
0\farcs3 companion. If all of these stars are physically bound, this
would be a sextuple system.

\vni{300.~HCG~384.} This is a 542~day SB2 with a faint secondary, for
which we report an orbit here. A wide 13\farcs9 companion listed in
Gaia EDR3 may not be physical: its parallax and p.m. are somewhat
different from those of the target.

\vni{303.~HII~2601.} A 1\farcs9 companion is listed in the Gaia EDR3
catalog.

\vni{316.~HII~2881.} \cite{Queloz:1998} reported this as a suspected
long-period double-lined binary. Our velocities show no significant
change, and the $e/i$ diagnostic incorporating the CORAVEL
measurements is below our threshold for variability.

\vni{328.~HII~3104.} Our RVs show this to be a 3.6~yr SB1. The
declination component of the p.m.\ from Gaia EDR3 is rather
different from the mean for the cluster, suggesting the object may not
be a member. However, the quality flags indicate the astrometric
solution may be severely disturbed by the companion, so membership
cannot yet be ruled out. The center-of-mass velocity is not far from
the expected value.

\begin{figure*}
\epsscale{1.15}
\plotone{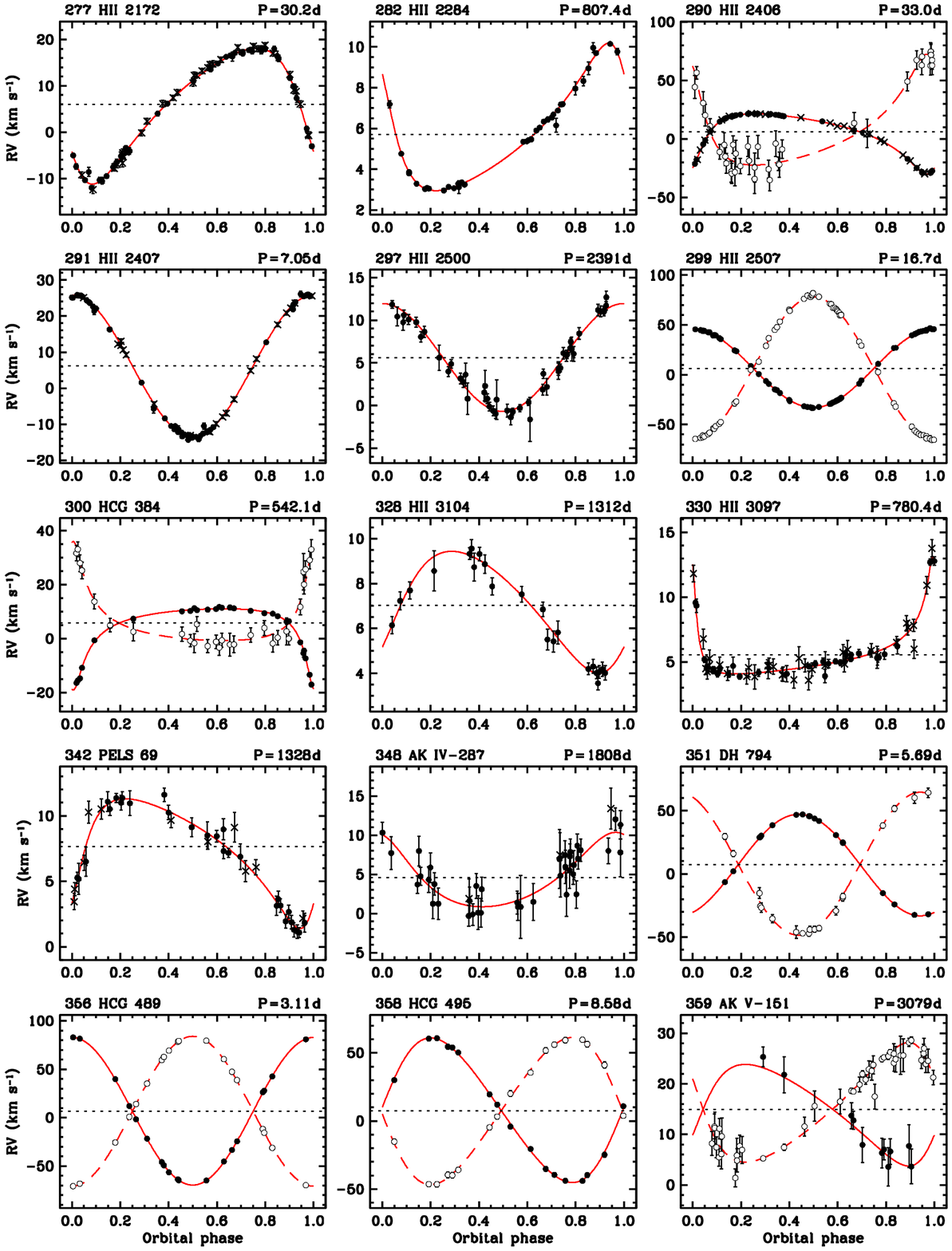}
\figcaption{Plots of orbital solutions (continued).\label{fig:sb.3}}
\end{figure*}

\vni{330.~HII~3097.} A 2~yr SB1 orbit was published by
\cite{Mermilliod:1992a} with a very high eccentricity of $e = 0.78$.
The better phase coverage of our observations improves the solution
significantly. Table~\ref{tab:sb1} and Figure~\ref{fig:sb.3} present
the combined fit using both data sets, augmented with additional
CORAVEL measures by \cite{Mermilliod:2009}.

\vni{332.~HII~3163.} The shape of the CCF suggests a blend of a
broad and narrow peak. Our RVs correspond to the narrow peak. No
astrometric companions have been reported.

\vni{338.~HII~3197.} This is a close visual binary with a 30.2~yr
astrometric orbit reported by \cite{Schaefer:2014}. The semimajor axis
is 0\farcs08. A wider companion at 0\farcs5 has also been found,
making this a triple system. Our spectroscopic measurements correspond
to the combined light.

\vni{342.~PELS~69.} This is a non-member according to Gaia. We find
it to be an SB1 with a period of 3.6~yr. Our solution combines our own
measurements with about a dozen older ones from the CORAVEL
\citep{Mermilliod:2009}.

\vni{348.~AK~IV-287.} Our observations show that this is an SB1 with a
5~yr period.

\vni{349.~TRU~S177.} The parallax and p.m.\ information in the Gaia
EDR3 catalog casts doubt on the membership of this object. However,
there are indications that the astrometric solution may have been
disturbed.  Our RVs display a slow downward drift of
4~\kms\ indicating binarity, which may be the cause of the excess
astrometric noise. We retain it as a possible member.

\vni{351.~DH~794.} We find this to be a 5.7~day SB2, with a secondary
that is too faint for us to establish its temperature and $v \sin i$
independently from our spectra. The corresponding template was
selected with help from the cluster isochrone.

\vni{352.~AK~IV-314.} The 1\farcs0 astrometric companion listed in the
WDS has an entry in the Gaia EDR3 catalog, but no parallax or
p.m.\ are reported.

\vni{356.~HCG~489.} We find this to be a 3~day double-lined binary.

\vni{358.~HCG~495.} Our observations show this is an 8.5~day SB2.

\vni{359.~AK~V-151.} We find this to be an 8.4~yr SB2, in which the
rapidly rotating primary is only detectable in our TRES spectra. Gaia
EDR3 indicates this is a background object; the center-of-mass
velocity from our orbital solution is far from the cluster mean.

\vni{366.~AK~V-198.} \cite{Mermilliod:1997} reported this as a
double-lined binary, but lacked enough observations for an orbit. We
now report an SB2 orbit with a period of 176 days. The parallax and
p.m.\ information in the Gaia EDR3 catalog are somewhat different than
the cluster mean, although evidence from a statistically significant
excess astrometric noise and a large value of 2.797 for the {\tt RUWE}
parameter indicate the astrometric solution may have been affected.
The rotation period measured by \cite{Rebull:2016a, Rebull:2016b} is
consistent with membership. We retain it as a possible member.

\vni{367.~TRU~S184x.} This is a non-member according to Gaia. It was
observed 25 times with the CORAVEL by \cite{Rosvick:1992}, and
although the measurements show considerable scatter, the authors were
unable to decide whether it is a binary. The effective temperature and
projected rotational velocity of the object are near the limit of the
instrumental capabilities of northern CORAVEL.  Our own observations
show less scatter, but have not clarified the picture. Using only our
TRES measurements, which have higher precision (except for the first,
with an uncertainty twice as large as the others), we are able to
obtain a marginally significant orbital solution with a period of
about 6~yr and a semiamplitude of only 1.5~\kms, which we consider too
tentative to report. The apparent variability could simply be due to
stellar activity.

\vni{369.~TRU~S185.} The Gaia EDR3 catalog lists a companion at
7\farcs4 with similar parallax and p.m.\ as the target.

\vni{372.~TRU~S194.} RVs along with a 10~yr SB1 orbit have been reported
by \cite{Torres:2020b} for this rapidly rotating B star.  Our new
effective temperature estimate is $10500 \pm 300$~K.

\vni{375.~PELS~173.} Information from Gaia EDR3 indicates this is a
non-member. Adding the CORAVEL observations to ours, there is a hint
of an upward drift in the velocities. The median RV from Gaia is
consistent with this.

\clearpage

\section{2. Graphical Representations of the Binary Orbital Solutions and Long-term Trends}

\begin{figure}
\epsscale{0.4}
\plotone{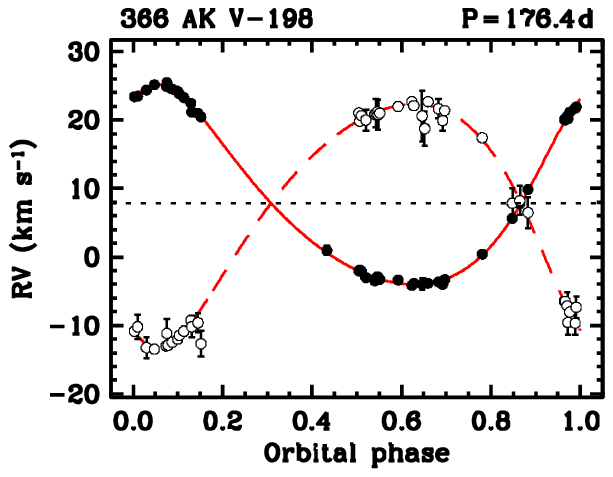}
\figcaption{Plots of orbital solutions (continued).\label{fig:sb.4}}
\end{figure}

\begin{figure*}
\epsscale{1.15}
\plotone{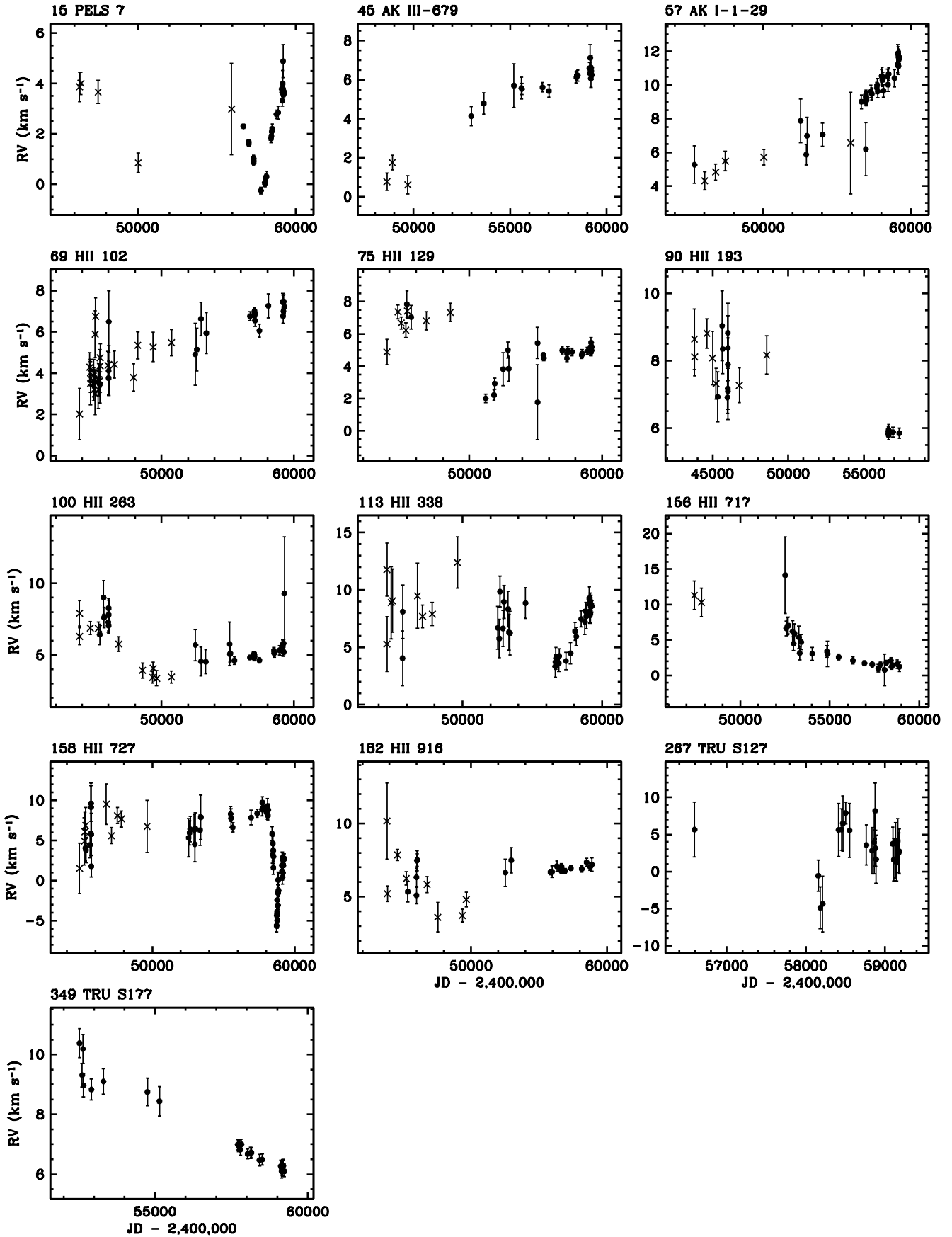}
\figcaption{Plots for objects with long-term RV trends. Filled circles
  are used for CfA measurements, and crosses for velocities from
  CORAVEL or other sources..\label{fig:long}}
\end{figure*}

\clearpage


\end{document}